\documentclass[12pt,twoside]{report}

    \makeatletter 
    \def\cleardoublepage{\clearpage\if@twoside \ifodd\c@page\else%
    \hbox{}%
    \thispagestyle{empty}
    \newpage%
    \if@twocolumn\hbox{}\newpage\fi\fi\fi} 
    \makeatother

\usepackage[spanish]{babel} 
\usepackage[latin1]{inputenc}
\usepackage[dvips]{graphicx}
\usepackage{mathrsfs}
\usepackage{latexsym}

\usepackage{amsmath}

\usepackage{psfrag}

\usepackage{amssymb}

\usepackage{natbib}
\bibpunct{(}{)}{;}{a}{}{,} 

\begin{document}
%


 \renewcommand{\listtablename}{{\'I}ndice de tablas}
 \renewcommand{\tablename}{Tabla}

\def \mnras {MNRAS}

\def \apj {ApJ}

\def \prd {PRD}

\def \pra {PRA}

\def \apjl {ApJL}

\def \apjs {ApJS}

\def \aap {A\&A}

\def \prl {PRL}


\pagestyle{empty}
\title{Variaci\'on de la constante de estructura fina y de la masa del electr\'on en el Universo primitivo}

\vspace{5.0cm}

\author{{\sc  Claudia Graciela Sc\'occola}\\ \\ \\ \\
Director: Dr. H\'ector Vucetich \\ \\
Co-directora: Dra. Susana J. Landau 
 \\ \\ \\ \\ 
\centerline{Tesis presentada para optar por el grado de DOCTOR EN ASTRONOMÍA}
\\ \\ \\
        Facultad de Ciencias Astron\'omicas y Geof\'{\i}sicas \\ \\
       Universidad Nacional de La Plata }

\date{15 de Mayo de 2009}
\maketitle

\thispagestyle{empty}



\vspace*{6.0cm}




\begin{abstract}

En esta tesis, presentamos un estudio de la variaci\'on de la masa del
electr\'on $m_e$, y de la constante de estructura fina $\alpha$, en
distintas etapas de la evoluci\'on del Universo.

Estudiamos la \'epoca de formaci\'on del hidr\'ogeno neutro.
Analizamos el escenario de recombinaci\'on incluyendo los detalles en
la recombinaci\'on del helio, para encontrar las dependencias de las
cantidades f{\'\i}sicas con las constantes fundamentales. Modificamos
las ecuaciones para las fracciones de ionizaci\'on, para
poder considerar valores arbitrarios de las constantes.

Utilizando los datos actualizados del fondo c\'osmico de radiaci\'on,
y el espectro de potencias de la distribuci\'on de galaxias del
cat\'alogo 2dFGRS, buscamos l{\'\i}mites a la posible variaci\'on de
las constantes en la \'epoca de recombinaci\'on. Estimamos el valor de
un conjunto de par\'ametros cosmol\'ogicos, junto con la variaci\'on
de las constantes. Analizamos la variaci\'on de $\alpha$ y $m_e$ por
separado, y el caso en que ambas  var{\'\i}an
simult\'aneamente. Presentamos resultados actualizados y comparamos
con las cotas existentes. Estimamos tambi\'en una relaci\'on
fenomenol\'ogica entre la variaci\'on de $\alpha$ y la variaci\'on de
$m_e$ entre el desacople y la actualidad.

Analizamos el modelo fenomenol\'ogico de Barrow-Magueijo, que propone
una variaci\'on de la masa del electr\'on inducida por cambios de un
campo escalar en el espacio-tiempo. Mejoramos las soluciones
originalmente presentadas por dichos autores. Estimamos los
par\'ametros del modelo usando los l{\'\i}mites para la variaci\'on de
la masa del electr\'on que se obtienen en esta tesis a partir de los
datos del fondo c\'osmico de radiaci\'on, y otros datos astron\'omicos
y de laboratorio. Por otra parte, utilizando resultados de
experimentos que testean la validez del Principio de Equivalencia
D\'ebil, encontramos una cota para el par\'ametro del modelo, que
viene de la variaci\'on en la masa del electr\'on inducida por cambios
espaciales del campo escalar. Comparando ambos l{\'\i}mites, se llega
a la conclusi\'on de que el modelo debe ser descartado.

Por \'ultimo, reanalizamos los efectos de la variaci\'on de $\alpha$ o
$m_e$ en los radios planetarios.  Explicamos por qu\'e la variaci\'on
de la masa del electr\'on no induce cambios en el radio factibles de
ser observados. Presentamos una cota local e independiente para la
variaci\'on de $\alpha$, obtenida a partir de principios muy generales.

\end{abstract}

\section*{Agradecimientos}

Quiero agradecer a mis directores, H\'ector Vucetich, y Susana Landau,
por el apoyo, la confianza y la libertad con la que me permitieron
trabajar durante el desarrollo de esta tesis.

Agradezco tambi\'en al Grupo de Evoluci\'on Estelar de la FCAG, en
especial a  Leandro Althaus y Alejandro C\'orsico, por haberme dado la
posibilidad de iniciar mi doctorado en su grupo, y la comprensi\'on y
apoyo cuando decid{\'\i} empezar de nuevo, para trabajar en
Cosmolog{\'\i}a.

Agradezco a Nelson Padilla, profesor en la PUC de Chile, y a Juan Veliz,
por facilitarme recursos computacionales con los que se hicieron gran
parte de los c\'alculos de esta tesis, y a los responsables del cluster
Kanbalam, en la UNAM de M\'exico, con la que se hicieron el resto de
los c\'alculos.  Mi agradecimiento es tambi\'en para Ariel S\'anchez,
por ayudarme a dar mis primeros pasos en el estudio del c\'odigo
{\sc cosmomc}. Las discusiones con Licia Verde durante la escuela 
"The Large Scale Structure-CMB Connection" en Santiago de Chile, en
Marzo de 2007, fueron fundamentales para mi entendimiento de las
cadenas de Markov y su utilizaci\'on en CMB.

Un agradecimiento especial a los geof{\'\i}sicos Claudia Ravazzoli y
Germ\'an Rubino, por ayudarme a buscar un modelo actualizado de la
estructura interna de la Tierra.

Gracias a mis amigos del Observatorio de La Plata, por la
compa\~n{\'\i}a, la paciencia para escuchar, las discusiones y el
apoyo: Anabella Araudo, Cecilia Fari\~na, Anahi Granada, Andrea
Fortier, Ver\'onica Firpo, Ileana Andruchow, Gonzalo De El{\'\i}a,
Marcelo Miller Bertolami, Jorge Panei...

Quiero agradecer a mi familia, por el apoyo incondicional. Y agradezco
principalmente a Mauricio Sturla, a quien le dedico esta tesis.

Finalmente, agradezco al CONICET por la beca doctoral otorgada durante
el per{\'\i}odo Abril 2004 - Marzo 2009.

\pagestyle{plain}
\pagenumbering{roman}


%
%
\newpage

\tableofcontents
\listoffigures                     

\newpage

\newpage
\pagenumbering{arabic}
%
%


\cleardoublepage
\chapter{Introducci\'on}
\label{chap:intro}

La  descripci\'on actual del Universo tiene su fundamento en dos
teor{\'\i}as bien establecidas: la relatividad general, y el modelo
est\'andar de f{\'\i}sica de part{\'\i}culas. En ambas existen
par\'ametros, llamados ``constantes fundamentales'', cuyos valores no
quedan determinados a partir de las teor{\'\i}as mismas, sino que
deben ser hallados experimentalmente. Tal vez, uno de los objetivos
m\'as ambiciosos de la f{\'\i}sica moderna sea formular una
teor{\'\i}a del ``todo'', una teor{\'i}a que unifique todas las
interacciones de la naturaleza, donde los valores de dichos
par\'ametros puedan ser deducidos a partir de primeros principios.
Hasta hoy, se han podido unificar las interacciones
electromagn\'eticas con la fuerza d\'ebil en las interacciones
electro-d\'ebiles, y \'estas con las interacciones fuertes, en lo que
se conoce como Teor{\'\i}as de Gran Unificaci\'on. Sin embargo, a\'un
no se ha logrado encontrar una teor{\'\i}a cu\'antica de la gravedad.

Las teor{\'\i}as que intentan unificar la gravedad con las dem\'as
interacciones son de variada naturaleza. Entre ellas se encuentran las
teor{\'\i}as de super-cuerdas
\citep{Wu86,Maeda88,Barr88,DP94,DPV2002a,DPV2002b}, teor{\'\i}as de
mundos brana \citep{Youm2001a,Youm2001b,branes03a,branes03b} y
teor{\'\i}as de Kaluza-Klein
\citep{Kaluza,Klein,Weinberg83,GT85,OW97}.
Las energ{\'\i}as involucradas en los fen\'omenos que permiten corroborar
la unificaci\'on son tan altas, que las posibles predicciones de las
teor{\'\i}as ser{\'\i}an muy
dif{\'\i}ciles de verificar. Sin embargo, algunas consecuencias a
bajas energ{\'\i}as son factibles de ser comprobadas a trav\'es de
experimentos u observaciones. Una de las predicciones m\'as
importantes de estas teor{\'\i}as es que el valor de las constantes
fundamentales puede variar en distintos puntos del espacio-tiempo. La
dependencia de las constantes con el tiempo es diferente para
distintas teor{\'\i}as. Es por ello que a trav\'es de la comparaci\'on
con resultados de experimentos y observaciones es posible poner cotas
a las mismas.

Por otra parte, existen teor{\'\i}as efectivas a bajas energ{\'\i}as,
formuladas para describir la variaci\'on de determinadas constantes
fundamentales. En este caso, se asocia el valor de las constantes a un
campo en el espacio-tiempo, cuyo Lagrangiano se propone de manera tal
de cumplir principios muy generales y las ecuaciones de movimiento
tienen soluciones que pueden ser comparadas con datos
observacionales. A modo de ejemplo, mencionamos la teor{\'\i}a de
Bekenstein para la variaci\'on temporal de la constante de estructura
fina \citep{Bekenstein82} y el modelo de Barrow \& Magueijo para la
variaci\'on de la masa del electr\'on \citep{BM05}.

Se ha estudiado tambi\'en la variaci\'on de otras constantes
fundamentales. \citet{AlbMag99} introdujeron un
modelo cosmol\'ogico en el cual la velocidad de la luz es variable,
con el fin de resolver los problemas b\'asicos de la cosmolog{\'\i}a
(horizonte, planaridad, constante cosmol\'ogica, entrop{\'\i}a y
homogeneidad) sin acudir al paradigma inflacionario. Este modelo fue
extendido por varios autores
\citep{ClaytonMoffat01,ClaytonMoffat00,ClaytonMoffat99,KehagiasKiritsis99}.
Asimismo, existen teor{\'\i}as que predicen la variaci\'on de la
constante de gravitaci\'on universal $G$ \citep{BraxDavis01}.

Una de las constantes fundamentales cuya variaci\'on ha despertado
mayor inter\'es es la constante de estructura fina $\alpha$, que
est\'a relacionada con la constante de acoplamiento de las
interacciones electromagn\'eticas. \'Esta es una de las constantes
estudiadas en esta tesis.  Tambi\'en ha suscitado inter\'es la posible
variaci\'on de la constante de las interacciones d\'ebiles,
parametrizada por la constante de acoplamiento de Fermi $G_F$.  Sin
embargo, como se enfatiza en \citet{dixit88}, la constante de Fermi no
es una constante fundamental en el mismo sentido que lo es la
constante de estructura fina. Dado que $G_F \propto \left<\phi \right
>^{-2}$, es m\'as apropiado examinar la variaci\'on del valor de
expectaci\'on de vac{\'\i}o del campo Higgs $\left<\phi \right>$. En
particular, nos hemos concentrado en la variaci\'on de la masa en
reposo del electr\'on $m_e$, ya que es la \'unica consecuencia que
tiene la variaci\'on del Higgs en las etapas del Universo en las que
esta tesis se focaliza.

Desde el punto de vista observacional, el tema de la variaci\'on de
las constantes es de gran inter\'es. Son muchos los esfuerzos para
poner cotas a la variaci\'on de constantes como $\alpha$ y $m_e$, o el
cociente $\mu= m_e/m_p$, entre la masa del electr\'on y la masa del
prot\'on. Existen m\'etodos locales, astron\'omicos y
cosmol\'ogicos. Cada m\'etodo considera el comportamiento temporal de las
constantes a distintas escalas de energ{\'\i}a y de distancia, y con
distinta precisi\'on, resultando todos de gran utilidad.  Localmente,
se ponen cotas a la tasa de variaci\'on de las constantes usando
relojes at\'omicos
\citep{Bize03,Fischer04,Peik04,PTM95,Sortais00,Marion03}.
 La variaci\'on temporal de constantes a corrimientos al rojo (en adelante, redshifts $z$) del orden de $z
\simeq 0.14$ se puede acotar con datos del reactor nuclear natural que
oper\'o en Oklo \citep{Fujii00,DD96}. Otras cotas se obtienen a partir
de la vida media de emisores de part{\'\i}culas $\beta$
\citep{Olive04b}. Por otra parte, datos astron\'omicos recientes,
basados en el an\'alisis de espectros de absorci\'on de quasares a
alto redshift, sugirieron una posible variaci\'on de $\alpha$ y $\mu$
\citep{Webb99,Webb01,Murphy01a,Murphy01b,Murphy03b,Ivanchik05,Tzana07}.
Sin embargo, otros estudios de datos similares, arrojan resultados
nulos \citep{MVB04,QRL04,Bahcall04,Srianand04}. 

Para estudiar el Universo temprano, y contrastar modelos alternativos a
la cosmolog{\'\i}a est\'andar, se pueden utilizar los datos de las
abundancias primitivas de elementos qu{\'\i}micos livianos, producidos
durante la nucleos{\'\i}ntesis primordial (BBN), cuando el Universo
ten{\'\i}a $\sim 3$ minutos de edad; y los datos del fondo c\'osmico de
radiaci\'on (CMB), que proveen informaci\'on de la \'epoca de
formaci\'on del hidr\'ogeno neutro, 400.000 a\~nos despu\'es del Big
Bang.  Dado que los procesos f{\'\i}sicos relevantes en ambas
\'epocas son sensibles al valor de las constantes fundamentales, estos
observables sirven como herramientas para poner l{\'\i}mites a su
variaci\'on temporal. A pesar de que las cotas obtenidas a partir de
BBN y de CMB son m\'as d\'ebiles que las mencionadas anteriormente,
resultan sumamente importantes debido a que se refieren a las primeras
etapas del Universo.

En esta tesis se realiz\'o un estudio a nivel cosmol\'ogico y local de
la variaci\'on temporal de dos constantes: la constante de estructura
fina y la masa del electr\'on.  Uno de los objetivos m\'as importantes
fue el estudio de la variaci\'on de estas constantes durante la
\'epoca en la cual el plasma de electrones e iones del Universo se
 recombin\'o. Esta \'epoca se conoce con el nombre
 de \emph{recombinaci\'on del Universo}.

 El estudio de las dependencias con las constantes fundamentales de
las cantidades f{\'\i}sicas relevantes durante la formaci\'on del
hidr\'ogeno neutro se detalla en el
cap{\'\i}tulo~\ref{ch:fisica_recomb}. All{\'\i}, adem\'as, se analiza
por primera vez la influencia las constantes fundamentales en el
escenario de recombinaci\'on actualizado, que tiene en cuenta la
recombinaci\'on detallada del helio, el efecto del hidr\'ogeno neutro
en la recombinaci\'on de \'este y la contribuci\'on de las
transiciones desde estados tripletes del hidr\'ogeno para la
recombinaci\'on.

Utilizando el conjunto m\'as completo y actualizado de datos del fondo
c\'osmico de radiaci\'on, junto con datos del cat\'alogo de galaxias
2dFGRS para el espectro de potencias de la distribuci\'on de la
materia, se han obtenido cotas para la variaci\'on de $\alpha$ y $m_e$
en la \'epoca de formaci\'on del hidr\'ogeno neutro. Se consider\'o la
variaci\'on de dichas constantes de manera separada, y tambi\'en se
realiz\'o un an\'alisis de la variaci\'on conjunta de ambas, de manera
totalmente fenomenol\'ogica, sin suponer ninguna relaci\'on entre las
variaciones, de manera tal que las conclusiones obtenidas en esta
tesis resultan independientes de cualquier modelo. El \'unico trabajo
anterior donde se consider\'o la variaci\'on conjunta de dichas
constantes es el de \citet{Ichi06}, en el cual las
variaciones est\'an relacionadas en el contexto de modelos
dilat\'onicos de teor{\'\i}as de cuerdas. En cambio, el resultado
mostrado en esta tesis se obtiene suponiendo que las constantes
var{\'\i}an en forma independiente una de la otra. A partir del
ajuste, obtenemos una relaci\'on fenomenol\'ogica entre las
variaciones de $\alpha$ y $m_e$ en la \'epoca de recombinaci\'on.
Comparamos con el resultado de \citet{Ichi06}, y conclu{\'\i}mos que la
propuesta de estos autores deber{\'\i}a ser descartada.

Los datos m\'as recientes del fondo c\'osmico de radiaci\'on que
hab{\'\i}an sido utilizados en esta tesis fueron los recolectados en
los tres primeros a\~nos de funcionamiento del sat\'elite WMAP
(WMAP3). En marzo de 2008, el sat\'elite liber\'o datos recolectados
durante los primeros 5 a\~nos (WMAP5). De manera simult\'anea, el
escenario de recombinaci\'on fue actualizado, considerando procesos
f{\'\i}sicos que hab{\'\i}an sido despreciados en el c\'alculo
est\'andar, y que afectan de manera sutil, aunque no despreciable, la
historia de recombinaci\'on del Universo. Por esta raz\'on, se
repiti\'o el an\'alisis estad{\'\i}stico para la variaci\'on de
$\alpha$ y $m_e$, incluyendo los datos de WMAP5, y el nuevo escenario
de recombinaci\'on, y se compar\'o con los resultados obtenidos con
los datos de WMAP3, y el escenario de recombinaci\'on est\'andar.

 En el Cap{\'\i}tulo~\ref{chap:CMB} se muestran los distintos ajustes
realizados. Dado que se han llevado a cabo an\'alisis estad{\'\i}sticos
variando no s\'olo las constantes fundamentales, sino tambi\'en el
conjunto b\'asico de par\'ametros cosmol\'ogicos, en esta tesis se ha
abarcado el estudio y adaptaci\'on de un programa num\'erico que
utiliza cadenas de Markov Monte Carlo para muestrear el espacio de
par\'ametros, y hacer un an\'alisis estad{\'\i}stico desde un enfoque
bayesiano. Este m\'etodo permite la exploraci\'on del espacio de
par\'ametros en un tiempo computacionalmente m\'as accesible que el
m\'etodo frecuentista del c\'alculo de la Likelihood en cada punto de
una grilla en el espacio de par\'ametros. En dicho cap{\'\i}tulo
tambi\'en se muestran y comparan los resultados obtenidos.

Las cotas observacionales para la variaci\'on de las constantes
fundamentales son una herramienta importante para evaluar teor{\'\i}as,
por ejemplo, aquellas que intentan unificar las cuatro interacciones
de la naturaleza. En el Cap{\'\i}tulo~\ref{chap:BM} se analiza el
modelo de Barrow \& Magueijo, que predice la variaci\'on de la masa
del electr\'on como funci\'on del tiempo, a partir de la existencia de
un campo escalar.  Se obtuvieron mejores soluciones aproximadas del modelo, lo
cual se logr\'o teniendo en cuenta de manera m\'as rigurosa la
evoluci\'on de la expansi\'on del Universo, y sin despreciar t\'erminos
en las ecuaciones diferenciales que describen la evoluci\'on del campo
escalar.  Luego, se utilizaron las cotas para la variaci\'on de $m_e$
obtenidas en esta tesis a partir de los datos del fondo c\'osmico de
radiaci\'on, junto con otras cotas dadas a diferentes tiempos
cosmol\'ogicos, para poner l{\'\i}mites a los par\'ametros del
modelo. Adicionalmente, se derivaron cotas a partir de resultados de
experimentos que testean la validez del Principio de Equivalencia
D\'ebil, y se compararon con las anteriores. A la luz de los
resultados, se lleg\'o a la conclusi\'on de que el modelo de Barrow \&
Magueijo debe ser descartado.

Con el objetivo de estudiar la variaci\'on de las constantes
fundamentales a nivel local, tanto en el espacio como en el tiempo, se
realiz\'o un estudio en escalas correspondientes al sistema solar. Se
analiz\'o las consecuencias que un valor diferente de $\alpha$ y/o
$m_e$ acarrean para el radio de la Tierra, la Luna y el planeta
Mercurio.  A partir de consideraciones generales, se deduce una
expresi\'on para la variaci\'on de este par\'ametro
planetario. Utilizando los modelos m\'as actuales de la estructura
interna de dichos cuerpos celestes y l{\'\i}mites existentes para la
variaci\'on del radio de estos planetas, se dan cotas a la variaci\'on
de las constantes fundamentales entre la actualidad y la \'epoca de
formaci\'on del sistema solar. Asimismo, se analiza cuales de las
constantes fundamentales estudiadas en esta tesis resultan relevantes
en este problema. Los resultados se presentan en el
Cap{\'\i}tulo~\ref{chap:radios}.

Finalmente, en el Cap{\'\i}tulo~\ref{chap:conclusiones}, se resumen
los resultados de esta tesis, y se destacan las conclusiones m\'as
relevantes.

\vspace{1cm}

Parte de los resultados de esta tesis fueron incluidos en las
siguientes publicaciones en revistas internacionales con referato:

\begin{itemize}

\item ``Time variation of the fine structure constant and the Bekenstein model''. Mosquera, M. E., Sc\'occola, C. G., Landau, S. J., and Vucetich, H.  {\it Astronomy and Astrophysics} {\bf 478}, 675 (2008).

\item ``Time variation of the electron mass and the Barrow-Magueijo model''. Sc\'occola, C. G., Mosquera, M. E., Landau, S. J., and Vucetich, H.  {\it Astrophysical Journal} {\bf 681}, 737 (2008).

\item  ``Early Universe constraints on time variation of fundamental constants''. Landau, S. J., Mosquera. M. E., Sc\'occola, C. G., and Vucetich, H. {\it Physical Review D} {\bf 78 (8)}, 083527 (2008).

\item ``WMAP 5-year constraints on time variation of $\alpha$ and $m_e$ in a detailed recombination scenario''.  Sc\'occola, C. G., Landau, S. J., and Vucetich, H. {\it Physics Letters B}  {\bf 669}, 212 (2008).

\end{itemize}


\cleardoublepage
\chapter{F{\'\i}sica de Recombinaci\'on}
\label{ch:fisica_recomb}

En el modelo cosmol\'ogico est\'andar, la \'epoca de recombinaci\'on es
aquella cuando el Universo se vuelve lo suficientemente fr{\'\i}o como
para que los protones, y n\'ucleos de elementos qu{\'\i}micos
livianos, sean capaces de retener electrones y se formen los \'atomos
neutros. Los electrones, capturados en distintos niveles de
energ{\'\i}a, decaen al nivel fundamental. Sin embargo, este proceso
se ve impedido por las ionizaciones y excitaciones que se producen
r\'apidamente gracias al enorme reservorio de fotones de bajas
energ{\'\i}as que hay en el Universo. Adem\'as, los decaimientos del
nivel $n=2$ al nivel $n=1$ y las recombinaciones directas desde el
continuo al nivel fundamental se ven fuertemente disminu{\'\i}dos
debido a que en ambos casos el fot\'on emitido tiene una energ{\'\i}a
en el rango donde existen muy pocos fotones de fondo, y por lo tanto
ser\'a inmediatamente absorbido por un \'atomo cercano que est\'e en
el nivel fundamental. Estos fen\'omenos traen como consecuencia que la
recombinaci\'on no sea instant\'anea.  Eventualmente, los \'atomos
alcanzan el estado fundamental por el corrimiento al rojo de los
fotones Ly$\alpha$ ($2p \rightarrow 1s$) causado por la expansi\'on, o
por la transici\'on $2s \rightarrow 1s$ via dos fotones. Sin embargo,
el Universo se expande y enfr{\'\i}a antes de que la recombinaci\'on
termine, y queda una peque\~na fracci\'on de electrones y n\'ucleos
libres remanente.

El elemento m\'as abundante en el Universo es el hidr\'ogeno, y en
menor proporci\'on el helio, siendo las abundancias de los dem\'as
elementos qu{\'\i}micos formados en la nucleos{\'\i}ntesis primordial
despreciables para el proceso de la recombinaci\'on. Los primeros
estudios en este campo fueron realizados por \citet{Peebles68} y \citet{zeldovich68}. La
metodolog{\'\i}a est\'andar era considerar un \'atomo de tres niveles
(fundamental, $n=2$ y continuo) con los niveles con $n>2$
representados por un coeficiente de recombinaci\'on.  Se derivaba una
ecuaci\'on diferencial ordinaria para la fracci\'on de ionizaci\'on,
pero deb{\'\i}an hacerse muchas suposiciones para llegar a \'esta: i)
el \'unico elemento en el Universo es el hidr\'ogeno, ii) sus estados
excitados est\'an en equilibrio con la radiaci\'on, iii) se puede
despreciar la emisi\'on estimulada para la transici\'on $2p\rightarrow
1s$ (l{\'\i}nea Ly$\alpha$), iv) se puede usar un coeficiente de
recombinaci\'on simple, v) cada recombinaci\'on neta resulta en un
\'atomo en el nivel fundamental, vi) el corrimiento al rojo de los
fotones Ly$\alpha$ se puede tener en cuenta usando una probabilidad de
escape simple, y vii) los procesos colisionales son despreciables. Este
tratamiento de la recombinaci\'on fue usado durante treinta a\~nos con
pocas modificaciones. Motivados por la expectativa de medir con gran
precisi\'on las anisotrop{\'\i}as de la radiaci\'on c\'osmica de fondo
con sat\'elites y otros experimentos dise\~nados a tal fin, y gracias
al poder de c\'omputo moderno,  \citet{seager00}
presentaron un c\'alculo detallado de la recombinaci\'on del
hidr\'ogeno y del helio. El tratamiento consiste en considerar
\'atomos con m\'ultiples niveles, y tener en cuenta todas las
transiciones entre ellos, y tambi\'en con el continuo, y tratar la
recombinaci\'on del hidr\'ogeno y del helio en forma simult\'anea,
teniendo en cuenta la evoluci\'on de la poblaci\'on de electrones
libres y de la temperatura de la materia. Adem\'as,
en \citet{seager99}, estos autores presentaron un conjunto de
ecuaciones que permiten reproducir los resultados de los c\'alculos
detallados via parametrizaciones en los coeficientes de
recombinaci\'on. En esta tesis llamaremos escenario de recombinaci\'on
est\'andar a este tratamiento de la misma.

M\'as recientemente, la recombinaci\'on del helio fue tratada con
m\'as precisi\'on. Switzer y Hirata en una serie de tres trabajos
\citep{SH08a,SH08b,SH08c}, discutieron varios aspectos de la
recombinaci\'on y estudiaron los efectos de la retro-alimentaci\'on
producida por las distorsiones espectrales, las l{\'\i}neas
semi-prohibidas y prohibidas, el transporte radiativo en las
l{\'\i}neas del He {\sc I} con redistribuci\'on parcial, y la opacidad
del continuo producida por la fotoionizaci\'on del hidr\'ogeno. El
efecto del hidr\'ogeno neutro sobre la recombinaci\'on del helio es
acelerar la recombinaci\'on de \'este respecto de los resultados
obtenidos en Seager et al, debido a que fotones resonantes del He {\sc
I} ionizan \'atomos neutros de hidr\'ogeno antes de ionizar otro
\'atomo de helio, resultando en una recombinaci\'on neta de
helio. Este \'ultimo efecto fue tratado tambi\'en por \citet{KIV07}, quienes propusieron adem\'as una f\'ormula aproximada \'util
para ser inclu{\'\i}da en c\'odigos como R{\sc ecfast}, que resuelve
num\'ericamente las ecuaciones de recombinaci\'on. Nos referiremos a
este tratamiento como escenario detallado.

\section{Ecuaciones de recombinaci\'on}
\label{sec:eq_recomb}

\def \H {\mathrm{H}}
\def \He {\mathrm{He}}
\def \HeI {\mathrm{HeI}}
\def \HeII {\mathrm{HeII}}

El conjunto de ecuaciones de recombinaci\'on, conformado por dos
ecuaciones diferenciales ordinarias para la fracci\'on de ionizaci\'on
de $\mathrm{H}$ y de $\mathrm{He\ I}$ y una ecuaci\'on para la
temperatura de la materia $T_M$, se resuelven simult\'aneamente en el
c\'odigo num\'erico R{\sc ecfast}. Los coeficientes de recombinaci\'on
en cada caso se parametrizan para reproducir los resultados de los
an\'alisis de m\'ultiples niveles realizados por Seager et al (para el
hidr\'ogeno) y Switzer \& Hirata (para el helio), este \'ultimo
teniendo en cuenta la aproximaci\'on de Kholupenko et al para la
opacidad del continuo del $\mathrm{H}$ sobre el $\mathrm{He\
  I}$. La ecuaci\'on de equilibrio de Saha es una buena aproximaci\'on
para la recombinaci\'on del $\mathrm{He\ II}$.  El conjunto de
ecuaciones es el siguiente:

\begin{equation}
H(z)(1+z)\frac{dx_p}{dz} = \left( x_e x_p n_\H \alpha_\H - \beta_\H
(1-x_p) e^{-h\nu_{\H2s}/kT_M}\right) C_\H \, ,\label{eq:newstandard_xe} 
\end{equation}

\begin{multline}
H(z)(1+z)\frac{dx_{\HeII}}{dz} = \left( x_{\HeII} x_e n_\H \alpha_{\HeI} - \beta_{\HeI}
(f_{\He}-x_{\HeII}) e^{-h\nu_{\HeI,2^1s}/kT_M}\right) C_{\HeI}\\
+  \left(
x_{\HeII} x_e n_\H \alpha^t_{\HeI} - \frac{g_{\HeI,2^3s}}{g_{\HeI,1^1s}}\beta^t_{\HeI}
(f_{\He}-x_{\HeII}) e^{-h\nu_{\HeI,2^3s}/kT_M}\right) C^t_{\HeI} \,,
\label{eq:HeI_xe}{  }
\end{multline}
donde 
\begin{eqnarray}
 C_\H &=& \frac{1 + K_{\H}\Lambda_{\H}n_{\H} (1-x_p)}{1 +
 K_{\H}(\Lambda_{\H}+\beta_{\H})n_{\H} (1-x_p)} \, ,\label{C_H} \\
 C_{\HeI} &=& \frac{1 + K_{\HeI}\Lambda_{\He}n_{\H}(f_{\He}-x_{\HeII})
 e^{h\nu_{ps}/kT_M}}{1 +
 K_{\HeI}(\Lambda_{\He}+\beta_{\HeI})n_{\H}(f_{\He}-x_{\HeII})
 e^{h\nu_{ps}/kT_M}} \, ,\label{C_He}\\
C^t_{\HeI} &=& \frac{1}{1 + K^t_{\HeI}\beta^t_{\HeI}n_{\H}(f_{\He}-x_{\HeII})
 e^{h\nu^t_{ps}/kT_M}} \, \label{C_Het}.
\end{eqnarray}

Las tres variables independientes son la fracci\'on de protones $x_p=
n_p/n_{\H}$, la fracci\'on de helio simplemente ionizado $x_{\HeII}=
n_{\HeII}/n_{\H}$, y la temperatura de la materia $T_M$. La variable
dependiente es la fracci\'on de electrones libres $x_e= n_e/n_{\H} =
x_p + x_{\HeII}$. Aqu{\'\i}, $n$ se refiere a la densidad num\'erica,
y $n_{\H}$ es la densidad num\'erica total de hidr\'ogeno.  Los
factores $C$ tienen en cuenta el efecto \emph{cuello de botella} de la
transici\'on $2p \rightarrow 1s$, que hace m\'as lenta la
recombinaci\'on.  El segundo t\'ermino de la
ecuaci\'on~(\ref{eq:HeI_xe}) se debe a la consideraci\'on de las
transiciones semi-prohibidas $2^3p\rightarrow 1^1s$ del $\HeI$ como
canal adicional para la recombinaci\'on al nivel fundamental. El
supra{\'\i}ndice $t$ indica que se trata de los estados
tripletes. Esta es una de las actualizaciones en el escenario de
recombinaci\'on. La ecuaci\'on para la temperatura de la materia es
\begin{equation}
H(z)(1+z)\frac{dT_M}{dz} = \frac{8 \sigma_T a_R T_R^4}{3 m_e c}
\frac{x_e}{1+f_{\He} + x_e} (T_M-T_R) + 2 T_M H(z) \, .
\end{equation}
Se utiliza esta ecuaci\'on durante todo el c\'alculo, debido al efecto
que resulta de la diferencia entre $T_M$ y $T_R$ a bajo redshift.

A continuaci\'on se listan todos los  par\'ametros que
entran en las ecuaciones de m\'as arriba. \'Estos son la constante
 de Boltzmann $k$, la constante de Planck $h$, la velocidad
de la luz $c$, la secci\'on eficaz de Thomson $\sigma_T$, la masa del
electr\'on $m_e$, y la constante de radiaci\'on $a_R$.  Con respecto a
los datos at\'omicos, la longitud de onda en reposo de la l{\'\i}nea
$\H$ Ly $\alpha$ es $\lambda_{\H 2p}= 121.5682\, \mathrm{nm}$.  La
frecuencia de la l{\'\i}nea H $2s$--$1s$, $\nu_{{\rm H}2s} =
c/\lambda_{{\rm H}2p}$, est\'a lo suficientemente cerca de
Ly$\,\alpha$ como para tomar el mismo valor de longitud de onda.  La
longitud de onda de $\HeI$ $2^1p$--$1^1s$ es $\lambda_{{\rm HeI}2^1p}
= 58.4334\,$nm. Notar que la frecuencia de $\HeI$ $2^1s$--$1^1s$ es
$\nu_{{\rm He}I2s} = c/60.1404\,$nm.  A diferencia de lo que pasa con
el hidr\'ogeno, la separaci\'on de los niveles $\HeI$ $2^1p$ y $2^1s$
es suficientemente grande como para distinguir entre $\lambda_{{\rm
He}2^1p}$ y $\lambda_{{\rm He}2^1s}$, por lo tanto aparece un factor
exponencial extra en las ecuaciones~(\ref{C_He}) y (\ref{C_Het}), que
no aparec{\'\i}a en la ecuaci\'on~(\ref{C_H}), con $\nu_{{\rm
He}I2^1p2^1s} =\nu_{{\rm He}I2^1p} - \nu_{{\rm
He}I2^1s}\equiv\nu_{ps}$, y equivalentemente para el t\'ermino que
considera la transici\'on via tripletes.  La tasa de la transici\'on
$2s$--$1s$ via dos fotones para el H es $\Lambda_{\rm H} =
8.22458\,{\rm s}^{-1}$ \citep{Goldman89}, mientras que la tasa para el
$\HeI$ es $\Lambda_{\rm He} = 51.3\,{\rm s}^{-1}$
\citep{Drake69}.

En las ecuaciones (\ref{eq:newstandard_xe}) y (\ref{eq:HeI_xe})
aparece  el coeficiente de recombinaci\'on Caso B, que es la suma de los
coeficientes de recombinaci\'on a todos los niveles excepto el nivel
fundamental. Para el $\H$, el coeficiente $\alpha_{\rm H}$ fue calculado por
\citet{Hummer94}. Efectivamente, los t\'erminos de dicha suma est\'an constitu{\'\i}dos
por las tasas de recombinaci\'on al nivel $n$:
\begin{equation}
\alpha_n(T_e,Z) = \frac{c \alpha^3}{\sqrt{\pi}} Z^3 \lambda^{3/2}
n^{-2} \int_0^\infty d\epsilon (1+n^2\epsilon)^2 {\rm e}^{-\lambda
 \epsilon} \tilde{a}_n(Z,\epsilon)
\label{eq:coefB_H}
\end{equation}
donde $\alpha$ es la constante de estructura fina,
$\tilde{a}_n(Z,\epsilon)$ es la secci\'on eficaz de fotoionizaci\'on
para el nivel $n$, $\epsilon$ es la energ{\'\i}a del electr\'on
eyectado en unidades de $Z^2$Ryd y $\lambda= Z^2 {\rm Ryd}/T_e$.  El
ajuste de esta cantidad, realizado por \citet{Pequignot91}, es
\begin{equation}
\alpha_{\H} = F 10^{-19} \frac{at^b}{1+ct^d} \mathrm{m^3s^{-1}} \, ,
\end{equation}
con $a=4.309$, $b=-0.6166$, $c=0.6703$,
$d=0.5300$ y $t= T_{\rm M}/10^{4}\,$K. El factor $F$ vale 1.14, y
permite que la ecuaci\'on~(\ref{eq:newstandard_xe}) coincida con el
c\'alculo obtenido con el c\'odigo de niveles m\'ultiples, acelerando
la recombinaci\'on. 

Para el helio, el  coeficiente de recombinaci\'on Caso B, 
 $\alpha_{\rm He}$, fue  obtenido por \citet{HummerStorey98},
siendo la expresi\'on de los t\'erminos individuales de la suma:
\begin{equation}
\alpha(T;nlS)= \frac{c \alpha^3}{\sqrt{\pi}} (2l+1) \frac{S}{4}
\lambda^{3/2} \nu^{-4} \int_0^\infty dE (1+\nu^2E)^2{\rm e}^{-\lambda
E} \sigma(nlS;E)
\label{eq:coefB_He}
\end{equation}
donde $\alpha$ es la constante de estructura fina, $\sigma(nlS;E)$ es
la secci\'on eficaz de fotoionizaci\'on para el estado $nlS$ a una
energ{\'\i}a del electr\'on eyectado de $E$ (en Ryd), $\nu$ es el
n\'umero cu\'antico efectivo del estado, $T$ es la temperatura
electr\'onica, y
\begin{equation}
 \lambda= \frac{1.5789 \times 10^5}{T}.
\end{equation}
El ajuste de este coeficiente se debe a \citet{VernerFerland96} y tiene
la siguiente expresi\'on:
\begin{equation}
\alpha_{\HeI} = q \, \left[\sqrt{\frac{T_M}{T_2}} \left( 1 + \sqrt{\frac{T_M}{T_2}}\right)^{1-p} \left( 1 + \sqrt{\frac{T_M}{T_2}}\right)^{1+p} \right]^{-1} \mathrm{m^3s^{-1}} 
\end{equation}

Los par\'ametros son $q=10^{-16.744}$, $p=0.711$, $T_1=10^{5.114}\,$K,
y $T_2$ fijado arbitrariamente a $3\,$K.  Este ajuste es bueno hasta
$<0.1\%$ sobre el rango de temperaturas relevante
($4{,}000$--$10{,}000\,$K), y sigue siendo bastante preciso sobre un
rango de temperatura a\'un mayor. Para los tripletes, el ajuste se
realiza con la misma forma funcional, pero con distintos valores para
los par\'ametros: $p=0.761$; $q=10^{-16.306}$; $T_1=10^{5.114}$\,K;
and $T_2=3$\,K.  Este ajuste es mejor que el $1\%$ para
temperaturas de entre $10^{2.8}$ y $10^{4}$\,K.

 Los coeficientes de fotoionizaci\'on $\beta_{\rm X}$, donde ${\rm X}$
representa a $\H$ o $\HeI$, se calculan a partir de los coeficientes
de recombinaci\'on haciendo: $\beta_{\rm X}=\alpha_{\rm X} (2\pi
m_{\rm e} k \ T_{\rm M}/h^2)^{3/2}
\exp(-h\nu_{2s}/kT_{\rm M})$. Aqu{\'\i} $\nu_{2s}$  y $\alpha$ son
diferentes para H y $\HeI$. Notar adem\'as que se utilizan $T_{\rm
M}$ y $\nu_{2s}$, y que el uso incorrecto de $T_{\rm R}$ o $\nu_{2p}$
causar\'a una diferencia peque\~na pero importante en modelos con
gran cantidad de bariones.  Por \'ultimo,  $\beta^{\rm
  t}_{\rm HeI}$ es el coeficiente de fotoionizaci\'on para los
tripletes, y se calcula a partir de $\alpha^{\rm t}_{\rm HeI}$ de esta
manera {\setlength\arraycolsep{1pt}
\begin{eqnarray}
\beta^{\rm t}_{\rm HeI} = \alpha^{\rm t}_{\rm HeI}
\left(\frac{2 \pi m_{\rm e} k_{\rm B} T_{\rm M}}{h^2}\right)^{3/2}
\frac{2 g_{\rm He^+}}{g_{\rm HeI, 2^3s}} e^{-h \nu_{\rm 2^3s,c}/kT_{\rm M}},
\end{eqnarray}}
\\
donde $g_{\rm He^+}$ y $g_{\rm HeI, 2^3s}$ son las degeneraciones de los
 \'atomos de He$^+$ y de He\,{\sc i} con electrones en el nivel $2^3$s,
y $h \nu_{\rm 2^3s,c}$ es la energ{\'\i}a de ionizaci\'on del estado $2^3$s  
\citep{wong08}.

Los par\'ametros cosmol\'ogicos son el redshift $z$, el factor de
Hubble $H(z)$, y la temperatura de la radiaci\'on $T_{\rm R} = T_0
(1+z)$.  La abundancia primordial de He se toma como $Y_{\rm P}=0.24$
por unidad de masa \citep{SchrammTurner98}, y el valor actual de la
temperatura del fondo c\'osmico de radiaci\'on $T_{0}$ se toma como
$2.728\,$K (el valor central determinado por el experimento FIRAS
\citep{Fixsen96}).

El corrimiento al rojo cosmol\'ogico de los fotones H Ly$\alpha$ es
\begin{equation}
K_{\rm H}\equiv\lambda_{\rm H_{2p}}^3/(8\pi H(z)).
\end{equation} 
Para el helio, relacionamos los factores $K$ de corrimiento al rojo de
las l{\'\i}neas He\,{\sc i} $2^1$p--$1^1$s y He\,{\sc i}
$2^3$p--$1^1$s con la probabilidad de escape de Sobolev $p_S$ de la
siguiente manera:

\begin{eqnarray}
&& K_{\rm HeI} = \frac{g_{{\rm HeI}, 1^1{\rm s}}}{g_{{\rm HeI}, 2^1{\rm p}}}
\frac{1}{ n_{{\rm HeI}, 1^1{\rm s}}  
A^{\rm HeI}_{ 2^1{\rm p}-1^1{\rm s}} p_{\rm S}} \quad {\rm y} \label{eqKHeIa}\\
&& K^{\rm t}_{\rm HeI} = \frac{g_{{\rm HeI}, 1^1{\rm s}}}{g_{{\rm HeI}, 2^3{\rm p}}}
\frac{1}{ n_{{\rm HeI}, 1^1{\rm s}}  
A^{\rm HeI}_{ 2^3{\rm p}-1^1{\rm s}} p_{\rm S}} \ , 
\label{eqKHeIb}
\end{eqnarray}
\\ donde $A_{{\rm HeI}, 2^1{\rm p}-1^1{\rm s}}$ y $A_{{\rm HeI},
2^3{\rm p}-1^1{\rm s}}$ son los coeficientes $A$ de Einstein de las
transiciones He\,{\sc I} $2^1$p--$1^1$s y He\,{\sc I} $2^3$p--$1^1$s,
respectivamente.  Para incluir el efecto de la opacidad en el
continuo debido al H, basados en la f\'ormula aproximada
de \citet{KIV07}, se reemplaza $p_{\rm S}$ por la nueva probabilidad de
escape $p_{esc}=p_s + p_{\rm con, H}$ con
\begin{equation}
p_{\rm con, H} = \frac{1}{1 + a_{\rm He} \gamma^{b_{\rm He}}}, \\
\end{equation}
donde $a_{\rm He}$ y $b_{\rm He}$ son par\'ametros de ajuste.
El factor $\gamma$ tiene la siguiente expresi\'on
\begin{equation}
 \gamma = \frac{\frac{g_{{\rm HeI}, 1^1{\rm s}}}{g_{{\rm HeI}, 2^1{\rm p}}}
A^{\rm HeI}_{2^1{\rm p}-1^1{\rm s}} (f_{\rm He} - x_{\rm HeII})c^2}
{8 \pi^{3/2} \sigma_{{\rm H},1{\rm s}}(\nu_{\rm HeI,2^1 p}) 
\nu_{\rm HeI,2^1{\rm p}}^2 \Delta \nu_{\rm D,2^1p} 
(1 - x_{\rm p})}\, \label{eq:gamma} 
\end{equation}
donde $\sigma_{{\rm H},1s}(\nu_{\rm HeI,2^1p})$ es la secci\'on eficaz
de ionizaci\'on del H a la frecuencia $\nu_{\rm HeI,2^1p}$ y $\Delta \nu_{\rm
D,2^1p} = \nu_{\rm HeI,2^1p} \sqrt{2 k_{\rm B} T_{\rm M}/m_{\rm He}
c^2}$ es el ancho t\'ermico de la l{\'\i}nea He\,{\sc i} $2^1$p--$1^1$s. 
La secci\'on eficaz de fotoionizaci\'on desde el nivel $n$ es \citep{seaton59}  
\begin{equation}
\sigma_n(Z,h\nu)= \frac{2^6\alpha\pi a_0^2}{3\sqrt{3}} \frac{n}{Z^2}
(1+n^2\epsilon)^{-3} g_{II}(n,\epsilon)
\label{eq:secc_ef_fotoioniz}
\end{equation}
donde $g_{II}(n,\epsilon) \simeq 1$ es el factor de Gaunt-Kramers,
$\alpha$ es la constante de estructura fina y $a_0= \hbar /(m_e c \alpha)$ es el radio de
Bohr.

\section{Dependencia con las constantes fundamentales}

Las constantes fundamentales $\alpha$ y $m_e$ est\'an involucradas en
cantidades f{\'\i}sicas relevantes durante la recombinaci\'on. En esta
secci\'on estudiamos los detalles de las dependencias con las
constantes en cada una de las cantidades f{\'\i}sicas de
inter\'es. Luego analizaremos c\'omo cambia la historia de
recombinaci\'on por variaciones temporales en los valores de las
constantes fundamentales que estamos estudiando. Los primeros trabajos
donde se analiz\'o la dependencia de las constantes fundamentales en
el escenario de recombinaci\'on est\'andar son los
de \citet{Turner}, \citet{Hannestad99}, y \citet{KS00}.

La dependencia de los coeficientes de recombinaci\'on Caso B puede
verse de las ecuaciones (\ref{eq:coefB_H}) y (\ref{eq:coefB_He}). La
dependencia del factor $\lambda$ aparece a trav\'es del Ryd, una
unidad de energ{\'\i}a cuya dependencia con las constantes
fundamentales coincide con la de la energ{\'\i}a de ligadura del
hidr\'ogeno $B_1= \frac{1}{2} \alpha^2 m_e c^2$, es decir $ \alpha^2
m_e$.  De la secci\'on eficaz de fotoionizaci\'on viene una
dependencia de $\alpha a_0^2 \sim \alpha^{-1} m_e^{-2}$, que se deduce
de la expresi\'on de la ecuaci\'on (\ref{eq:secc_ef_fotoioniz}). Por
lo tanto, la dependencia total de los coeficientes de recombinaci\'on
Caso B con las constantes fundamentales es $\alpha^3 m_e^{-3/2}$. Por
su parte, los coeficientes de fotoionizaci\'on $\beta$ tienen una
dependencia con las constantes que se deduce de la dependencia de los
coeficientes de recombinaci\'on m\'as el factor extra $m_e^{3/2}$, y
la dependencia en el exponencial que va como $ \alpha^2 m_e$, como
todos los t\'erminos proporcionales a la energ{\'\i}a que aparecen en
las ecuaciones.

La dependencia con las constantes fundamentales que viene del redshift
de los fotones Ly$\alpha$ se deduce de la expresi\'on $K_H =
\lambda_\alpha^3 /8 \pi H(z)$, con $\lambda_\alpha = 8 \pi \hbar c/3
B_1$, es decir, $\alpha^{-6}m_e^{-3}$.

En esta tesis se estudi\'o la influencia de las constantes en el
escenario actualizado de la recombinaci\'on. Para ello, buscamos la
dependencia que viene de los factores de corrimiento al rojo de las
l{\'\i}neas del helio, analizando los factores de las ecuaciones
(\ref{eqKHeIa}) y (\ref{eqKHeIb}). En primera instancia, buscamos
c\'omo dependen los coeficientes $A$ de Einstein de las constantes
fundamentales.
Las tasas de probabilidad de transici\'on $A_{{\rm HeI}, 2^1{\rm p}-1^1{\rm
s}}$ y $A_{{\rm HeI}, 2^3{\rm p}-1^1{\rm s}}$ se pueden expresar como
\citep{DM07}:
\begin{equation}
 A^{\rm HeI}_{i-j} = \frac{4 \alpha}{3 c^2} \omega_{ij}^3 \left|\left<\psi_i|r_1 + r_2|\psi_j\right>\right|^2
\label{rate}
\end{equation}
donde $\omega_{ij}$ es la frecuencia de la transici\'on, y $i$($j$)
hace referencia a los estados inicial (final) del \'atomo. Primero
analizamos el braket. A primer orden en teor{\'\i}a de perturbaciones,
todas las funciones de onda se pueden aproximar a la correspondiente
funci\'on de onda del hidr\'ogeno. Estas \'ultimas se pueden analizar
como $P(r/a_0)*\exp(-qr/a_0)$ en donde $P$ es un polinomio, $a_0$ es
el radio de Bohr y $q$ es un n\'umero. Se puede mostrar que cualquier
integral del tipo de la ecuaci\'on (\ref{rate}) puede ser resuelta con
un cambio de variables $x=r/a_0$. Si las funciones de onda est\'an
apropiadamente normalizadas, la dependencia con las constantes
fundamentales viene del operador, en este caso $r_1 + r_2$. Por ende,
la dependencia del braket va como $a_0$. Por otra parte, $\omega_{ij}$
es proporcional a la diferencia de los niveles de energ{\'\i}a y por
lo tanto, dependen de las constantes fundamentales como
$\omega_{ij} \simeq m_e \alpha^2$. En consecuencia, la dependencia de
las probabilidades de transici\'on de ${\rm HeI}$ con $\alpha$ y $m_e$
se puede expresar como:
\begin{equation}
 A^{\rm HeI}_{i-j} \simeq  \alpha^5 m_e
\end{equation} 

En segunda instancia, buscamos la dependencia que viene de la
modificaci\'on de la probabilidad de escape. En la expresi\'on de
$\gamma$ (ver ecuaci\'on (\ref{eq:gamma})) adem\'as de los
coeficientes $A$ de Einstein, aparece la secci\'on eficaz de
fotoionizaci\'on del hidr\'ogeno, cuya dependencia con las constantes
es $\alpha^{-1} m_e^{-2}$ como vimos anteriormente. S\'olo queda
analizar la dependencia de los factores $a_{\rm He}$ y $b_{\rm He}$
que aparecen en $p_{\rm con,H}$.  El c\'alculo detallado de estos
par\'ametros de ajuste no est\'a disponible a\'un, por lo que no es
posible determinar c\'omo se ven afectados por variaciones de $\alpha$
y/o $m_e$. \citet{wong08} han mostrado que estos
par\'ametros deben conocerse con un $1\%$ de precisi\'on para los
datos que se obtendr\'an con el sat\'elite Planck. Sin embargo, en
esta tesis se ha trabajado con los datos del sat\'elite WMAP, para el
cual esta precisi\'on no es necesaria. Para llegar a esta conclusi\'on
hemos calculado los espectros de potencia de la temperatura, la
polarizaci\'on y la correlaci\'on cruzada del fondo c\'osmico de
radiaci\'on, permitiendo que los par\'ametros $a_{\rm He}$ y $b_{\rm
He}$ var{\'\i}en un 50\%. Hemos encontrado que para los espectros de
temperatura y polarizaci\'on la variaci\'on es siempre menor que el
error observacional (1\% para la temperatura y casi 40 \% para la
polarizaci\'on). Las mayores variaciones aparecen en el espectro de
correlaci\'on cruzada. Para \'este, hemos calculado los errores
observacionales para
todos los $\ell$'s medidos y los hemos comparado con la variaci\'on
inducida cuando cambiamos $a_{\rm He}$ y $b_{\rm He}$ en un $50\%$. En
todos los casos, el error observacional es varios \'ordenes de
magnitud mayor que la variaci\'on de los $C_{\ell}$'s por el cambio en
los par\'ametros. Por lo tanto, para analizar los datos de WMAP no es
necesario la modificaci\'on de los mismos. Todo este an\'alisis y
conclusiones son contribuciones originales de esta tesis y fueron
publicadas en \citet{Scoccola08b}.

Para encontrar la dependencia de las tasas de recombinaci\'on via dos
fotones, $\Lambda_{\rm H}$ y $\Lambda_{\rm He}$, con las constantes
fundamentales, consideremos una transici\'on donde uno de los fotones
tiene una frecuencia $y\,\nu_{12}$ y el otro tiene frecuencia
$(1-y)\nu_{12}$, siendo $\nu_{12}$ la frecuencia correspondiente a la
diferencia de energ{\'\i}a entre el nivel $2s$ y el nivel $1s$
\citep{BreitTeller40,SpitzerGreenstein51}. La probabilidad de emisi\'on
de un fot\'on de frecuencia $\nu_{12}$ es
\begin{equation}
A(y)=  \frac{9\alpha^6 c R}{2^{10}} \psi(y)
\end{equation}
donde $\alpha$ es la constante de estructura fina, R es el Rydberg ($1
{\rm Ryd}= \frac{1}{2} m_e c^2 \alpha^2$), y $\psi(y)$ es una
combinaci\'on lineal de funciones de onda normalizadas.  De aqu{\'\i}
se ve que dichas tasas de recombinaci\'on dependen de las constantes
fundamentales como $\alpha^8m_e$.

En la Tabla~(\ref{resumen_depend}), se presenta un resumen de las
dependencias con las constantes $\alpha$ y $m_e$, para las cantidades
f{\'\i}sicas relevantes en los procesos que ocurren durante la
recombinaci\'on.

\begin{table}[!ht]
\begin{center}
\renewcommand{\arraystretch}{1.3}
\begin{tabular}{|l|l|l|}
\hline
Descripci\'on & Magnitud f{\'\i}sica  & Dependencia  \\
\hline
Energ{\'\i}a de ligadura del hidr\'ogeno & $B_1$ & {\ \ \ }$\alpha^2 m_e$ \\
\hline 
Secci\'on eficaz de fotoionizaci\'on &  $\sigma_n(Z,h\nu)$   & {\ \ \ }$\alpha^{-1}m_e^{-2} $  \\
\hline 
Secci\'on eficaz de Thomson &  $\sigma_T$   & {\ \ \ }$\alpha^2m_e^{-2} $  \\
\hline 
Coeficientes de recombinaci\'on Caso B & $\alpha_{\rm H}$, $\alpha_{\rm HeI}$, $\alpha^t_{\rm HeI}$ &  {\ \ \ }$\alpha^3 m_e^{-3/2}$ \\
\hline 
Coeficientes de fotoionizaci\'on & $\beta_{\rm H}$, $\beta_{\rm HeI}$ &{\ \ \ }$\alpha^3$
\\ \hline
Redshift de los fotones Ly$\alpha$ & $K_{\rm H}$, $K_{\rm HeI}$, $K^t_{\rm HeI}$ & {\ \ \ }$\alpha^{-6}m_e^{-3}$ \\
\hline
Coeficiente $A$ de Einstein & $A^{\rm HeI}_{i-j}$ & {\ \ \ }$\alpha^5 m_e$ \\
\hline
Tasa de recombinaci\'on via 2 fotones & $\Lambda_{\rm H}$, $\Lambda_{\rm HeI}$ & {\ \ \ }$\alpha^8 m_e$ \\
\hline
\end{tabular}
\caption[Dependencias de magnitudes f{\'\i}sicas con $\alpha$ y $m_e$.]{ Dependencias con $\alpha$ y $m_e$ de las magnitudes f{\'\i}sicas relevantes en los procesos que tienen lugar durante la recombinaci\'on.}
 \label{resumen_depend}
\end{center}
\end{table}

\begin{figure}[p]
\begin{center}
\includegraphics[scale=1.8,angle=0]{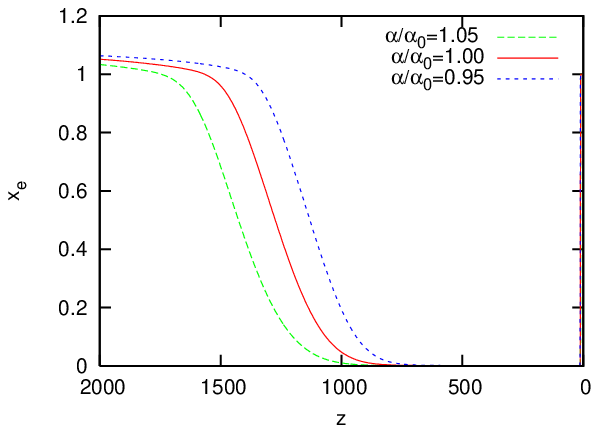}
\includegraphics[scale=1.8,angle=0]{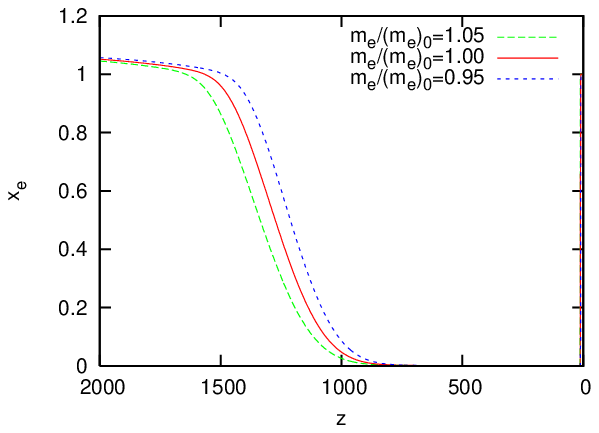}
\end{center}
\caption[Historia de ionizaci\'on, para distintos valores de $\alpha$ y $m_e$.]{Fracci\'on  de ionizaci\'on como funci\'on del redshift, para distintos valores de  $\alpha$ (panel superior) y $m_e$ (panel inferior) en la \'epoca de recombinaci\'on. $\alpha_0$ y $(m_e)_0$ son los valores actuales de las constantes. Si  $\alpha$ o $m_e$ ten{\'\i}an en la \'epoca de recombinaci\'on un valor mayor que el actual, la recombinaci\'on ocurre m\'as temprano.  }
\label{ionization_history_plot}
\end{figure}

La variaci\'on temporal de las constantes fundamentales afecta la
f{\'\i}sica durante la recombinaci\'on. El principal efecto es el
corrimiento de la \'epoca de recombinaci\'on a redshifts mayores
cuando aumenta $\alpha$ o $m_e$. Esto se entiende f\'acilmente dado
que la energ{\'\i}a de ligadura $B_n$ escalea como $\alpha^2 m_e$, de
manera tal que los fotones deber{\'\i}an tener mayor energ{\'\i}a para
poder ionizar \'atomos de hidr\'ogeno.  En las
Figs.~({\ref{ionization_history_plot}}) se muestra c\'omo se ve afectada la
historia de ionizaci\'on por cambios en $\alpha$ y en $m_e$, en un
Universo plano con par\'ametros cosmol\'ogicos $(\Omega_bh^2;
\Omega_{CDM}h^2; h; \tau) = (0.0223; 0.1047; 0.73; 0.09)$.
Cuando $\alpha$ y/o $m_e$ tienen valores mayores que los actuales, la
recombinaci\'on ocurre m\'as temprano  (a redshifts m\'as grandes). La
historia de ionizaci\'on  es m\'as sensible a variaciones en  $\alpha$
que a variaciones en $m_e$ por las dependencias en $B_n$.

El mecanismo m\'as eficiente para alcanzar el equilibrio t\'ermico en
el gas de fotones del Universo temprano es el scattering de Thomson
por electrones libres, por lo tanto, otro efecto importante es el
corrimiento en la secci\'on eficaz de Thomson $\sigma_{T}$, definida
seg\'un:
\begin{equation}
\sigma _T = \frac{8\pi \ \hbar ^2}{3\
m_e^2 c^2}\alpha^2
\end{equation}
de donde se ve su dependencia con las constantes fundamentales.

\begin{figure}[p]
\begin{center}
\includegraphics[scale=1.8,angle=0]{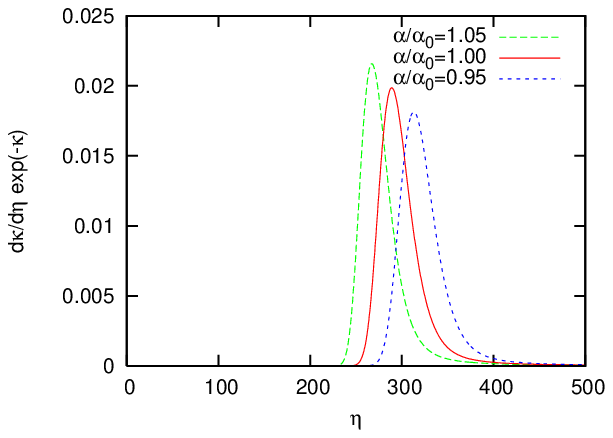}
\includegraphics[scale=1.8,angle=0]{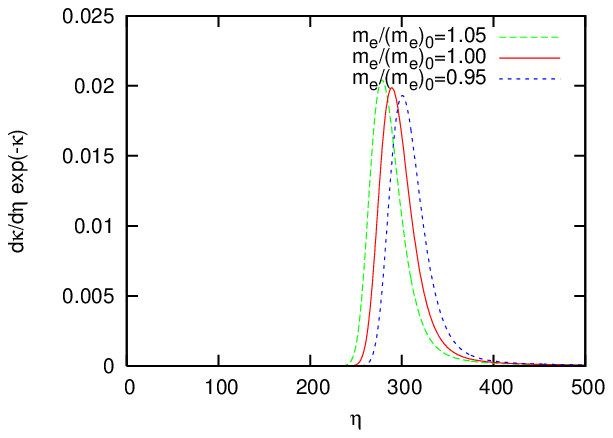}
\end{center}
\caption[Funci\'on visibilidad, para distintos valores de $\alpha$ y $m_e$.]{Funci\'on visibilidad como funci\'on del tiempo conforme,
  para diferentes valores de $\alpha$ (panel superior) y $m_e$
  (panel inferior). Si  $\alpha$ o $m_e$ ten{\'\i}an en la \'epoca de recombinaci\'on un valor mayor que el actual, la recombinaci\'on ocurre m\'as temprano, es decir, a $\eta$ menor.} 
\label{visibility_function}
\end{figure}

La funci\'on visibilidad, la cual mide la probabilidad diferencial de
 que un fot\'on se disperse por \'ultima vez al tiempo conforme
 $\eta$,  depende de $\alpha$ y de $m_e$. Esta funci\'on se define como
\begin{equation}
g(\eta)=e^{-\kappa}\frac{d \kappa}{d\eta} \, ,\qquad \mathrm{donde}
\qquad \frac{d \kappa}{d\eta}=x_{e}n_{p} a  \sigma_{T} \,
\end{equation}
es la profundidad \'optica diferencial de los fotones debido a la
dispersi\'on de Thomson. La densidad num\'erica total de protones es
$n_{p}$ (tanto libres como ligados), $x_{e}$ es la fracci\'on de
electrones libres, y $a$ es el factor de escala. El efecto m\'as
fuerte de variar $\alpha$ y $m_e$ sobre la funci\'on visibilidad
ocurre debido a la alteraci\'on de la historia de ionizaci\'on
$x_{e}(\eta)$.  En la Fig.~(\ref{visibility_function}) se muestra que si
el valor de $\alpha$ o $m_e$ fuera m\'as peque\~no (grande) en
la \'epoca de recombinaci\'on que su valor actual, el m\'aximo en la
funci\'on visibilidad se correr{\'\i}a a valores de redshift m\'as
peque\~nos (grandes), y su ancho crecer{\'\i}a (decrecer{\'\i}a)
levemente.

\cleardoublepage
\chapter{Fondo c\'osmico de radiaci\'on}
\label{chap:CMB}

\section{Consideraciones generales}

El Universo primordial est\'a compuesto por un plasma de fotones,
electrones y n\'ucleos de elementos qu{\'\i}micos livianos. A medida
que el Universo se expande, la temperatura baja, y es posible la
formaci\'on de
\'atomos neutros. La fracci\'on de electrones libres se hace
pr\'acticamente despreciable, por lo que la dispersi\'on de Thomson
entre fotones y electrones deja de tener lugar.  La opacidad del medio
cae bruscamente, y los fotones comienzan a viajar libremente. Estos
fotones constituyen el fondo c\'osmico de radiaci\'on 
(CMB por \emph{cosmic microwave background}), y
traen informaci\'on directa de c\'omo era el Universo en el momento
del desacople de la materia y la radiaci\'on, aproximadamente 400.000
a\~nos despu\'es del Big Bang.

Los fotones del fondo c\'osmico de radiaci\'on fueron detectados por
primera vez de manera accidental por \citet{PW68}. M\'as adelante, el
instrumento FIRAS a bordo del sat\'elite COBE (1992) confirm\'o que
dicha radiaci\'on tiene un espectro de cuerpo negro, con una
temperatura de 2.728 K
\citep{FIRAS}. Adem\'as, con el instrumento DMR (differential microwave
radiometer) se obtuvo el primer mapa de las anisotrop{\'\i}as en CMB
en todo el cielo \citep{DMR}. Luego de COBE, distintos experimentos en
tierra y en globos fueron lanzados para medir con mayor precisi\'on la
distribuci\'on de dichas anisotrop{\'\i}as: ARCHEOPS \citep{ARCHEOPS},
BOOMERANG
\citep{BOOMERANG}, DASI \citep{DASI}, MAXIMA \citep{MAXIMA}, VSA
\citep{VSA}, y con mejor resoluci\'on angular: CBI \citep{CBI}, 
y ACBAR \citep{ACBAR}. En Junio de 2001, NASA lanz\'o el sat\'elite WMAP,
que en el año 2003 liber\'o un primer conjunto de datos que
constituyeron un mapa completo del cielo con una precisi\'on mucho
mayor a la de COBE \citep{wmap1}. En el 2005, WMAP liber\'o un conjunto
de datos mayor, que llamamos WMAP3
\citep{wmap3a}, y en marzo de 2008 public\'o los datos recolectados durante 5
a\~nos, conjunto que llamaremos WMAP5 \citep{wmap5}.  La Agencia
espacial europea (ESA) planea lanzar el sat\'elite Planck durante el
a\~no 2009, para realizar el mapa definitivo de las anisotrop{\'\i}as
en la temperatura y polarizaci\'on de CMB en todo el cielo, con una
resoluci\'on angular de un d\'ecimo de grado \citep{Planck_mission}.

La distribuci\'on de la temperatura de CMB en el cielo es una
funci\'on definida sobre una esfera, por lo tanto, es natural
analizarla con un desarrollo en arm\'onicos esf\'ericos

\begin{equation}
T(\theta,\phi)= \sum_{\ell,m} a_{\ell m} Y_{\ell m} (\theta,\phi).
\end{equation}

En esta combinaci\'on lineal, el t\'ermino del monopolo da la temperatura media
de CMB, $T=2.728$K. El t\'ermino dipolar, con $\ell=1$, se interpreta
como el resultado del corrimiento Doppler causado por el movimiento
del sistema solar relativo a CMB.  Una vez que se remueven estos dos
t\'erminos, lo que queda son las anisotrop{\'\i}as intr{\'\i}nsecas de
CMB, que son del orden de $10^{-5}$ en todas las escalas angulares, y
contienen la informaci\'on de la f{\'\i}sica del Universo en la
\'epoca del desacople de la materia y la radiaci\'on. La mayor parte
de la informaci\'on cosmol\'ogica est\'a contenida en la funci\'on de
correlaci\'on de dos puntos de la temperatura. Esta funci\'on se
define como el promedio en el cielo del producto de la desviaci\'on
fraccional de la temperatura en las direcciones  ${\bf n}$ y ${\bf
n'}$, desarrollando el resultado en polinomios de Legendre

\begin{equation}
C(\theta) = \left< \frac{\Delta T({\bf n})}{T}  \frac{\Delta T({\bf
n'})}{T} \right> = \sum_{\ell=2}^{\infty} \frac{2\ell+1}{4\pi} C_\ell
P_{\ell}(\cos{\theta}).
\end{equation}
Los coeficientes del desarrollo, $C_\ell$, son las varianzas de los
coeficientes $a_{\ell m}$, de manera tal que
\begin{equation}
\left<a_{\ell m} a^\ast_{\ell' m'} \right> =  \delta_{\ell \ell'}
\delta_{mm'} C_\ell.
\end{equation}
Esta funci\'on, llamada \emph{espectro angular de potencias de las
anisotrop{\'\i}as en el fondo c\'osmico de radiaci\'on},  es la
herramienta m\'as importante al comparar teor{\'\i}a con observaciones.

Las anisotrop{\'\i}as en la temperatura est\'an {\'\i}ntimamente
relacionadas con las fluctuaciones de la densidad en la \'epoca de
recombinaci\'on. Los modelos inflacionarios predicen que el campo de
inhomogeneidades en la densidad, $\delta(x)$, es un campo aleatorio
gaussiano. Estas inhomogeneidades son la semilla para la formaci\'on
de estructura a gran escala, y se supone que fueron generadas a partir de
fluctuaciones cu\'anticas en el campo del inflat\'on, que crecieron
con un factor de expansi\'on exponencial. Definiendo la desviaci\'on
media cuadr\'atica de las inhomogeneidades como $\sigma^2
= \left< \delta (x)^2 \right>$, la funci\'on de autocorrelaci\'on se
define como
\begin{equation}
\xi(|{\mathbf x_2} - {\mathbf x_1}|) \equiv \xi(x) = \frac{1}{\sigma^2} \left< \delta({\mathbf x_2})\delta({\mathbf x_1})\right>.
\end{equation}
En la aproximaci\'on donde las fases son aleatorias, todos los modos
de Fourier $\delta (k)$ est\'an no correlacionados, lo que significa
que
\begin{equation}
\left< \delta^\ast ({\mathbf k})  \delta ({\mathbf k'})\right> = (2 \pi)^3 \delta^3 ({\mathbf k} - {\mathbf k'}) P({\mathbf k}).
\end{equation}
La funci\'on $P({\mathbf k})$ es el espectro de potencias de las
fluctuaciones. En la mayor{\'\i}a de los modelos, el espectro inicial
toma la forma
\begin{equation}
P(k) = A_s \left(\frac{k}{k_0}\right)^{n_s}
\end{equation}
donde $A_s$ es la amplitud y $n_s$ es el {\'\i}ndice espectral. El
valor $n_s=1$ corresponde a las llamadas fluctuaciones invariantes de
escala, dado que para tal espectro, las fluctuaciones $\delta$ tienen
la misma amplitud para todas las escalas de longitud. La teor{\'\i}a
inflacionaria predice perturbaciones que sean casi invariantes de
escala.

Los fotones que fueron dispersados por \'ultima vez en el desacople,
nos llegan de la \'epoca cuando se formaron los \'atomos neutros en el
Universo. A esa superficie en el cielo, que se encuentra a una
distancia correspondiente al redshift del desacople, se la conoce como
superficie de \'ultimo scattering. Sin embargo, la recombinaci\'on no
fue instant\'anea, por lo que la superficie de
\'ultimo scattering tiene cierto espesor en redshift.
La dispersi\'on de Thomson entre fotones y electrones en la superficie
de \'ultimo scattering genera polarizaci\'on en los fotones de CMB.
Por lo tanto, se pueden definir nuevas funciones de correlaci\'on
angular a partir de la polarizaci\'on, y as{\'\i} tener m\'as
herramientas con las cuales extraer informaci\'on acerca de las
condiciones f{\'\i}sicas del Universo.  La manera m\'as conveniente de
describir la polarizaci\'on es a trav\'es de sus modos $E$ (escalar) y
$B$ (pseudo-escalar). La funci\'on de correlaci\'on cruzada entre
temperatura y polarizaci\'on se define como
\begin{equation}
\left<a^{T}_{\ell m} a^{E \ast}_{\ell' m'} \right> =  \delta_{\ell \ell'}
\delta_{mm'} C_\ell^{TE}
\end{equation}
y la funci\'on de autocorrelaci\'on de los modos $E$ de polarizaci\'on
se define como
\begin{equation}
\left<a^{E}_{\ell m} a^{E \ast}_{\ell' m'} \right> =  \delta_{\ell \ell'}
\delta_{mm'} C_\ell^{EE}
\end{equation}
donde $a^{E}_{\ell m}$ son los coeficientes del desarrollo en
arm\'onicos esf\'ericos de la polarizaci\'on modo $E$. Dado que las
perturbaciones iniciales escalares no generan modos $B$ de
polarizaci\'on, no estudiamos las correlaciones que involucren estos
modos.

\begin{figure}[!ht]
\psfrag{angularl}{\small{$\ell$}}
\psfrag{powerCl}{\small{$\ell (\ell +1) C_{\ell}/2\pi$}}
\begin{center}
\includegraphics[scale=1.8,angle=0]{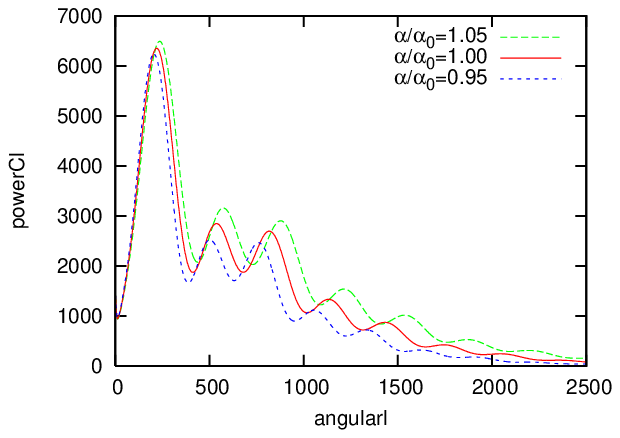}
\includegraphics[scale=1.8,angle=0]{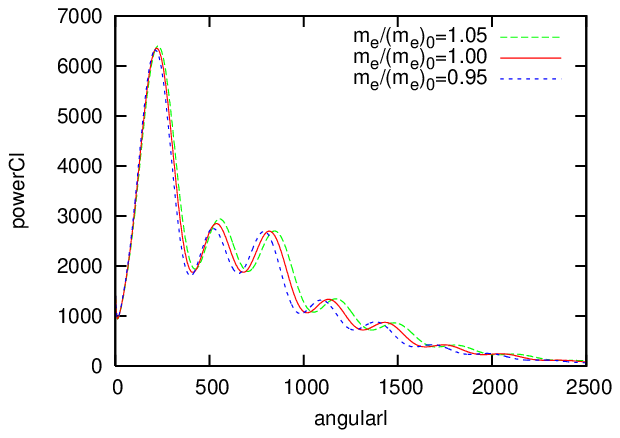}
\end{center}
\caption[Espectro de anisotrop{\'\i}as de la temperatura de CMB para distintos valores de $\alpha$ y $m_e$.]{Espectro de anisotrop{\'\i}as de la temperatura de CMB para diferentes valores de
  $\alpha$ (panel superior) y $m_e$ (panel inferior).}
\label{Cls_alfa_emasa}
\end{figure}

 Tanto los par\'ametros cosmol\'ogicos como el valor de las constantes
fundamentales $\alpha$ y $m_e$ afectan la forma de las funciones de
correlaci\'on, y de esta manera sus valores pueden estimarse a partir
de observaciones de CMB.  Los efectos en el espectro angular de
potencias de CMB debido a variar las constantes fundamentales son
similares a los producidos por cambios en los par\'ametros
cosmol\'ogicos, es decir, cambios en las amplitudes relativas de los
picos, y un corrimiento en sus posiciones. Efectivamente, el aumento
de $\alpha$ o $m_e$ aumenta el redshift de la superficie de \'ultimo
scattering, lo que corresponde a un horizonte de sonido m\'as
peque\~no. La posici\'on del primer pico ($\ell_{1}$) es inversamente
proporcional a \'este \'ultimo, por lo que resulta un $\ell_{1}$
mayor.  Tambi\'en se produce un efecto ISW (integrated Sachs-Wolfe)
mayor, haciendo que el primer pico Doppler sea m\'as alto. M\'as
a\'un, un aumento en $\alpha$ o $m_e$ decrece el damping de difusi\'on
para $\ell$'s grandes, el que se produce por el ancho finito de la
superficie de \'ultimo scattering, y por lo tanto, aumenta la potencia
en peque\~nas escalas. Todos estos efectos se ilustran en la
Fig.~(\ref{Cls_alfa_emasa}), donde  $\alpha_0$ y $(m_e)_0$ son los valores
actuales de las constantes fundamentales $\alpha$ y $m_e$. En este
ejemplo, los par\'ametros cosmol\'ogicos fueron fijados a:

\begin{displaymath}
\begin{tabular}{ccccc}
$\Omega_b h^2$ = 0.0223 & {} & $\Omega_{CDM} h^2$ =  0.1047& {} & $H_0$ = 73 {\rm km/s/Mpc}\\
 $\tau_{re}$ = 0.09  & {} & $n_s$ = 0.951 & {} &   $A_s$ = 2.3$\times 10^{-9}$
\end{tabular}
\end{displaymath}

donde $\Omega_{b} h^2$ es la densidad de materia bari\'onica, en
unidades de la densidad cr{\'\i}tica, $\Omega_{CDM} h^2$ es la
densidad de materia oscura, en las mismas unidades, $H_0$ es la tasa
actual de expansi\'on de Hubble, $\tau_{re}$ es la profundidad
\'optica de reionizaci\'on, $n_s$ es el {\'\i}ndice espectral de
perturbaciones escalares, y $A_s$ es la amplitud de las fluctuaciones
escalares.

\begin{figure}[!ht]
\psfrag{angularl}{\small{$\ell$}}
\psfrag{powerCl}{\small{$\ell (\ell +1) C_{\ell}^{TE}/2\pi$}}
\begin{center}
\includegraphics[scale=1.8,angle=0]{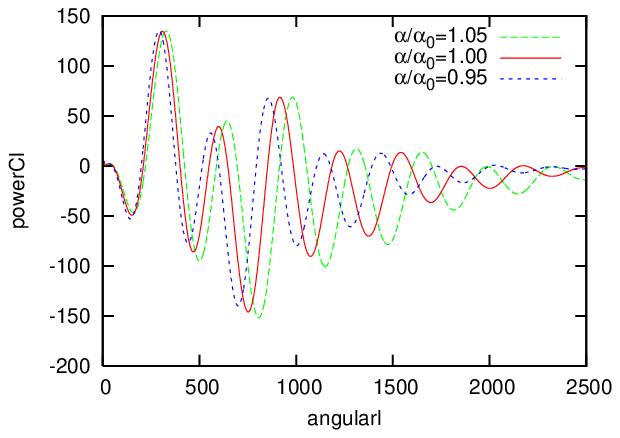}
\includegraphics[scale=1.8,angle=0]{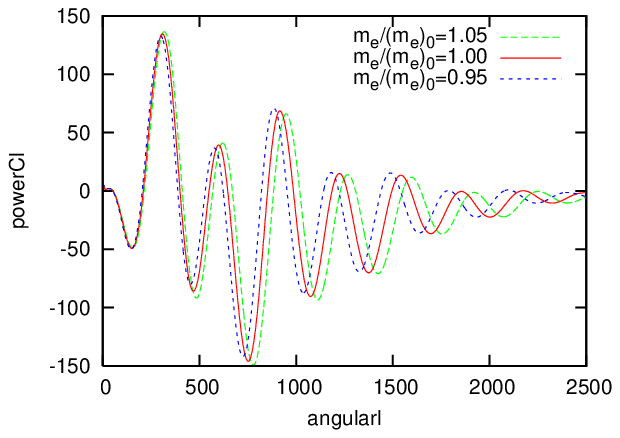}
\end{center}
\caption[Espectro de correlaci\'on cruzada entre $T$ y polarizaci\'on modo $E$ de CMB para distintos valores de $\alpha$ y $m_e$.]{Espectro de correlaci\'on cruzada entre $T$ y polarizaci\'on modo $E$ de CMB para diferentes valores de   $\alpha$ (panel superior) y $m_e$ (panel inferior).}
\label{Cls_cross_alfa_emasa}
\end{figure}

\begin{figure}[!ht]
\psfrag{angularl}{\small{$\ell$}}
\psfrag{powerCl}{\small{$\ell (\ell +1) C_{\ell}^{EE}/2\pi$}}
\begin{center}
\includegraphics[scale=1.8,angle=0]{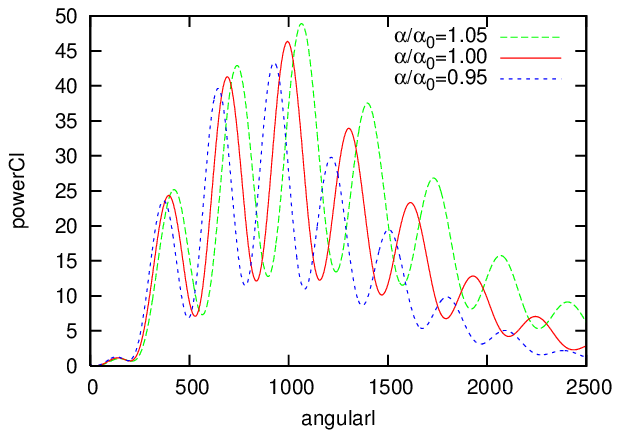}
\includegraphics[scale=1.8,angle=0]{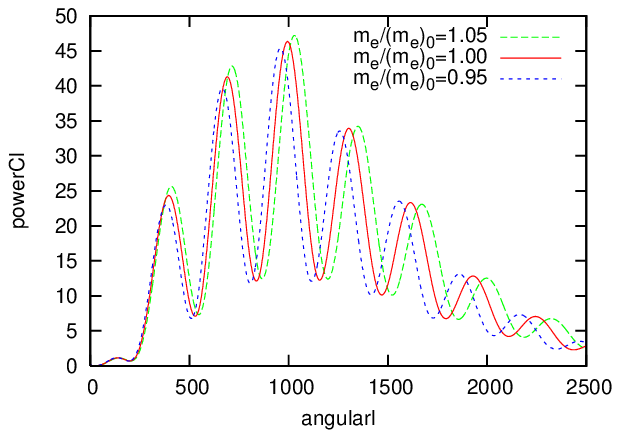}
\end{center}
\caption[Espectro de potencias de la polarizaci\'on modo $E$ de CMB, para distintos valores de $\alpha$ y $m_e$.]{Espectro de potencias de la polarizaci\'on modo $E$ de CMB para diferentes valores de
  $\alpha$ (panel superior) y $m_e$ (panel inferior).} 
\label{Cls_polar_alfa_emasa}
\end{figure}

La manera en la que cambia el espectro de correlaci\'on cruzada entre
temperatura y polarizaci\'on para distintos valores de las constantes
fundamentales, se ve en la Fig.~(\ref{Cls_cross_alfa_emasa}). El
mismo efecto, para el espectro de potencias de la polarizaci\'on,
est\'a ilustrado en la Fig.~(\ref{Cls_polar_alfa_emasa}).

\section{Estad{\'\i}sticas}
\label{sec:estadisticas}

En general, es de inter\'es en estad{\'\i}stica usar una muestra de
datos para hacer inferencias sobre un modelo probabil{\'\i}stico, por
ejemplo  para evaluar si el modelo es v\'alido o bien  para estimar
los valores de sus par\'ametros.

Hay dos enfoques principales para la inferencia estad{\'\i}stica, que
llamaremos \emph{frecuentista} y \emph{bayesiano}. En la
estad{\'\i}stica frecuentista, la probabilidad se interpreta como la
frecuencia de aparici\'on de un resultado en un experimento que se
repite. Las herramientas m\'as importantes en este enfoque son la
estimaci\'on de par\'ametros, y los tests estad{\'\i}sticos. La
estad{\'\i}stica frecuentista provee las herramientas usuales para
informar objetivamente acerca del resultado de un experimento sin
necesidad de incorporar creencias previas concernientes al par\'ametro
a ser medido o a la teor{\'\i}a a ser testeada.

En la estad{\'\i}stica bayesiana, la interpretaci\'on de probabilidad
es m\'as general, e incluye el grado de creencia. Se puede entonces
hablar de la funci\'on densidad de probabilidad para un par\'ametro,
lo cual expresa el grado de conocimiento que tenemos acerca de d\'onde
se encuentra su valor verdadero. Los m\'etodos bayesianos permiten
agregar de manera natural informaci\'on extra tal como limitaciones
f{\'\i}sicas o informaci\'on subjetiva. De hecho, requieren la
imposici\'on de un \emph{prior} para los par\'ametros, es decir, el
grado de creencia sobre los valores de los par\'ametros antes de
realizar la medida. Usando el Teorema de Bayes, el grado de creencia
\emph{a priori} se actualiza con  los datos obtenidos en el  experimento.

Para muchos problemas de inferencia, los enfoques frecuentista y
bayesiano dan los mismos resultados num\'ericos, a pesar de estar
basados en interpretaciones diferentes de la probabilidad. Sin
embargo, para muestras de datos peque\~nas, o para medidas de un
par\'ametro cerca de un l{\'\i}mite f{\'\i}sico, los distintos
enfoques pueden dar resultados diferentes, y es necesario hacer una
elecci\'on.  Elegimos el enfoque bayesiano porque es el adoptado en la
literatura y, por lo tanto, podremos comparar nuestros resultados con
los obtenidos por otros autores.

\subsection{El enfoque bayesiano}

Desde la perspectiva bayesiana, no hay una distinci\'on fundamental
entre observables y par\'ametros de un modelo estad{\'\i}stico: todos
son considerados cantidades ``desconocidas'', y se les puede asociar
un concepto de probabilidad. Sea $D$ los datos observados, y sea
$\theta$ los par\'ametros del modelo y los datos
faltantes. Formalmente, se puede construir una distribuci\'on de
probabilidad conjunta $P(D,\theta)$ sobre todas las cantidades
aleatorias y/o desconocidas. Esta distribuci\'on conjunta consta de dos 
partes: una distribuci\'on a priori $P(\theta)$ y una funci\'on Likelihood
$P(D|\theta)$.  La Likelihood es una funci\'on que expresa, luego de
que el resultado de un experimento es conocido, que tan probable era
obtener ese resultado. Es una funci\'on de los par\'ametros y no debe
confundirse con una densidad de probabilidad.  Especificar $P(\theta)$
y $P(D|\theta)$ da un modelo probabil{\'\i}stico completo, en el cual
$P(D,\theta)= P(D|\theta)P(\theta)$. Habiendo observado $D$, se usa el
teorema de Bayes para determinar la distribuci\'on de $\theta$
condicionada a $D$:
\begin{equation}
P(\theta|D) = \frac{P(\theta)P(D|\theta)}{\int P(\theta)P(D|\theta) d\theta }
\end{equation}
esto se llama la distribuci\'on \emph{posteriori} de $\theta$, y es el
objeto de la inferencia bayesiana.  A partir de esta distribuci\'on
pueden calcularse los valores esperados \emph{a posteriori} de
cualquier funci\'on $f(\theta)$:
\begin{equation}
E[f(\theta)|D] = \frac{\int f(\theta) P(\theta) P(D|\theta)
d\theta}{\int P(\theta) P(D|\theta) d\theta}.
\label{posterior_expection}
\end{equation}
Las integraciones en esta expresi\'on eran la dificultad m\'axima de
la inferencia bayesiana, y el m\'etodo de las cadenas de Markov Monte
Carlo (MCMC) surge como una manera viable de evaluar este tipo de
expresiones. Seguimos el tratamiento de \citet{GRS96}.

\subsection{Cadenas de Markov Monte Carlo}

Las cadenas de Markov Monte Carlo surgen como un m\'etodo para evaluar
expresiones del tipo  de la Ec.~(\ref{posterior_expection}) y est\'an
compuestas de dos partes: la integraci\'on Monte Carlo y la cadena de
Markov. 

\subsubsection{Integraci\'on Monte Carlo}

La integraci\'on Monte Carlo eval\'ua $E[f(X)]$ tomando muestras
$\{X_t, t=1, \ldots,n \}$ a partir de la funci\'on $P(\theta|D)$, y
luego aproximando
\begin{equation}
E[f(X)] \simeq \frac{1}{n} \sum_{t=1}^{n} f(X_t).
\end{equation}
Por lo tanto, la media de la poblaci\'on se estima con la media de la
muestra. Cuando las muestras $\{ X_t\}$ son independientes, la Ley de
los grandes n\'umeros de la estad{\'\i}stica asegura que la
aproximaci\'on se hace cada vez m\'as precisa al aumentar $n$. Es
importante notar que  $n$ no es el n\'umero de datos, sino el tama\~no de
la muestra, que est\'a controlado por el analista.

En general, tomar muestras $\{X_t\}$ a partir de $P(\theta|D)$ no es
posible dado que esta funci\'on puede ser muy poco com\'un. Sin
embargo, se pueden generar por cualquier proceso que tome muestras
dentro del soporte de $P(\theta|D)$, en las proporciones
correctas. Una manera de hacer esto es a trav\'es de una cadena de
Markov, que tenga a $P(\theta|D)$ como distribuci\'on estacionaria. En
ese caso, se trata de una \emph{cadena de Markov Monte Carlo}.

\subsubsection{Cadenas de Markov}

Supongamos que generamos una secuencia de variables aleatorias 
\begin{displaymath}
\{X_0, X_1, X_2, \ldots\}
\end{displaymath}
 de tal manera que, para cada $t \geq 0$, el siguiente estado $X_{t+1}$
 se muestree de una distribuci\'on $P(X_{t+1}|X_t)$ que dependa s\'olo
 del estado actual de la cadena, $X_t$. Esta secuencia se llama
 \emph{cadena de Markov}, y $P(\cdot| \cdot)$ se llama el \emph{kernel
 de transici\'on} de la cadena. La cadena ir\'a perdiendo
 informaci\'on de cual era el $X_0$ inicial, y la distribuci\'on de
 la cual se muestrean los valores tender\'a a una \emph{distribuci\'on
 estacionaria} $\phi(\cdot)$.

Luego de un tiempo de ``burn-in'' suficientemente largo, digamos de
$m$ iteraciones, los puntos $\{X_t; t= m+1, \ldots, n\}$ ser\'an
extra{\'\i}dos de la distribuci\'on $\phi(\cdot)$. Hay m\'etodos para
estimar el valor de $m$. Adem\'as, es necesaria una manera de asegurar
que la cadena constru{\'\i}da es tal que su distribuci\'on
estacionaria $\phi(\cdot)$ coincide con la posteriori $P(\theta|D)$.
Una de las posibilidades es utilizar el algoritmo de Metropolis-Hastings.
Luego de descartar las muestras del burn-in, el estimador de $E[f(X)]$ es:
\begin{equation}
\bar f = \frac{1}{n-m} \sum_{t=m+1}^{n} f(X_t)
\end{equation}
que se conoce como promedio erg\'odico. La convergencia al valor de
expectaci\'on requerido se asegura mediante el teorema erg\'odico.

\subsubsection{Algoritmo de Metropolis-Hastings}

Para este algoritmo, en cada paso $t$, el siguiente estado $X_{t+1}$
se elige primero muestreando un punto candidato $Y$, de la
distribuci\'on \emph{propuesta} $q(\cdot|X_t)$, que por ejemplo, puede
ser una distribuci\'on normal de variables m\'ultiples, con valor
medio $X$ y una matriz de covarianza fija. Para el punto candidato $Y$
se calcula una probabilidad $\alpha(X_t,Y)$ tal que
\begin{equation}
\alpha(X,Y) = min\left(1, \frac{P(Y)q(X|Y)}{P(X)q(Y|X)} \right).
\end{equation}
A la vez, se elige un n\'umero aleatorio entre $0$ y $1$. Luego se
compara dicho n\'umero con $\alpha(X_t,Y)$; si es menor, se toma
$X_{t+1}=Y$, sino, se toma $X_{t+1}=X_t$. Finalmente se incrementa
$t$.
Cabe destacar que la distribuci\'on propuesta $q(\cdot|\cdot)$
puede tener cualquier forma, y la distribuci\'on estacionaria de la
cadena ser\'a $P(\theta|D)$.

\subsubsection{Convergencia}

El problema fundamental de la inferencia a partir de simulaciones de
cadenas de Markov es que siempre habr\'a regiones del espacio de
par\'ametros que no ser\'an cubiertas por una cadena de longitud
finita. A medida que la simulaci\'on avanza, la propiedad erg\'odica
de la cadena de Markov hace que eventualmente se cubra toda la
regi\'on, pero en el corto plazo, las simulaciones no pueden decir
nada acerca de regiones que la cadena a\'un no haya explorado.  Esto
es un problema cuando la convergencia es lenta. La mejor manera de
notar la convergencia lenta es examinando m\'ultiples simulaciones
independientes.  Tambi\'en es importante que los puntos iniciales de
cada una de las cadenas est\'en suficientemente separados \citep{Gelman96}.

La manera recomendada para monitorear la convergencia se basa en
detectar cu\'ando las cadenas de Markov se han ``olvidado'' de sus
puntos de inicio, comparando varias secuencias largadas desde diferentes
puntos iniciales y chequeando que sean indistinguibles. Una forma
cuantitativa utiliza el an\'alisis de las varianzas: se diagnostica
convergencia aproximada cuando la varianza entre cadenas diferentes no
es mayor que la varianza dentro de una secuencia individual.

De manera independiente, se monitorea la convergencia de todos los
``res\'umenes escalares'' de inter\'es. Por ejemplo, podemos estar
interesados en todos los par\'ametros de la distribuci\'on, y varias
cantidades predichas. Cada uno de esos par\'ametros se considera un
resumen escalar $\psi$. Consideraremos uno solo, para ilustrar el
m\'etodo. Supongamos que tenemos $m$ simulaciones paralelas, cada una
de longitud $n$. Queremos una manera cuantitativa de determinar si
``las diferentes secuencias se apartan entre s{\'\i} m\'as de lo
esperado, respecto de su variabilidad interna''. Llamamos $\psi_{ij}$
con $i=1,\ldots,m$ y $j=1,\ldots,n$ a los eslabones de cada cadena.
Se define la varianza entre cadenas $B$, y la varianza dentro de las
cadenas $W$:
\begin{eqnarray}
B &=& \frac{n}{m-1} \sum_{i=1}^{m}(\bar \psi_i - \bar \psi)^2, \qquad
{\rm donde}  \qquad \bar\psi_i = \frac{1}{n} \sum_{j=1}^{n} \psi_{ij},
\qquad \bar \psi = \frac{1}{m}  \sum_{i=1}^{m}\bar \psi_i \nonumber \\
W &=& \frac{1}{m} \sum_{i=1}^{m} s_i^2 \qquad {\rm donde}  \qquad
s_i^2=\frac{1}{n-1} \sum_{j=1}^{n} (\psi_{ij} - \bar \psi_i)^2.
\end{eqnarray}

A partir de estas dos cantidades, construimos dos estimadores de la
varianza de $\psi$.  El primero es
\begin{equation}
\widehat{var}(\psi) = \frac{n-1}{n} W + \frac{1}{n} B
\end{equation}
es un estimador de la varianza  ``no-sesgado'' si los puntos
iniciales de las simulaciones fueron verdaderamente extra{\'\i}dos de la
distribuci\'on $P(\theta|D)$, pero es ``sobre-estimado'' si los puntos
iniciales est\'an apropiadamente dispersados.
 
Por otra parte, para cualquier $n$ finito, la varianza dentro de la
secuencia, $W$, debe subestimar la varianza de $\psi$ porque las
secuencias individuales no han tenido tiempo de recorrer toda la
distribuci\'on, y como resultado tendr\'an menos variabilidad. En el
l{\'\i}mite $n\rightarrow \infty$  tanto $\widehat{var}(\psi)$ como $W$ se
acercan a $var(\psi)$ pero desde direcciones opuestas.

Existe un estimador $R$ definido como 
\begin{equation}
\sqrt{R} = \sqrt{\frac{\widehat{var}(\psi)}{W}}
\end{equation}
que a medida que la secuencia converge, va acerc\'andose a $1$,
reflejando que las cadenas de Markov se est\'an superponiendo. Se
considera que la cadena ha convergido cuando $R$ es $1.2$ o $1.1$. Una
pr\'actica m\'as conservadora es tomar $R-1 \lesssim 0.03$ .

\subsection{Funci\'on Likelihood para CMB}

Las teor{\'\i}as m\'as simples que explican la generaci\'on de
inhomogeneidades en el Universo, tales como la inflacionaria, predicen
que la se\~nal de la temperatura en cierta direcci\'on del cielo es
una muestra de una distribuci\'on gaussiana. A partir de all{\'\i} se
puede deducir la funci\'on Likelihood $L$ para $N_p$ pixeles de datos
de la anisotrop{\'\i}a en la temperatura del fondo c\'osmico de
radiaci\'on como
\begin{equation}
L = \frac{1}{\left(2 \pi \right)^{N_p/2} \left( {\rm det}\, C\right)^{1/2}} \exp \left\{ -\frac{1}{2} \Delta C^{-1} \Delta \right\}
\end{equation}
donde $\Delta$ es el vector de datos que consiste en las $N_p$ medidas
de la anisotrop{\'\i}a de la temperatura en distintas direcciones en
el cielo y $C$ es la matriz de covarianza total, suma de la matriz de
covarianza del ruido m\'as la matriz de covarianza de la se\~nal.

\section{Ajustes}
\label{sec:ajustes_cmb}


Uno de los objetivos principales de esta tesis fue encontrar
l{\'\i}mites a la variaci\'on de las constantes fundamentales entre
la \'epoca de la recombinaci\'on del Universo y la actualidad.  Hemos
estudiado los casos donde s\'olo una de las constantes fundamentales
var{\'\i }a con el tiempo (ya sea $\alpha$ o $m_e$), y el caso donde
la variaci\'on es simult\'anea, aunque independiente.

Para el an\'alisis estad{\'\i}stico, hemos utilizado distintos
conjuntos de datos y comparado c\'omo cambian las cotas obtenidas en
cada caso. Asimismo, cotejamos nuestros resultados con los disponibles
en la literatura.
Hemos utilizado el conjunto m\'as completo de datos de las
anisotrop{\'\i}as en la temperatura y polarizaci\'on del Fondo
C\'osmico de Radiaci\'on, que incluye los resultados de
CBI \citep{CBI04}, ACBAR \citep{ACBAR02}, y
BOOMERANG \citep{BOOM05_polar,BOOM05_temp}. Utilizamos tambi\'en los
datos del sat\'elite WMAP.  Para \'estos, tenemos dos conjuntos de
datos, que fueron liberados en distintos momentos durante la
realizaci\'on de este trabajo de tesis: los recolectados durante los
primeros tres a\~nos de funcionamiento del sat\'elite, liberados en el
a\~no 2005, que llamaremos WMAP3 \citep{wmap3a}, y los publicados en el
a\~no 2008, que llamaremos WMAP5 \citep{wmap5}, y son el resultado de cinco
a\~nos de observaci\'on.

Dada la fuerte degeneraci\'on que existe entre la variaci\'on de las
constantes y el factor de expansi\'on de Hubble $H_0$, hemos realizado
un ajuste con los datos de CMB y un prior gaussiano para $H_0$, el
cual fue obtenido del Proyecto Clave del Telescopio espacial
Hubble \citep{hst01}. No utilizamos el ajuste final para $H_0$ que
presentan en dicho trabajo, dado que algunos de los objetos utilizados
est\'an a alto redshift. En cambio, utilizamos los valores obtenidos a
partir de los objetos celestes m\'as cercanos, de manera tal de poder
despreciar cualquier posible diferencia entre el valor de $\alpha$ y
$m_e$ a ese redshift determinado y el valor actual. El valor medio de
la gaussiana considerada en esta tesis es de 71 km/s/Mpc, y la
desviaci\'on est\'andar es de 10 km/s/Mpc, teniendo en cuenta errores
aleatorios y sistem\'aticos. Mencionamos que el valor final que
presentan en \citet{hst01} es de $H_0 = 72 \pm 8$ km/s/Mpc.

El espectro de potencias de las galaxias (o de la materia) $P(k)$ es
creciente para valores peque\~nos de $k$ (grandes escalas) y
decreciente para $k$ grandes. La escala para la cual se presenta el
quiebre corresponde a aquella que entra al horizonte en la \'epoca de
igualdad de materia y radiaci\'on. El factor de escala correspondiente
es $a_{eq}= 4,15\times10^{-5}/\Omega_mh^2$. Esta dependencia con
$\Omega_m h^2$ nos permite romper la degeneraci\'on que existe con
$\alpha$ y $m_e$, y obtener cotas m\'as restrictivas para sus
variaciones. Por lo tanto, hemos utilizado los datos del espectro de
potencias de la distribuci\'on de galaxias del cat\'alogo
2dFGRS \citep{2dF05}, junto con los de CMB para obtener cotas a la
variaci\'on de las constantes fundamentales.

Clasificamos los ajustes realizados seg\'un el conjunto de datos que
se haya utilizado:

\begin{itemize}

\item Todas las observaciones de CMB, incluyendo los datos de WMAP3.

\item Todas las observaciones de CMB, incluyendo los datos de WMAP3, junto
con el prior para $H_0$ del HST Key Project, obtenido a partir de
objetos cercanos.

\item Todas las observaciones de CMB, incluyendo los datos de WMAP3, junto
 con el espectro de potencias $P(k)$ del cat\'alogo 2dFGRS.

\item Todas las observaciones de CMB, incluyendo los datos de WMAP5, junto
con el espectro de potencias $P(k)$ del cat\'alogo 2dFGRS.

\end{itemize}

En los an\'alisis estad{\'\i}sticos, hemos considerado un modelo
cosmol\'ogico plano con fluctuaciones en densidad adiab\'aticas, con
el siguiente conjunto de par\'ametros cosmol\'ogicos, que tambi\'en
son ajustados:

\begin{equation}
P=(\Omega_b h^2, \Omega_{CDM} h^2, \Theta, \tau_{re}, n_s, A_s)
\label{conj_parametros}
\end{equation}
donde $\Omega_{b} h^2$ es la densidad de materia bari\'onica, en
unidades de la densidad cr{\'\i}tica, $\Omega_{CDM} h^2$ es la
densidad de materia oscura, en las mismas unidades, $\Theta$ da el
cociente entre el horizonte com\'ovil de sonido en la \'epoca de
desacople y la distancia angular a la superficie de \'ultimo
scattering, $\tau_{re}$ es la profundidad \'optica de reionizaci\'on,
$n_s$ es el {\'\i}ndice espectral de perturbaciones escalares, y $A_s$
es la amplitud de las fluctuaciones escalares.

Para llevar a cabo el an\'alisis estad{\'\i}stico, utilizamos el
m\'etodo de las cadenas de Markov para explorar el espacio de
par\'ametros porque el uso de grillas cuando el n\'umero de
par\'ametros es tan grande demanda un tiempo de c\'omputo
excesivamente largo. Estas t\'ecnicas est\'an implementadas en el
c\'odigo CosmoMC de \citet{LB02}. El mismo, utiliza el c\'odigo
CAMB \citep{LCL00} para calcular los espectros de potencias angulares
de CMB, y el c\'odigo R{\sc ecfast} \citep{recfast} para resolver las
ecuaciones de recombinaci\'on. Modificamos dichos programas para
incluir variaciones en el valor de $\alpha$ y de $m_e$ durante la
recombinaci\'on. En estos an\'alisis, hemos considerado la
recombinaci\'on est\'andar, seg\'un el tratamiento de
\citet{seager00,seager99}. Hemos realizado el an\'alisis variando
 las constantes por separado, y en forma conjunta, y los resultados se
muestran en las secciones subsiguientes. Se han considerado priors
planos para todos par\'ametros (excepto cuando se tiene en cuenta los
datos del HST, en cuyo caso se toma un prior gaussiano para $H_0$,
como se mencion\'o anteriormente).  Para cada uno de los an\'alisis
llevados a cabo, se han corrido 8 cadenas de Markov, y se ha utilizado
el criterio de convergencia de \citet{Raftery&Lewis} para detener las
cadenas cuando el estimador $R-1$ toma un valor bajo (que
especificaremos en cada caso).  En general hemos trabajado con cadenas
largas ya que, para asegurar la convergencia, no hemos cortado el
proceso hasta que el valor del estimador $R-1$ fuera m\'as bajo que
los valores que se consideran habitualmente.

Finalmente, comentamos que {\it Planck} ser\'a la primera misi\'on en
realizar un mapa de todo el cielo con una sensibilidad del orden del mJy
y una resoluci\'on mejor que 10' \citep{Villa03}. Tal resoluci\'on
permitir\'a ver dentro de la cola de decaimiento (damping) del espectro de
anisotrop{\'\i}as, cerca del tercer y cuarto pico, con una precisi\'on
limitada exclusivamente por la varianza c\'osmica
\citep{White06}. Esto permitir\'a una determinaci\'on muy precisa del
cociente bari\'on-fot\'on a partir de la altura relativa de los picos
del espectro.

\subsection{Variaci\'on de $\alpha$}
\label{variacion_alfa}

Para poner cotas a la variaci\'on temporal de $\alpha$ durante la
recombinaci\'on, al conjunto de par\'ametros (\ref{conj_parametros})
le agregamos $\Delta \alpha/\alpha_0$, que tiene en cuenta posibles
variaciones en el valor de la constante de estructura fina respecto de
su valor actual $\alpha_0$.

Existen algunos an\'alisis previos de los datos de CMB incluyendo una
posible variaci\'on de la constante de estructura fina $\alpha$.  Los
primeros trabajos que presentan cotas a partir de datos del fondo
c\'osmico de radiaci\'on son previos al sat\'elite WMAP. Ajustes a la
variaci\'on de $\alpha$ usando datos de BOOMERANG y MAXIMA son
presentados en \citet{AV00,BCW01}, y usando BOOMERANG, COBE y DASI
en \citet{av01}.  En los trabajos de \citet{Martins02,Rocha03} se
utilizan los datos de WMAP1 (recolectados durante el primer a\~no de
funcionamiento del sat\'elite) y se obtiene
\begin{itemize}
\item $-0.06 < \Delta \alpha / \alpha_0 < 0.01$ 
\end{itemize}
con un 95\% de confianza.  Por otra parte, en \citet{Ichi06} utilizan
los datos de WMAP1 y el prior gaussiano para $H_0$ dado por el HST Key
Project ($H_0= 72 \pm 8$ km/s/Mpc).  El m\'etodo estad{\'\i}stico que
utilizan es el c\'alculo del m{\'\i}nimo del $\chi^2$ como funci\'on
de $\alpha$. Para minimizar sobre los otros par\'ametros
cosmol\'ogicos, utilizan iterativamente el m\'etodo de Brent, de la
interpolaci\'on sucesiva por par\'abolas.  Las cotas que obtienen son
\begin{itemize}
\item $-0.107 < \Delta \alpha/ \alpha_0 < 0.043$, usando los datos de WMAP1, y
\item$-0.048 < \Delta \alpha/ \alpha_0 < 0.032$, cuando se agrega el prior
de HST.
\end{itemize}
  Stefanescu \citep{stefanescu07} encontr\'o l{\'\i}mites a la
variaci\'on de $\alpha$ utilizando los datos de WMAP3 junto con el
prior para $H_0$ dado por el HST Key Project. Por otra parte, en dicho
trabajo se desprecia el efecto de la variaci\'on de $\alpha$ sobre la
recombinaci\'on del helio. El l{\'\i}mite presentado por esta autora
es
\begin{itemize}
\item $-0.039 < \Delta \alpha/\alpha_0 < 0.010$,
\end{itemize}
 con un 95\% de confianza.  Por \'ultimo, \citet{nakashima08}
utilizan los datos de WMAP5 para obtener cotas sobre la variaci\'on de
$\alpha$. Los resultados que obtienen son 
\begin{itemize}
\item $-0.028 < \Delta \alpha/ \alpha_0 < 0.026$, si se usa el prior del HST, y 
\item $-0.050 < \Delta \alpha / \alpha_0 < 0.042$, si no se lo utiliza.
\end{itemize}
 Resaltamos que estos autores tambi\'en usan el resultado final para
$H_0$, calculado a partir de objetos a distintos redshifts.

En nuestro an\'alisis considerando s\'olo datos de WMAP3 encontramos 
la siguiente cota
\begin{equation}
-0.065 < \Delta \alpha / \alpha_0 < 0.019
\end{equation}
con una confianza del 95\%.  Para estas cadenas, el estimador de
convergencia vale $R-1= 0.0266$.  Cuando incorporamos el prior para
$H_0$ del HST, calculado a partir de objetos cercanos ($H_0= 71 \pm
10$ km/s/Mpc), el intervalo de confianza se hace m\'as peque\~no
\begin{equation}
-0.047 < \Delta \alpha / \alpha_0 < 0.014.
\end{equation}
La diferencia entre nuestro resultado y el presentado
por \citet{stefanescu07} se debe al prior elegido para $H_0$. En esta
tesis preferimos considerar el prior obtenido a partir de objetos
cercanos.  Si a las cadenas correspondientes al an\'alisis realizado
con los datos de WMAP3 se les realiza un post-procesamiento para agregar
el prior para $H_0$ dado por el valor final del HST Key Project, el
intervalo del 95\% de confiabilidad que obtenemos es
\begin{equation}
-0.040 < \Delta \alpha / \alpha_0 < 0.014.
\end{equation}
Tanto Stefanescu como Nakashima et al, no han considerado la
posibilidad de que la cota sobre $H_0$ pueda verse afectada si
$\alpha$ tiene un valor diferente al actual en las \'epocas
correspondientes a los redshifts de los objetos utilizados en el HST
Key Project.

\begin{figure}[!ht]
\begin{center}
\includegraphics[scale=2.,angle=0]{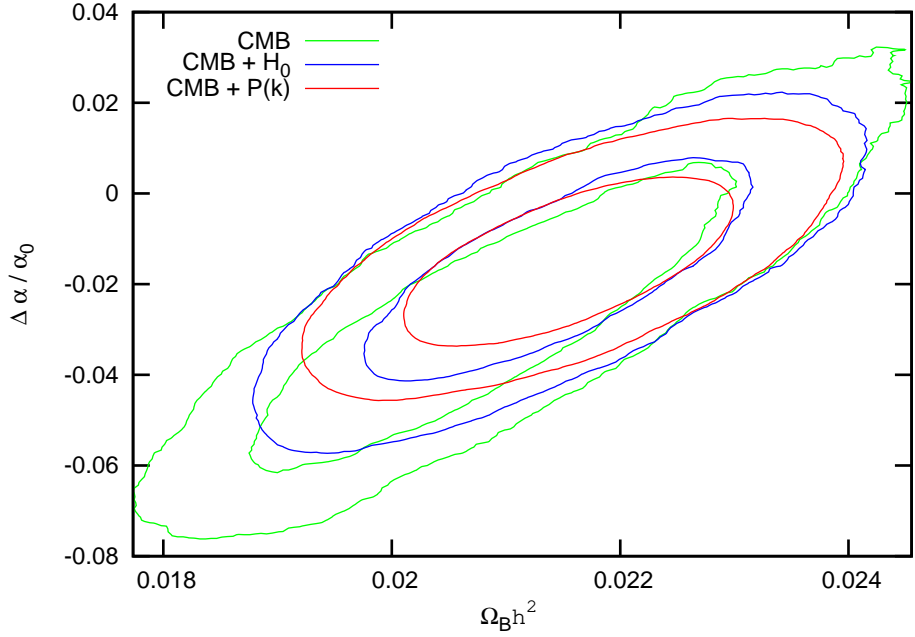}
\end{center}
\caption[Contornos de confianza de $\Delta \alpha / \alpha_0$ y $\Omega_bh^2$ para distintos conjuntos de datos.]{Contornos de confianza del 68\% y 95\%  para los par\'ametros $\Delta \alpha / \alpha_0$ y $\Omega_bh^2$ obtenidos con  datos de WMAP3, con datos de WMAP3 m\'as el espectro de potencias  $P(k)$ del 2dFGRS, y con WMAP3 y el prior para $H_0$ calculado a  partir de objetos cercanos, para el an\'alisis con variaci\'on de  $\alpha$.}
\label{individuales_alfa_ob}
\end{figure}

\begin{figure}[!ht]
\begin{center}
\includegraphics[scale=2.,angle=0]{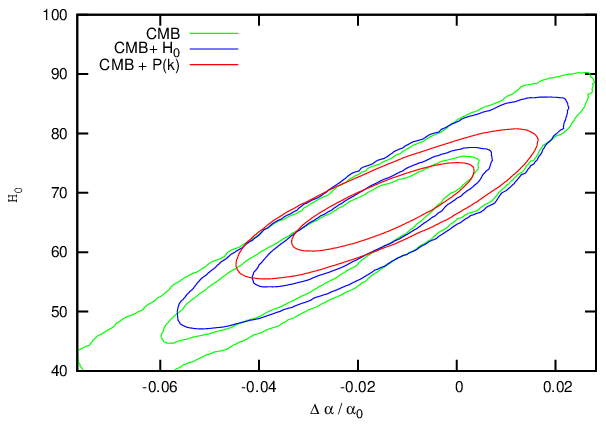}
\end{center}
\caption[Contornos de confianza para $\Delta \alpha / \alpha_0$ y  $H_0$ para distintos conjuntos de datos.]{Contornos de confianza del  68\% y 95\%  para los par\'ametros $\Delta \alpha / \alpha_0$ y  $H_0$  obtenidos con  datos de WMAP3, con y sin datos del espectro  de potencias del   2dFGRS, para el an\'alisis con variaci\'on de $\alpha$.}
\label{individuales_alfa_H0}
\end{figure}

Las cotas m\'as fuertes se obtienen al combinar los datos de WMAP
con los del $P(k)$ del 2dFGRS (ver Figs.~(\ref{individuales_alfa_ob})
y (\ref{individuales_alfa_H0})).

Cuando utilizamos los datos de WMAP3 y el espectro del 2dFGRS,
cortamos las cadenas cuando $R-1 < 0.003$.  En este caso, el intervalo
de confianza del 95\% es
\begin{equation}
-0.038 < \Delta \alpha / \alpha_0 <0.009.
\end{equation}
Estos resultados est\'an publicados en \citet{Mosquera07}.

Finalmente, realizamos el ajuste con los datos de WMAP5 y 2dFGRS. Las
cadenas se cortaron cuando $R-1= 0.0213$.  En este caso, el intervalo
de confianza del 95\% es
\begin{equation}
 -0.019 < \Delta \alpha / \alpha_0 < 0.017.
\label{cota_nuestra_alfa}
\end{equation}
Esta cota es m\'as fuerte que la presentada en \citet{nakashima08}.
Efectivamente, en las Figs.~(\ref{individuales_alfa_ob}) y
(\ref{individuales_alfa_H0}) se puede ver que los contornos son m\'as
peque\~nos cuando se agregan a CMB los datos del $P(k)$ que cuando se
agrega el prior para $H_0$.

\begin{figure*}[p]
\begin{center}
\includegraphics[scale=1.2,angle=0]{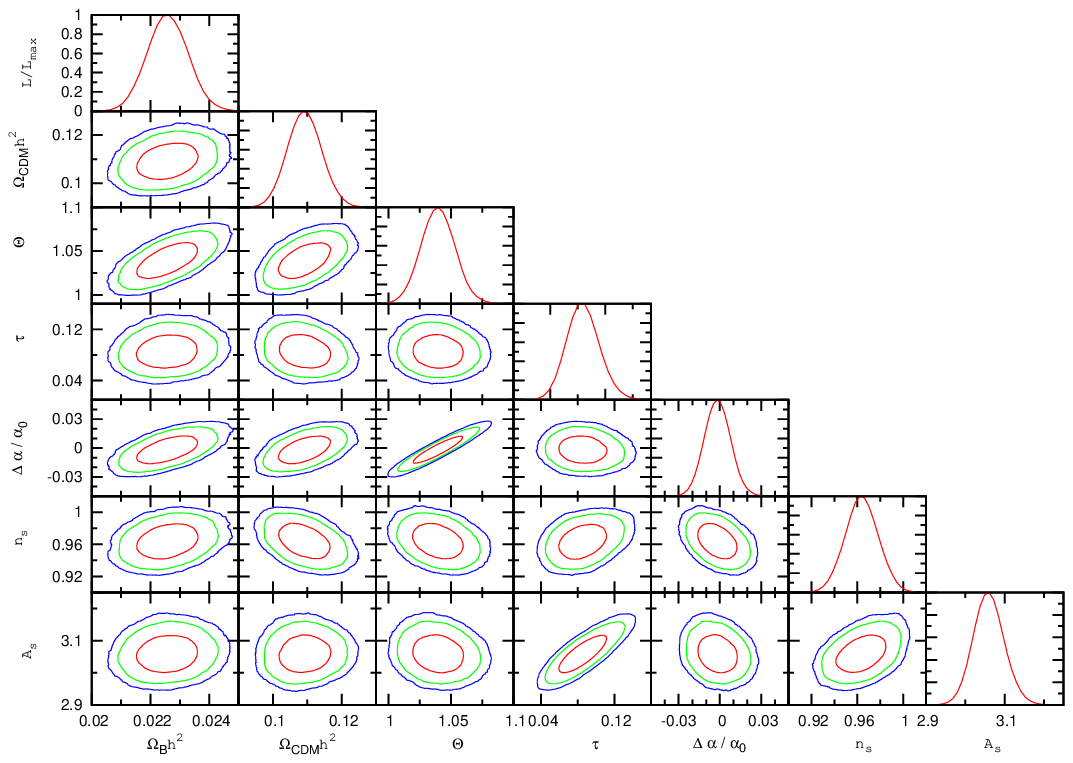}
\end{center}
\caption[Distribuciones de probabidad y contornos de confianza. Variaci\'on de $\alpha$.]{Distribuciones
a posteriori marginalizadas, para el caso de variaci\'on de $\alpha$
obtenidas con los datos de CMB, incluyendo los datos de WMAP5 y el
espectro de potencias del 2dFGRS. En la diagonal se muestran las
distribuciones de densidad de probabilidad a posteriori para los
par\'ametros individuales indicados en el eje $x$. Estas
distribuciones est\'an marginalizadas sobre los otros par\'ametros y
normalizadas a $1$ en el m\'aximo. Los otros paneles muestran los
contornos de confianza para pares de par\'ametros. Los contornos
representan 68\%, 95\% y 99\% de confiabilidad, desde el m\'as interno
al m\'as externo, respectivamente.  An\'alisis estad{\'\i}stico con
variaci\'on de $\alpha$.}
\label{resulcmb_alfa_wmap5}
\end{figure*}

En la Fig.~(\ref{resulcmb_alfa_wmap5}) se muestran las distribuciones
a posteriori marginalizadas, para el caso de variaci\'on de $\alpha$
obtenidas con los datos de CMB, incluyendo los datos de WMAP5 y el
espectro de potencias del 2dFGRS. En la diagonal se muestran las
distribuciones de densidad de probabilidad a posteriori para los
par\'ametros individuales indicados en el eje $x$. Estas
distribuciones est\'an marginalizadas sobre los otros par\'ametros y
normalizadas a $1$ en el m\'aximo. Los otros paneles muestran los
contornos de confianza para pares de par\'ametros. Los contornos
representan 68\%, 95\% y 99\% de confiabilidad, desde el m\'as interno
al m\'as externo, respectivamente. En dicha figura se puede ver una
degeneraci\'on fuerte entre $\alpha$ y $\Theta$, esta \'ultima
directamente relacionada con $H_0$, y tambi\'en entre $\alpha$ y
$\Omega_b h^2$. Los valores obtenidos para $\Omega_b h^2$, $h$,
$\Omega_{CDM} h^2$, $\tau_{re}$, y $n_s$ est\'an en acuerdo, dentro
del intervalo a $1\sigma$, con los correspondientes valores cuando no
se incluye variaci\'on de $\alpha$, obtenidos por el equipo de
WMAP \citep{wmap3a}.  Los resultados aqu{\'\i} presentados son
consistentes a $2 \sigma$ con variaci\'on nula de $\alpha$ en
la \'epoca de recombinaci\'on. Para el ajuste con los datos de WMAP3 se
obtienen resultados similares, aunque las cotas son m\'as precisas
cuando se utilizan los datos de WMAP5.

Los resultados de los ajustes, con WMAP3 y WMAP5, para los
par\'ametros cosmol\'ogicos y $\Delta \alpha / \alpha_0$ se muestran
en la Tabla~(\ref{tablacmb_al}).
\begin{table}[!ht]
\begin{center}
\renewcommand{\arraystretch}{1.3}
\begin{tabular}{|c|c|c|}
\hline
Par\'ametro & WMAP3 + $P(k)$  &  WMAP5 + $P(k)$ \\
\hline 
$\Omega_b h^2$ & $0.0216 \pm 0.0009$ & $0.0226\pm 0.0007$   \\ 
\hline 
$\Omega_{CDM} h^2$ & $0.102 \pm 0.006$ & $0.110 \pm 0.005$  \\ 
\hline
 $\Theta$ & $1.021 \pm 0.017$ & $1.040 \pm 0.013$\\ 
\hline
 $\tau_{re}$ & $0.092 \pm 0.014$  & $0.086_{-0.0008}^{+0.007}$ \\
 \hline
 $\Delta \alpha / \alpha_0$  & $-0.015 \pm 0.012$  & $-0.002\pm 0.009$ \\ 
\hline 
$n_s$ &  $0.965 \pm 0.016$  & $0.964 \pm 0.014$ \\
 \hline 
$\log\left(10^{10}A_s\right)$ & $3.039_{-0.065}^{+0.064}$  &  $3.060 \pm 0.037$  \\ 
\hline 
$H_0$ &  $67.7_{-4.6}^{+4.7}$  &  $71.7_{-4.1}^{+4.0}$ 
\\ \hline
\end{tabular}
\caption{Valores medios y errores a  1$\sigma$ para los par\'ametros
  incluyendo la variaci\'on de $\alpha$, para el an\'alisis con WMAP3 y el an\'alisis con WMAP5.
 Las unidades de $H_0$ son ${\rm km \, \, s^{-1} \, \, Mpc^{-1}}$.}
 \label{tablacmb_al}
\end{center}
\end{table}
A modo de comparaci\'on, en la Tabla~(\ref{tabla_parametrosCMB}) se
muestran los ajustes para los par\'ametros cosmol\'ogicos, publicados
por el equipo del sat\'elite WMAP, en el modelo cosmol\'ogico
can\'onico donde no hay variaci\'on de las constantes. Vemos que los
valores medios de los par\'ametros cosmol\'ogicos pr\'acticamente no
cambian si se considera variaci\'on de $\alpha$, a excepci\'on de
$\Omega_bh^2$, que en caso de variar $\alpha$ prefiere valores m\'as
bajos que en el caso can\'onico, para ambos conjuntos de datos. Otra
diferencia que se observa es en los valores de $\tau_{re}$ y $H_0$,
para el ajuste con WMAP3. En este caso, cuando hay variaci\'on de
$\alpha$, se prefiere un valor ligeramente m\'as alto para
$\tau_{re}$, y un valor m\'as bajo para $H_0$.

\begin{figure}[!ht]
\begin{center}
\includegraphics[scale=1.5,angle=0]{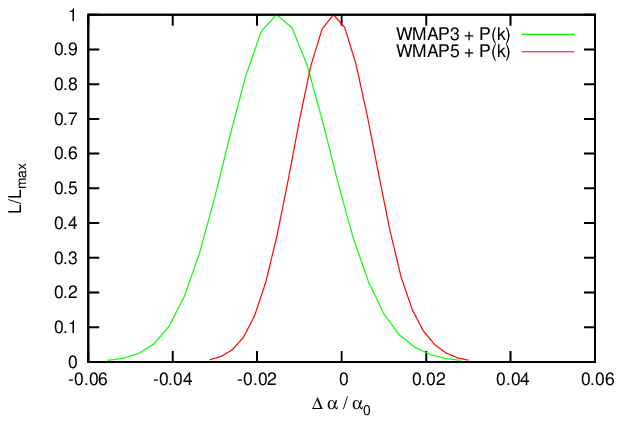}
\end{center}
\caption[Distribuci\'on de probabilidad para $\Delta \alpha / \alpha_0$, comparando los ajustes con WMAP3 y WMAP5.]{Distribuci\'on a posteriori unidimensional para $\Delta \alpha / \alpha_0$, comparando los ajustes con WMAP3 y WMAP5. An\'alisis con variaci\'on s\'olo de $\alpha$.}
\label{compara_wmap3_wmap5_alfa}
\end{figure}

\begin{table}[!ht]
\begin{center}
\renewcommand{\arraystretch}{1.3}
\begin{tabular}{|c|c|c|}
\hline
Par\'ametro &   WMAP3  &  WMAP5 \\
\hline
100 $\Omega_b h^2$ &   $2.229 \pm 0.073$ & $2.273 \pm 0.062$ \\
\hline
$\Omega_{CDM}h^2$ &  $0.1054 \pm 0.0087$ &  $0.1099 \pm 0.0062$ \\
\hline
$\tau_{re}$ &  $0.089 \pm 0.030$ & $0.087 \pm 0.017$   \\
\hline
$n_s$ & $0.958 \pm 0.016$ & $0.963_{-0.015}^{+0.014}$ \\
\hline
$H_0$ & $73.2_{-3.2}^{+3.1}$ & $71.9_{-2.7}^{+2.6}$ \\
\hline
\end{tabular}
\caption{Valores medios y errores a  1$\sigma$ para los par\'ametros cosmol\'ogicos a partir de los datos de   WMAP, sin variaci\'on de las constantes. Las unidades de $H_0$ son ${\rm km \, \,
 s^{-1} \, \, Mpc^{-1}}$.}
\label{tabla_parametrosCMB}
\end{center}
\end{table}
En la Fig.~(\ref{compara_wmap3_wmap5_alfa}) se compara c\'omo
cambia la distribuci\'on de densidad de probabilidad para el
par\'ametro $\Delta \alpha / \alpha_0$ cuando se consideran los datos
de WMAP3 y los de WMAP5. En ambos casos, tambi\'en se utilizaron los
datos del espectro de potencias del 2dFGRS.

\begin{figure}[!ht]
\begin{center}
\includegraphics[scale=2,angle=0]{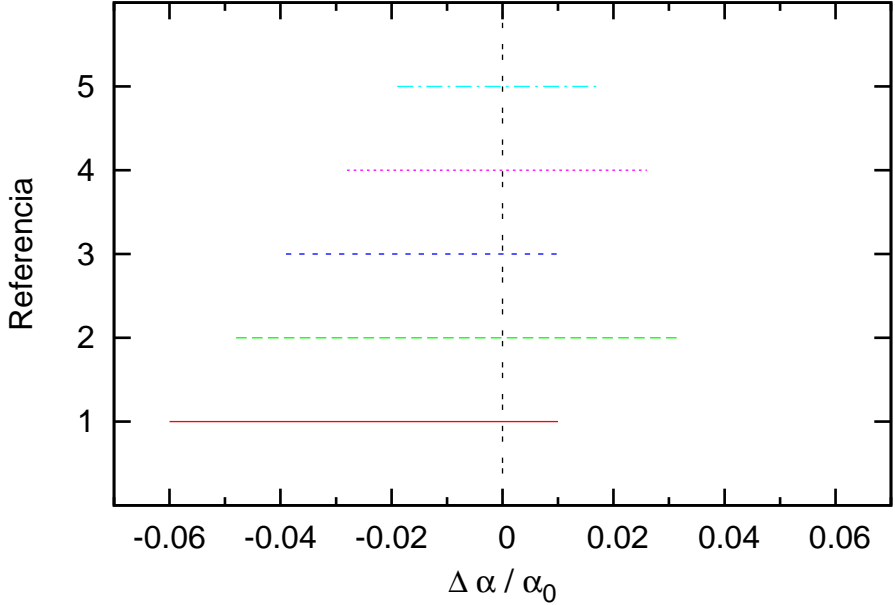}
\end{center}
\caption[Cotas existentes para la variaci\'on de $\alpha$.]{Comparaci\'on de las cotas existentes para la variaci\'on de $\alpha$ entre la recombinaci\'on y la \'epoca actual. La l{\'\i}nea vertical indica variaci\'on nula. Referencias: 1).\ Martins et al. 2002, Rocha et al. 2003 -- 2).\ Ichikawa et al. 2006 -- 3).\ Stefanescu. 2007 -- 4).\ Nakashima et al. 2008. -- 5).\ Resultado de esta tesis (Ec.~(\ref{cota_nuestra_alfa})).}
\label{todas_cotas_alfa}
\end{figure}

En la Fig.~(\ref{todas_cotas_alfa}) se resume cuales son las cotas
m\'as actuales para la variaci\'on de $\alpha$, entre la \'epoca de
recombinaci\'on y la \'epoca actual, existentes en la literatura.

\subsection{Variaci\'on de $m_e$}
\label{variacion_me}

Para encontrar cotas a la variaci\'on de $m_e$ entre la \'epoca de
recombinaci\'on y la actualidad, al conjunto de par\'ametros
(\ref{conj_parametros}) le agregamos la variable $\Delta m_e
/(m_e)_0$.

Existen algunos trabajos previos que estudian la variaci\'on de la
masa del electr\'on usando datos de CMB. En \citet{KS00} realizan un
an\'alisis de c\'omo afecta al espectro de potencias de las
anisotrop{\'\i}as de CMB un valor distinto de la masa del electr\'on, y
estiman la sensibilidad de los experimentos a la variaci\'on de $m_e$,
pero no se llevan a cabo estudios estad{\'\i}sticos.  En \citet{YS03} se realiza
un estudio de los l{\'\i}mites que imponen los datos de
nucleos{\'\i}ntesis primordial y CMB sobre la variaci\'on del valor
medio de vac{\'\i}o del campo Higgs (que, como mencionamos en la
Introducci\'on, durante la recombinaci\'on s\'olo afecta a la masa del
electr\'on). Estos autores utilizan datos de VSA, BOOMERANG, DASI,
MAXIMA y CBI, todos previos a WMAP. Presentan gr\'aficos con los
contornos de probabilidad, y mencionan que las cotas obtenidas con CMB
son m\'as d\'ebiles que las obtenidas con los datos de BBN, pero no
publican expl{\'\i}citamente los valores que obtienen.

En esta tesis se presenta la \'unica cota que existe en la literatura para la
variaci\'on de la masa del electr\'on encontrada a partir de datos del
fondo c\'osmico de radiaci\'on.  En los an\'alisis estad{\'\i}sticos
que realizamos para este estudio, nuevamente hemos utilizado los datos
de CMB, el prior de HST y el espectro de potencias del 2dFGRS para
poner cotas a los par\'ametros cosmol\'ogicos y al valor de la masa
del electr\'on cuando se permite variar esta \'ultima durante
la \'epoca de recombinaci\'on. A continuaci\'on detallamos los
resultados.

Para el an\'alisis utilizando todos los datos de CMB, incluyendo los
datos de WMAP3, obtuvimos el siguiente intervalo de confianza de 95\%
\begin{equation}
-0.154 < \Delta m_e / (m_e)_0 < 0.091
\end{equation}
aunque la degeneraci\'on entre $m_e$ y $H_0$ es tan grande que las
cotas no resultan confiables, ya que las cadenas llegan a los extremos
del prior plano para $H_0$ ($40<H_0<100$). Al agregar a este conjunto
de datos el prior para $H_0$ del HST Key Project obtenido a partir de
objetos cercanos, se rompe la fuerte degeneraci\'on, y se obtiene un
intervalo m\'as estricto. El intervalo de confianza del 95\% es
\begin{equation}
-0.104 < \Delta m_e / (m_e)_0 < 0.060.
\end{equation}
En este caso, las cadenas fueron cortadas cuando el valor del
estimador de convergencia fue $R-1= 0.029$.

\begin{figure}[!ht]
\begin{center}
\includegraphics[scale=2.,angle=0]{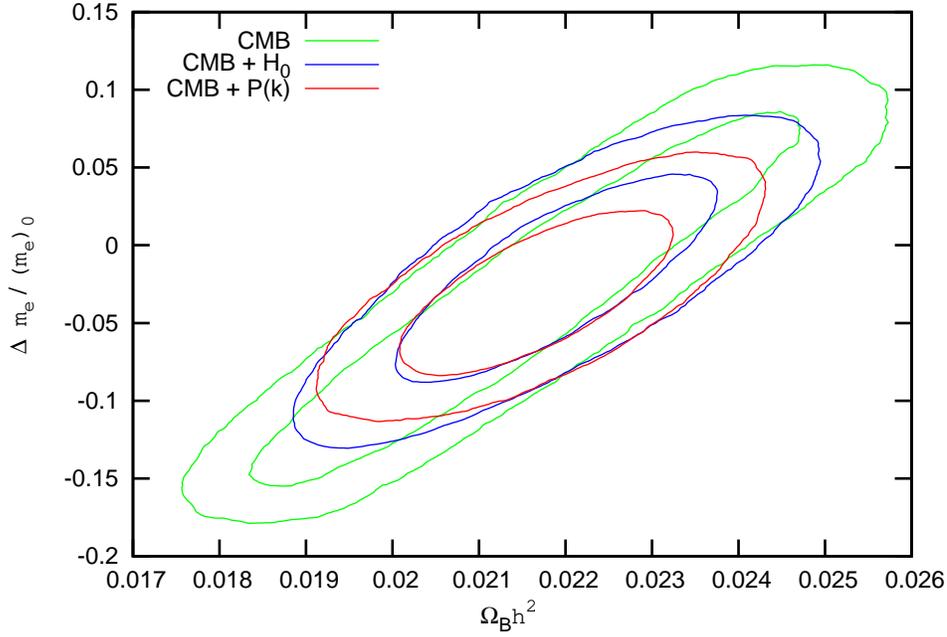}
\end{center}
\caption[Contornos de confianza para $\Delta m_e / (m_e)_0$ y $\Omega_bh^2$ para distintos conjuntos de datos. ]{Contornos de confianza del 68\% y 95\%  para los par\'ametros $\Delta m_e / (m_e)_0$ y $\Omega_bh^2$ obtenidos con
  datos de WMAP3, con datos de WMAP3 m\'as el espectro de potencias
  $P(k)$ del 2dFGRS, y con WMAP3 y el prior para $H_0$ calculado a
  partir de objetos cercanos, para el an\'alisis con variaci\'on de
  $m_e$.}
\label{individuales_emasa_ob}
\end{figure}

\begin{figure}[!ht]
\begin{center}
\includegraphics[scale=2.,angle=0]{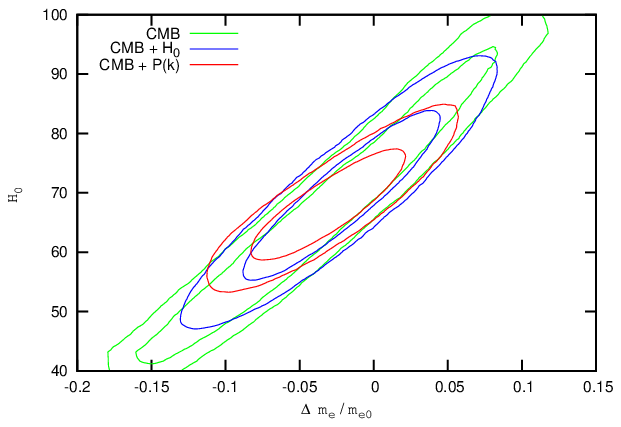}
\end{center}
\caption[Contornos de confianza para $\Delta m_e / (m_e)_0$ y $H_0$  para distintos conjuntos de datos. ]{Contornos de confianza del 68\% y 95\%  para los par\'ametros $\Delta m_e / (m_e)_0$ y $H_0$  obtenidos con
  datos de WMAP3, con y sin datos del espectro de potencias del
  2dFGRS, para el an\'alisis con variaci\'on de $m_e$.}
\label{individuales_emasa_H0}
\end{figure}

Cuando realizamos el an\'alisis utilizando los mismos datos de CMB, es
decir incluyendo WMAP3, m\'as el espectro $P(k)$ del 2dFGRS,
obtenemos cotas consistentes, aunque m\'as estrictas:
\begin{equation}
-0.093 < \Delta m_e / (m_e)_0 < 0.040.
\end{equation}
 Esto muestra tambi\'en la consistencia del an\'alisis. En las
Figs.~(\ref{individuales_emasa_ob}) y (\ref{individuales_emasa_H0}) se
muestra c\'omo los contornos se hacen m\'as peque\~nos cuando se
agrega los datos del 2dFGRS al an\'alisis que cuando se agrega el
prior para $H_0$.  En este caso, las cadenas se cortaron cuando se
alcanz\'o el valor $R-1 < 0.0044$. Los resultados se muestran en la
Tabla~(\ref{tablacmb_emasa}).

Finalmente, realizamos el ajuste con los datos de WMAP5 y 2dFGRS. Las
cadenas se cortaron cuando $R-1= 0.0169$.  En este caso, el intervalo
de confianza del 95\% es
\begin{equation}
 -0.071 <  \Delta m_e / (m_e)_0 < 0.053.
\end{equation}

\begin{figure*}[p]
\begin{center}
\includegraphics[scale=1.2,angle=0]{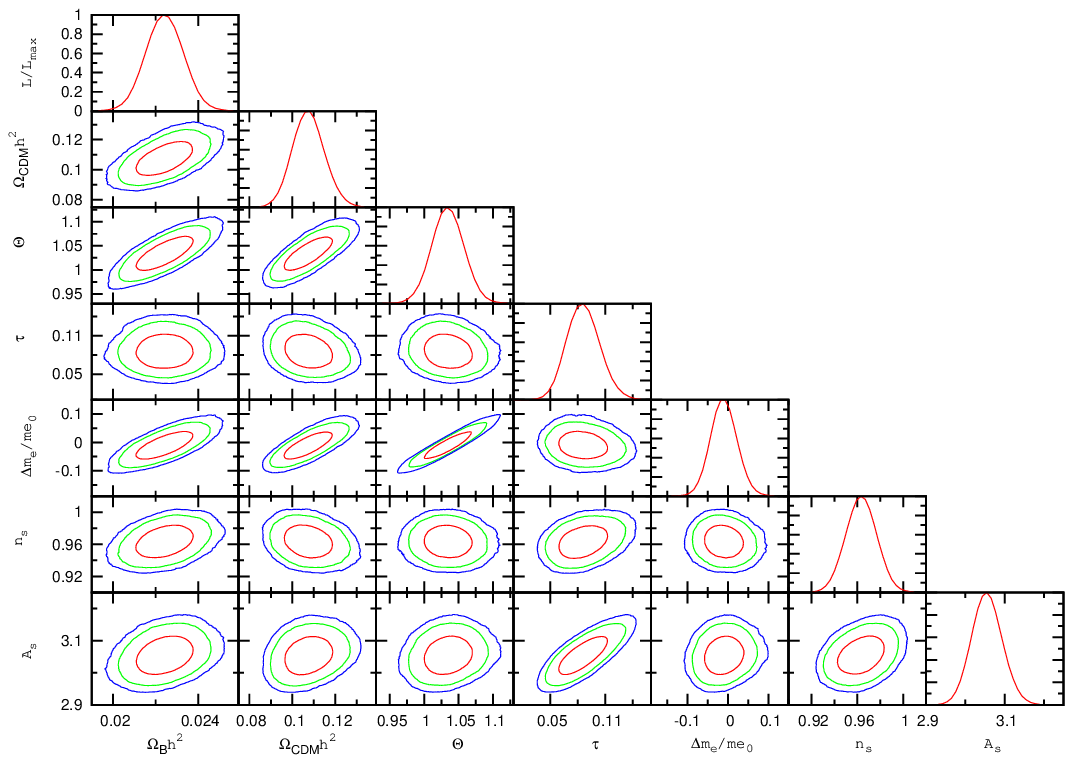}
\end{center}
\caption[Distribuciones de probabilidad y contornos de confianza. Variaci\'on de $m_e$.]{Distribuciones 
a posteriori marginalizadas, para el caso de variaci\'on de $m_e$
obtenidas con los datos de CMB, incluyendo los datos de WMAP5 y el
espectro de potencias del 2dFGRS. En la diagonal se muestran las
distribuciones de densidad de probabilidad a posteriori para los
par\'ametros individuales indicados en el eje $x$. Estas
distribuciones est\'an marginalizadas sobre los otros par\'ametros y
normalizadas a $1$ en el m\'aximo. Los otros paneles muestran los
contornos de confianza para pares de par\'ametros. Los contornos
representan 68\%, 95\% y 99\% de confiabilidad, desde el m\'as interno
al m\'as externo, respectivamente.  An\'alisis estad{\'\i}stico con
variaci\'on de $m_e$.}
\label{resulcmb_emasa_wmap5}
\end{figure*}

En la Fig.~(\ref{resulcmb_emasa_wmap5}) se muestran las distribuciones
a posteriori marginalizadas, para el caso de variaci\'on de $m_e$
obtenidas con los datos de CMB, incluyendo los datos de WMAP5 y el
espectro de potencias del 2dFGRS. En la diagonal se muestran las
distribuciones a posteriori para los par\'ametros individuales. Los
otros paneles muestran los contornos 2D para pares de par\'ametros,
marginalizando sobre los dem\'as. En dicha figura se puede ver una
fuerte degeneraci\'on entre $m_e$ y $\Theta$, y tambi\'en entre $m_e$
y $\Omega_b h^2$, y entre $m_e$ y $\Omega_{CDM} h^2$. En el caso de
utilizar los datos de WMAP3, en lugar de WMAP5, se observan las mismas
degeneraciones, s\'olo que en el \'ultimo caso, los ajustes son m\'as
precisos.

Los resultados de los ajustes, con WMAP3 y WMAP5, para los
par\'ametros cosmol\'ogicos y $\Delta m_e /(m_e)_0$ se muestran
en la Tabla~(\ref{tablacmb_emasa}).
\begin{table}[!ht]
\begin{center}
\renewcommand{\arraystretch}{1.3}
\begin{tabular}{|c|c|c|}
\hline
Par\'ametro & WMAP3 + $P(k)$  &  WMAP5 + $P(k)$ \\
\hline 
$\Omega_b h^2$ & $0.0217 \pm 0.0010$ & $0.0224\pm 0.0008$   \\ 
\hline 
$\Omega_{CDM} h^2$ & $0.101 \pm 0.009$ & $0.108 \pm 0.007$  \\ 
\hline
 $\Theta$ & $1.020 \pm 0.025$ & $1.034 \pm 0.022$\\ 
\hline
 $\tau_{re}$ & $0.091_{-0.014}^{+0.013}$  & $0.086_{-0.008}^{+0.007}$ \\
 \hline
 $\Delta m_e / (m_e)_0$  & $-0.029 \pm 0.034$  & $-0.010\pm 0.030$ \\ 
\hline 
$n_s$ &  $0.960 \pm 0.015$  & $0.963 \pm 0.013$ \\
 \hline 
$\log\left(10^{10}A_s\right)$ & $3.020 \pm 0.064$  &  $3.055 \pm 0.037$  \\ 
\hline 
$H_0$ &  $68.1_{-6.0}^{+5.9}$  &  $70.6\pm 5.8$ 
\\ \hline
\end{tabular}
\caption{Valores medios y errores a  1$\sigma$ para los par\'ametros
  incluyendo la variaci\'on de $m_e$, para el an\'alisis con WMAP3 y el an\'alisis con WMAP5.
 Las unidades de $H_0$ son ${\rm km \, \, s^{-1} \, \, Mpc^{-1}}$.}
 \label{tablacmb_emasa}
\end{center}
\end{table}
Comparando con los valores de los par\'ametros en el caso can\'onico,
se observa que el valor de $\Omega_bh^2$ es menor cuando se permite
que $m_e$ var{\'\i}e, para ambos conjuntos de datos. Por otra parte,
para el ajuste con WMAP3, el valor preferido para $H_0$ es menor
cuando $m_e$ var{\'\i}a que en el caso can\'onico.

\begin{figure}[!ht]
\begin{center}
\includegraphics[scale=1.5,angle=0]{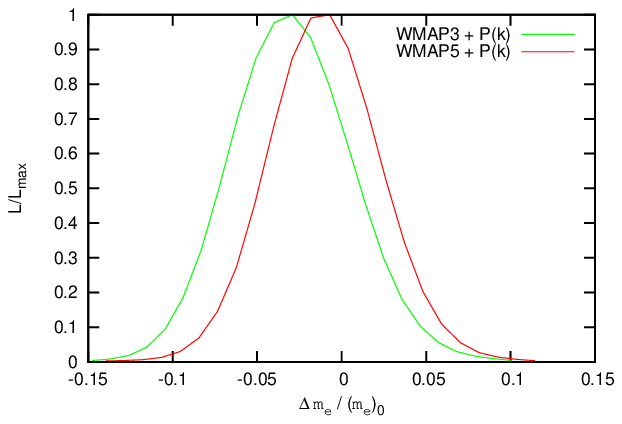}
\end{center}
\caption[Distribuci\'on de probabilidad para $\Delta m_e / (m_e)_0$, comparando los ajustes con WMAP3 y WMAP5.]{Distribuci\'on a posteriori unidimensional para  $\Delta m_e / (m_e)_0$, comparando los ajustes con WMAP3 y WMAP5. An\'alisis con variaci\'on s\'olo de $m_e$.}
\label{compara_wmap3_wmap5_emasa}
\end{figure}

En la Fig.~(\ref{compara_wmap3_wmap5_emasa}) se compara c\'omo
cambia la distribuci\'on de densidad de probabilidad para el
par\'ametro $\Delta m_e / (m_e)_0$ cuando se consideran los datos
de WMAP3 y los de WMAP5. En ambos casos, tambi\'en se utilizaron los
datos del espectro de potencias del 2dFGRS.

\subsection{Variaci\'on conjunta de $\alpha$ y $m_e$}
\label{variacion_al_me}

Para poner cotas a la variaci\'on conjunta de $\alpha$ y
$m_e$ durante la recombinaci\'on, al conjunto de par\'ametros
(\ref{conj_parametros}) le agregamos las variables
$\Delta \alpha/\alpha_0$ y $\Delta m_e / (m_e)_0$.

El \'unico trabajo en la literatura que presentaba cotas para la
variaci\'on conjunta de $\alpha$ y $m_e$ usando datos de CMB
es el de \citet{Ichi06}. En dicho trabajo se propone una relaci\'on funcional
para las variaciones de $\alpha$ y $m_e$ que estar{\'\i}a motivada en
la teor{\'\i}a de cuerdas. Los resultados que presentan son los
l{\'\i}mites sobre la variaci\'on temporal de la constante de
estructura fina $\alpha$ utilizando los datos de WMAP1 y el prior
gaussiano para $H_0$ dado por el HST Key Project. Como no se conoce la
relaci\'on entre las variaciones porque la teor{\'\i}a est\'a en
construcci\'on, suponen que $m_e$ var{\'\i}a con una ley de potencias
de $\alpha$.  A partir de la acci\'on a bajas energ{\'\i}as derivada
de la Teor{\'\i}a de cuerdas heter\'otica en el marco de Einstein,
derivan la relaci\'on
\begin{equation}
\frac{m_e + \Delta m_e}{(m_e)_0} = \left(\frac{\alpha + \Delta \alpha}{\alpha} \right)^{1/2}.
\end{equation}
Tambi\'en consideran otras posibilidades, fenomenol\'ogicamente, pero
con forma de ley de potencias:
\begin{equation}
\frac{m_e + \Delta m_e}{(m_e)_0} = \left(\frac{\alpha + \Delta \alpha}{\alpha} \right)^{p}
\end{equation}
con $p= 0, 2, 4$. Notar que el caso $p=0$ corresponde a variaci\'on
s\'olo de $\alpha$. La cota que obtienen en este caso fue comparada en
la secci\'on~\ref{variacion_alfa}.

El m\'etodo estad{\'\i}stico que utilizan es el c\'alculo del
m{\'\i}nimo del $\chi^2$ como funci\'on de $\alpha$. Para minimizar
sobre los otros par\'ametros cosmol\'ogicos, utilizan iterativamente
el m\'etodo de Brent, de la interpolaci\'on sucesiva por par\'abolas.
 La cota que obtienen es 
\begin{itemize}
\item $-0.097 < \Delta \alpha/ \alpha < 0.034$ al 95\% usando s\'olo los datos de WMAP1, y 
\item $-0.042 < \Delta \alpha / \alpha < 0.026$ al combinar estos datos con la cota
para $H_0$ que se obtiene del HST Key Project.
\end{itemize}

En esta tesis se realiz\'o el \'unico an\'alisis que hay en la
literatura sobre la variaci\'on conjunta de $\alpha$ y $m_e$,
permitiendo que las mismas sean independientes. Parte de estos
resultados est\'an publicados en \citet{landau08}. Se utilizaron datos
actualizados de CMB, y en distintos estudios se combin\'o esta
informaci\'on con el prior para $H_0$ calculado a partir de los
objetos m\'as cercanos, y con el espectro de potencias $P(k)$ del
cat\'alogo 2dFGRS.  Nuestros l{\'\i}mites no pueden ser comparados con
los de \citet{Ichi06} dado que estos autores realizan el ajuste
fijando una relaci\'on determinada entra las variaciones de ambas
constantes. Los resultados se detallan a continuaci\'on.

En el an\'alisis estad{\'\i}stico que realizamos con los datos de CMB
incluyendo WMAP3, encontramos los siguientes intervalos de 95\% de confianza
\begin{equation}
-0.077 < \Delta \alpha /\alpha_0 < 0.024 \qquad -0.135< \Delta m_e/(m_e)_0 < 0.237
\end{equation}
y al combinar estos datos con el prior para $H_0$ calculado a partir
del HST con objetos cercanos, los intervalos son
\begin{equation}
-0.078 < \Delta \alpha / \alpha_0 < 0.023 \qquad -0.096 < \Delta m_e / (m_e)_0 < 0.215.
\end{equation}
Vemos que no se modifican demasiado los intervalos de confianza al
agregar el prior para $H_0$.

En el an\'alisis con los datos de WMAP3 m\'as el $P(k)$ del cat\'alogo
 2dFGRS, cortamos las cadenas cuando $R-1<0.015$.  Los intervalos del
 95\% de confianza en este caso son
\begin{equation}
-0.068 < \Delta \alpha /\alpha_0 < 0.031 \qquad -0.116 < \Delta m_e/(m_e)_0 < 0.198.
\end{equation}

Los resultados se muestran en la Tabla~(\ref{tablacmb}).  Los valores
 de los par\'ametros cosmol\'ogicos son similares a aquellos obtenidos
 considerando la variaci\'on de una sola constante a la vez
 \citep{Mosquera07,Scoccola08}. Adem\'as, los resultados son
 consistentes dentro de 1$\sigma$ con variaci\'on nula de $\alpha$ y
 $m_e$ durante la recombinaci\'on.

\begin{figure}[!ht]
\begin{center}
\includegraphics[scale=1.,angle=0]{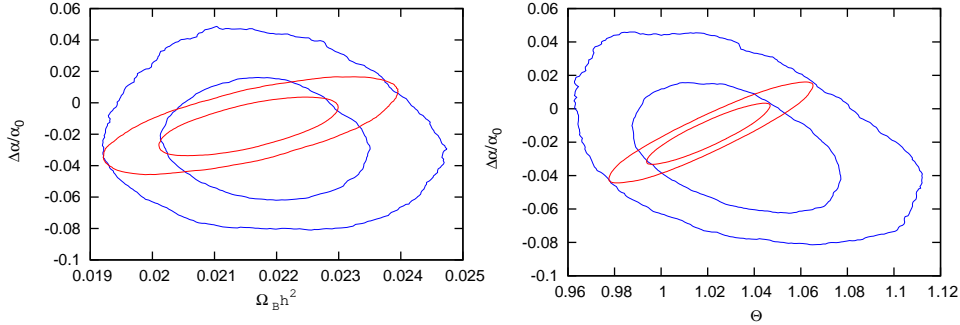}
\end{center}
\caption[Comparaci\'on de los contornos cuando se var{\'\i}an ambas constantes, o s\'olo $\alpha$.]{Contornos de confianza a 1$\sigma$ y 2$\sigma$. L{\'\i}nea azul:
variaci\'on de $\alpha$ y $m_e$; l{\'\i}nea roja: S\'olo
variaci\'on de $\alpha$. Los par\'ametros cosmol\'ogicos son libres de
variar en ambos casos.} \label{multi_alfa}
\end{figure}

\begin{figure}[!ht]
\begin{center}
\includegraphics[scale=1.,angle=0]{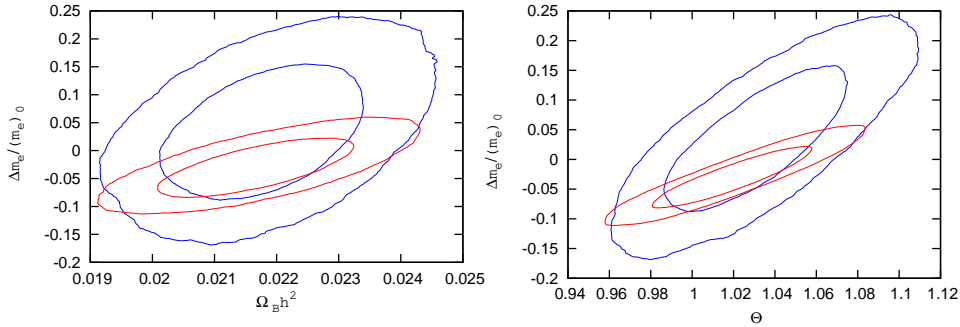}
\end{center}
\caption[Comparaci\'on de los contornos cuando se var{\'\i}an ambas constantes, o s\'olo $m_e$.]{Contornos de confianza a 1$\sigma$ y 2$\sigma$. L{\'\i}nea azul:
variaci\'on de $\alpha$ y $m_e$; l{\'\i}nea roja: S\'olo
variaci\'on de  $m_e$. Los par\'ametros cosmol\'ogicos son libres de
variar en ambos casos.}
\label{multi_emasa}
\end{figure}

En las Fig.~(\ref{multi_alfa}) y (\ref{multi_emasa}) comparamos las
degeneraciones que existen entre diferentes par\'ametros
cosmol\'ogicos y las constantes fundamentales cuando una o ambas
constantes pueden variar. En ambos casos, la regi\'on permitida en el
espacio de par\'ametros es mayor cuando las dos constantes pueden
variar. Esto es esperable dado que en ese caso, la dimensi\'on del espacio de
par\'ametros es mayor y por lo tanto, las incertezas en los
par\'ametros son mayores. Las correlaciones de $\frac{\Delta
  \alpha}{\alpha_0}$ con los otros par\'ametros cosmol\'ogicos cambia
el signo cuando se permite variar tambi\'en $m_e$. Por el contrario,
las correlaciones de  $\frac{\Delta m_e}{(m_e)_0}$ con los par\'ametros
cosmol\'ogicos no cambian el signo cuando se permite tambi\'en variar $\alpha$.

Cuando s\'olo se permite variar una constante fundamental, la
correlaci\'on entre esa constante y cualquier par\'ametro
cosmol\'ogico particular tiene el mismo signo, sin importar si la
constante es $\alpha$ o $m_e$. Esto puede ser entendido sobre la base
de que ambas constantes entran en las mismas cantidades
f{\'\i}sicas. Sin embargo, dado que la forma funcional de la
dependencia es diferente, los valores medios del mejor ajuste para las
variaciones temporales de las constantes fundamentales son diferentes
y la distribuci\'on de probabilidad es m\'as extendida en un caso que
en el otro. De todas maneras, en los casos en que se permite variar
s\'olo una constante, \'esta prefiere un valor m\'as bajo que el valor
actual.

\begin{figure}[p]
\begin{center}
\includegraphics[scale=1.1,angle=0]{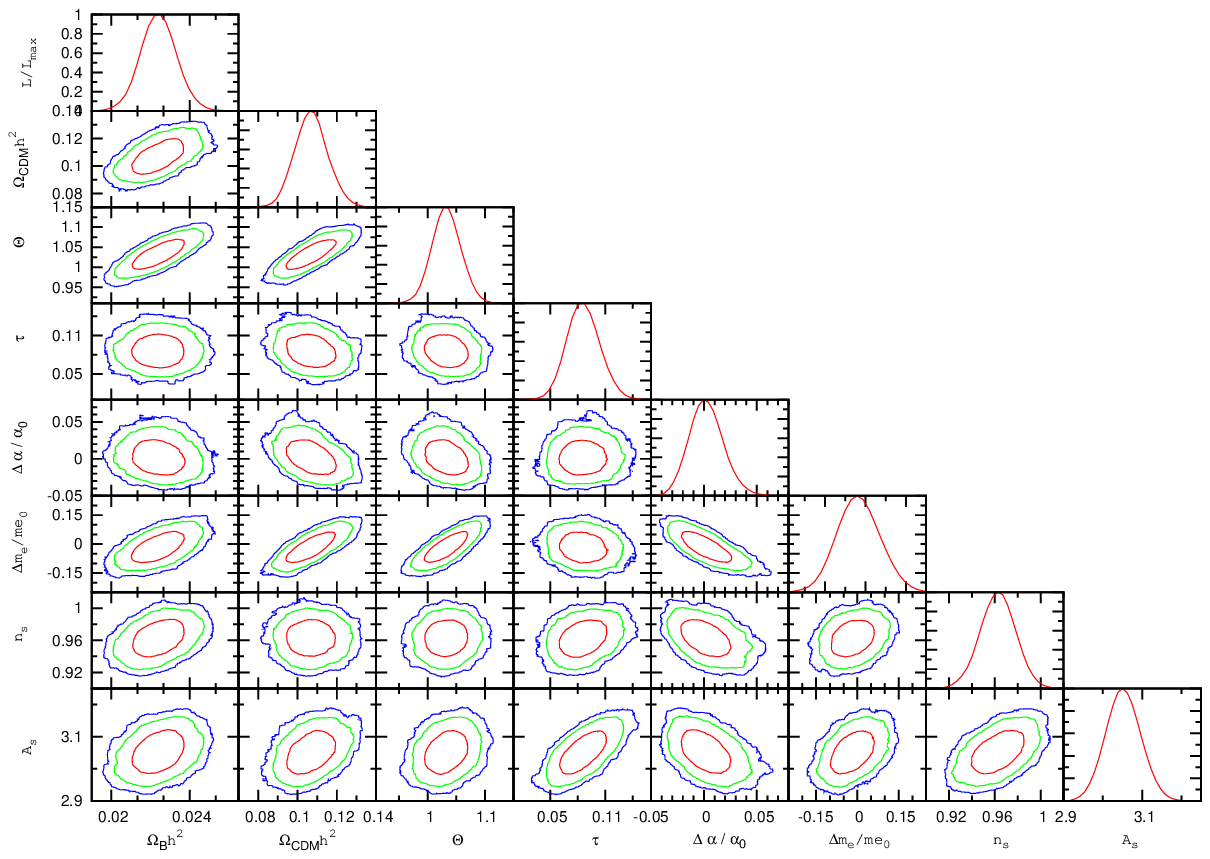}
\end{center}
\caption[Distribuciones de probabidad y contornos de confianza. Variaci\'on de $\alpha$ y $m_e$.]{Distribuciones 
a posteriori marginalizadas, para el caso de variaci\'on de $\alpha$ y
$m_e$ obtenidas con los datos de CMB, incluyendo los datos de WMAP5 y
el espectro de potencias del 2dFGRS. En la diagonal se muestran las
distribuciones de densidad de probabilidad a posteriori para los
par\'ametros individuales indicados en el eje $x$. Estas
distribuciones est\'an marginalizadas sobre los otros par\'ametros y
normalizadas a $1$ en el m\'aximo. Los otros paneles muestran los
contornos de confianza para pares de par\'ametros. Los contornos
representan 68\%, 95\% y 99\% de confiabilidad, desde el m\'as interno
al m\'as externo, respectivamente.  An\'alisis estad{\'\i}stico con
con variaci\'on conjunta de $\alpha$ y $m_e$.}
\label{resulcmb_al_me}
\end{figure}

Finalmente, realizamos el ajuste con los datos de WMAP5 y 2dFGRS. Las
cadenas se cortaron cuando $R-1= 0.018$. Los resultados se muestran
en la Tabla~(\ref{tablacmb}) y en la Fig.~(\ref{resulcmb_al_me}).
En este caso, los intervalos de confianza del 95\% son
\begin{equation}
-0.026 <\Delta \alpha / \alpha_0 < 0.036 \qquad  -0.115 < \Delta m_e / (m_e)_0 < 0.086.
\end{equation}

\begin{table}
\begin{center}
\begin{tabular}{|c|c|c|}
\hline
Par\'ametro &  wmap5 & wmap3 \\
\hline
 $\Omega_b h^2$   & $0.0224 \pm 0.0009 $ & $0.0218\pm 0.0010$ \\
\hline
 $\Omega_{CDM} h^2$   & $0.107 \pm 0.008$ & $0.106 \pm 0.011$ \\
\hline 
$\Theta$ & $1.033_{-0.023}^{+0.024}$ & $1.033_{-0.029}^{+0.028}$ \\
\hline
$\tau$ &  $0.086 \pm 0.008$ & $0.090 \pm 0.014$ \\
\hline
$\Delta \alpha / \alpha_0$   & $0.003\pm 0.015$  & $-0.023 \pm 0.025$\\
\hline 
$\Delta m_e /(m_e)_0$   & $-0.017 \pm 0.051$ & $0.036 \pm 0.078$\\
\hline
$n_s$   & $0.963 \pm 0.015$  & $0.970 \pm 0.019$\\
\hline
$A_s$  & $3.052 \pm 0.043$  & $3.054 \pm 0.073$ \\
\hline
$H_0$  &   $70.3_{-6.0}^{+6.1}$  &  $70.4_{-6.8}^{+6.6}$  \\
\hline
\end{tabular}
\caption{Valores medios y errores a 1$\sigma$ para los par\'ametros,
incluyendo variaci\'on de $\alpha$ y $m_e$.  Comparaci\'on de resultados entre los an\'alisis con WMAP3 y WMAP5.}
\label{tablacmb}
\end{center}
\end{table}

Son notables las fuertes degeneraciones que existen entre
$\frac{\Delta m_e}{(m_e)_0}$ y $\Omega_{CDM} h^2$, $\frac{\Delta
m_e}{(m_e)_0}$ y $\Theta$, $\frac{\Delta \alpha}{\alpha_0}$ y $n_s$, y
finalmente entre $\frac{\Delta \alpha}{\alpha_0}$ y $\frac{\Delta
m_e}{(m_e)_0}$. Los valores obtenidos para $\Omega_b h^2$, $h$,
$\Omega_{CDM} h^2$, $\tau_{re}$, y $n_s$ concuerdan, a 1$\sigma$, con
aquellos del equipo de WMAP \citep{wmap3a}, donde no se considera
variaci\'on de $\alpha$ ni $m_e$.  Los resultados para los valores de
los par\'ametros cosmol\'ogicos son similares a aquellos obtenidos
considerando la variaci\'on de una sola constante a la vez. Los
resultados son consistentes dentro de 1$\sigma$ con variaci\'on nula
de $\alpha$ y $m_e$ durante la recombinaci\'on.

\begin{figure}[!ht]
\begin{center}
\includegraphics[scale=2.,angle=0]{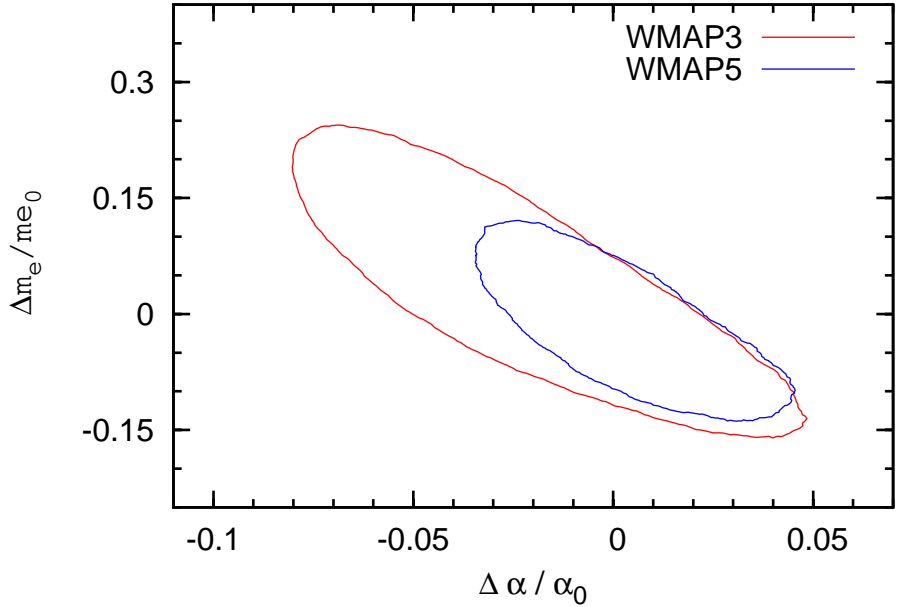}
\end{center}
\caption[Contornos de confianza de $\Delta \alpha / \alpha_0$ y  $\Delta m_e / (m_e)_0$, con WMAP3 o WMAP5.]{Contornos de confianza del 95\%, para los par\'ametros $\Delta \alpha / \alpha_0$ y  $\Delta m_e / (m_e)_0$, correspondientes a los ajustes realizados con WMAP3 y con WMAP5.}
\label{contorno_wmap3_wmap5}
\end{figure}

Al comparar los contornos de las distribuciones marginalizadas para
 $\Delta \alpha / \alpha_0$ versus $\Delta m_e / (m_e)_0$, que se
 obtienen de los an\'alisis con los datos de WMAP3 y WMAP5, se
 distingue que los contornos no s\'olo se hacen m\'as peque\~nos para
 los nuevos datos, sino que tambi\'en cambian los valores medios de los par\'ametros (ver
 Fig.~(\ref{contorno_wmap3_wmap5})). Los valores medios de
 $\Delta \alpha / \alpha_0$ y de $\Delta m_e / (m_e)_0$ cambian de
 signo, como puede verse en la Tabla~(\ref{tablacmb}).

\section{Comparaci\'on entre escenarios de recombinaci\'on}


Dado que los datos del fondo c\'osmico de radiaci\'on son cada vez
m\'as precisos, es necesario tener en cuenta fen\'omenos cada vez
m\'as sutiles que puedan afectar el escenario de recombinaci\'on. En
el a\~no 2008, la recombinaci\'on del helio fue estudiada con sumo
cuidado, como se explic\'o en el Cap.~\ref{ch:fisica_recomb}. En esta
tesis hicimos un an\'alisis detallado de c\'omo entran las constantes
fundamentales en la recombinaci\'on del helio, y modificamos la
versi\'on actualizada del c\'odigo de evoluci\'on R{\sc ecfast} para
introducir dichas dependencias. En esta secci\'on mostramos cuales son
las cotas para la variaci\'on de $\alpha$ y $m_e$ cuando se tiene en
cuenta la recombinaci\'on detallada del helio, y comparamos con las
cotas obtenidas en la secci\'on anterior, en la cual se consider\'o el
escenario est\'andar de la recombinaci\'on, descripto
en \citet{seager00}. Los resultados de este an\'alisis est\'an
publicados en \citet{Scoccola08b}.

Realizamos el an\'alisis estad{\'\i}stico teniendo en
cuenta la posible variaci\'on de $\alpha$, de $m_e$, y de $\alpha$ y
$m_e$ simult\'aneamente, en la \'epoca de la recombinaci\'on en el
escenario descripto en la Sec.~\ref{sec:eq_recomb}. Para este
an\'alisis utilizamos los datos de CMB incluyendo WMAP5, y el espectro
de potencias $P(k)$ del 2dFGRS.

Exploramos el espacio de par\'ametros con 8 cadenas de Markov, y
seguimos el criterio de convergencia de  \citet{Raftery&Lewis} para
detener las cadenas cuando $R-1< 0.0180$.
 Las cotas sobre los par\'ametros no se ven mayormente afectadas cuando
se cambia el escenario de recombinaci\'on, como puede verse en la
Tabla~(\ref{cotas_distintos_escenarios}).
\begin{table}
\renewcommand{\arraystretch}{1.3}
\begin{center}
\begin{tabular}{|c|c|c|c|}
\hline
An\'alisis & $\Delta \alpha / \alpha_0$ & $\Delta m_e /(m_e)_0$ \\
\hline
Variaci\'on conjunta - PS & $0.003 \pm 0.015$ & $-0.017 \pm 0.051$ \\
\hline
Variaci\'on conjunta - NS & $0.004 \pm 0.015$ & $-0.019 \pm 0.049$  \\
\hline
99999Variaci\'on $\alpha$ - PS & $-0.002\pm 0.009$ & --- \\
\hline
Variaci\'on $\alpha$ - NS & $-0.001 \pm 0.009$ & --- \\ 
\hline
Variaci\'on $m_e$ - PS &---  &$-0.010\pm 0.030$ \\
\hline
Variaci\'on $m_e$ - NS & --- & $-0.010_{-0.030}^{+0.031}$ \\
\hline
\end{tabular}
\caption{Cotas para la variaci\'on de las constantes en distintos 
escenarios de recombinaci\'on. NS significa el nuevo escenario de
recombinaci\'on, y PS significa el escenario est\'andar. Se presentan
los resultados para todos los an\'alisis.}
\label{cotas_distintos_escenarios}
\end{center}
\end{table}
Los valores de los par\'ametros cosmol\'ogicos pr\'acticamente no
cambian entre este ajuste y el realizado teniendo en cuenta el
escenario est\'andar.

\begin{figure}[!ht]
\begin{center}
\includegraphics[scale=2.0,angle=0]{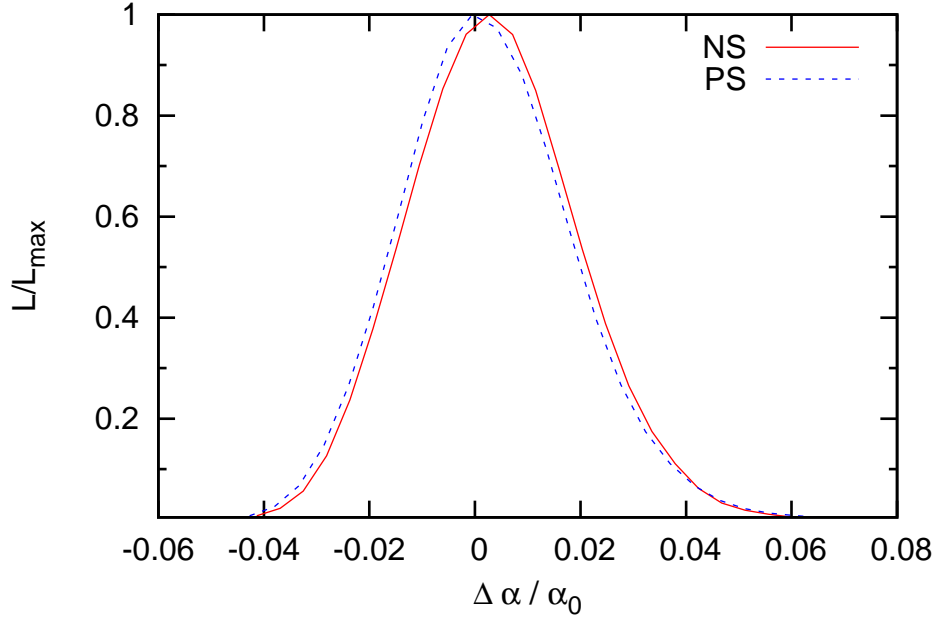}
\end{center}
\caption[Distribuci\'on de probabilidad para $\alpha$ en dos escenarios de recombinaci\'on diferentes.]{Likelihood unidimensional para $\frac{\Delta
\alpha}{\alpha_0}$ para  dos
escenarios de recombinaci\'on diferentes.}
\label{compara_recomb_alfa}
\end{figure}

\begin{figure}[!ht]
\begin{center}
\includegraphics[scale=2.0,angle=0]{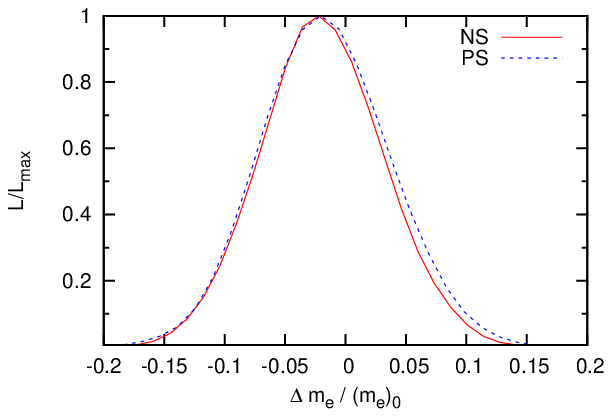}
\end{center}
\caption[Distribuci\'on de probabilidad para $m_e$ en dos escenarios
de recombinaci\'on diferentes.]{Likelihood unidimensional para
$\frac{\Delta m_e}{(m_e)_0}$ para dos escenarios de recombinaci\'on diferentes.}
\label{compara_recomb_emasa}
\end{figure}

En la Fig.~(\ref{compara_recomb_alfa}), comparamos la distribuci\'on
de probabilidad para $\Delta \alpha/\alpha_0$ en diferentes
escenarios. En la Fig.~(\ref{compara_recomb_emasa}), hacemos lo mismo
para $\Delta m_e/(m_e)_0$. Puede verse que la diferencia es
 despreciable.

\begin{figure}[!ht]
\begin{center}
\includegraphics[scale=2,angle=0]{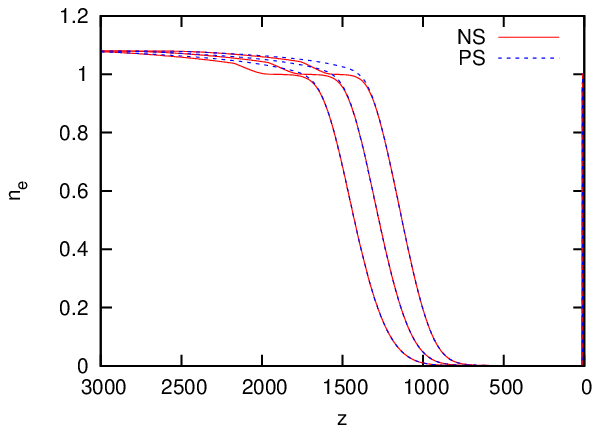}
\end{center}
\caption[Historia de ionizaci\'on para $\alpha$ variable, en dos escenarios de recombinaci\'on diferentes.]{Historia de ionizaci\'on permitiendo que varie $\alpha$ con
  el tiempo. De izquierda a derecha, los valores de $\alpha/\alpha_0$
  son 1.05, 1.00, y 0.95, respectivamente. Las l{\'\i}neas azules (rayas)
  corresponden al escenario de recombinaci\'on est\'andar, y las
  l{\'\i}neas rojas (s\'olidas), corresponden al escenario actualizado.}
\label{ionization_history_comparacion}
\end{figure}

En la Fig.~(\ref{ionization_history_comparacion}) mostramos c\'omo una variaci\'on
en el valor de $\alpha$ en la \'epoca de recombinaci\'on afecta la
historia de ionizaci\'on, moviendo el redshift al cual ocurre la
recombinaci\'on a tiempos anteriores para mayores valores de
$\alpha$. La diferencia entre las funciones cuando se consideran los
dos escenarios de recombinaci\'on, para un dado valor de $\alpha$, es
menor que la diferencia que existe cuando se cambia el valor de
$\alpha$. Algo similar ocurre cuando se var{\'\i}a $m_e$.

\section{Relaci\'on fenomenol\'ogica}

Como mencionamos en la Introducci\'on, el ajuste de la variaci\'on
conjunta de $\alpha$ y $m_e$ se ha realizado en esta tesis de manera
totalmente independiente de cualquier modelo que relacione dichas
variaciones. Por lo tanto, la correlaci\'on encontrada entre las
variaciones de estas constantes nos permite estimar una relaci\'on
fenomenol\'ogica entre ambas.  A partir del contorno de confiabilidad
para $\Delta \alpha / \alpha_0$ y $\Delta m_e / (m_e)_0$ del ajuste
m\'as actualizado (es decir, el conjunto de datos que incluye WMAP5),
se puede estimar una relaci\'on entre las variaciones de $\alpha$ y
$m_e$ entre la \'epoca de recombinaci\'on y la actualidad.

 Una manera de hacerlo es proponer una relaci\'on lineal del
tipo
\begin{equation}
\frac{\Delta m_e}{(m_e)_0} = a_{CMB} \frac{\Delta \alpha}{\alpha_0} + b_{CMB}
\label{relac_fenomenologica}
\end{equation}
y buscar valores apropiados para  $a_{CMB}$ y $b_{CMB}$.

Para encontrar estos coeficientes podemos hacer una regresi\'on
ortogonal, que consiste en ajustar una elipse al contorno 2D de la
Likelihood marginalizada sobre todos los par\'ametros excepto
$\Delta \alpha / \alpha_0$ y $\Delta m_e / (m_e)_0$, y la relaci\'on
entre ambas variaciones estar\'a dada por el semieje mayor de la
recta.

La ecuaci\'on de una elipse rotada un \'angulo $\phi$ respecto de
los ejes $(x',y')$ (cuyo origen se encuentra en el punto $(a_1,a_2)$
del sistema $(x,y)$) es 

\begin{equation}
C_1 \, (x-a_1)^2 + C_2 \, (y-a_2)^2 + C_3 \, (x-a_1)(y-a_2) = 1 
\label{elipserotada}
\end{equation}
donde 
\begin{eqnarray}
C_1 &=& \left( \frac{1}{(1+\theta^2)a^2} + \frac{\theta^2}{(1+\theta^2)b^2}\right) \nonumber\\
C_2 &=&  \left( \frac{\theta^2}{(1+\theta^2)a^2} + \frac{1}{(1+\theta^2)b^2}\right)\nonumber\\
C_3 &=& -2\,\left( \frac{1}{b^2} -\frac{1}{a^2} \right) \,\frac{\theta}{1+\theta^2}  
\end{eqnarray}
y $\theta = \tan(\phi)$. Ajustando el contorno de confiabilidad a una
elipse, encontramos valores para los par\'ametros de la misma, es
decir, para $a$, $b$, $\theta$, $a_1$, y $a_2$.  Luego, dado que la
recta que coincide con el semieje mayor de la elipse tiene la
ecuaci\'on
\begin{equation}
y= \theta\,(x-a_1) + a_2 = \theta\, x + (a_2 - \theta a_1)
\end{equation}
podemos encontrar los valores $a_{CMB}$ y $b_{CMB}$ y sus errores, los
cuales calculamos haciendo
\begin{eqnarray}
\sigma_{a_{CMB}}&=& \sigma_\theta \\
\sigma_{b_{CMB}} &=& \sigma_{a_2} - a_1\sigma_\theta -\theta \sigma_{a_1}
\end{eqnarray}
Encontramos que $a_{CMB}= -4.2_{-0.45}^{+0.27}$ y $b_{CMB}= -0.007_{-0.03}^{+0.01}$.
Es decir, la relaci\'on entre las variaciones de $\alpha$
 y $m_e$ es
\begin{equation}
\frac{\Delta m_e}{(m_e)_0} = -4.2\, \frac{\Delta \alpha}{\alpha_0} -0.007.
\label{relac_fenomenologica_metodo1}
\end{equation}

Otra manera de encontrar una relaci\'on entre $\Delta \alpha
/ \alpha_0$ y $\Delta m_e / (m_e)_0$ es a trav\'es de la regresi\'on
m{\'\i}nimo cuadr\'atica. En este m\'etodo, dada la distribuci\'on
conjunta de dos variables aleatorias $X$ e $Y$, que puede expresarse
como una distribuci\'on de ``masa'' en el plano $(X,Y)$, se determina
una recta en el plano con la condici\'on de que dicha recta
proporcione el mejor ajuste posible a la masa de la distribuci\'on.
Se puede demostrar \citep{cramer} que la recta tiene la expresi\'on 
\begin{equation}
\frac{y- a_2}{\sigma_2} = \rho \frac{x - a_1}{\sigma_1}
\end{equation}
es decir
\begin{equation}
y = a_2 + \frac{\rho \sigma_2}{\sigma_1} (x - a_1) =  \frac{\rho \sigma_2}{\sigma_1} x + \left(a_2 -  \frac{\rho \sigma_2}{\sigma_1} a_1\right).
\label{regresion_lineal}
\end{equation}

Para encontrar los valores de $\rho$, $\sigma_1$, y $\sigma_2$,
consideramos que para una distribuci\'on normal de 2 variables, la
matriz de covarianza es $C$, tal que sus elementos son

\begin{eqnarray}
 c_{11}= \sigma_1^2 && \qquad c_{12}= \mathrm{cov}(x_1,x_2) \nonumber\\
c_{21}= \mathrm{cov}(x_1,x_2) && \qquad c_{22}= \sigma_2^2 
\end{eqnarray}

Sea $\rho=\mathrm{cov}(x_1,x_2)/(\sigma_1\sigma_2)$. Cuando se
calculan las l{\'\i}neas de probabilidad constante (es decir, los
contornos de probabilidad), se llega a la ecuaci\'on de una elipse
(centrada en $a_1$ y $a_2$):

\begin{equation}
\frac{(x_1 - a_1)^2}{\sigma_1^2} - 2\rho \,
\frac{x_1-a_1}{\sigma_1} \,\frac{x_2-a_2}{\sigma_2} + \frac{(x_2 -
  a_2)^2}{\sigma_2^2} = 1 - \rho^2
\end{equation}

o lo que es lo mismo:

\begin{equation}
\frac{(x_1 - a_1)^2}{\sigma_1^2 \,(1 - \rho^2)} - 2\frac{\rho}{(1 - \rho^2)} \,
\frac{x_1-a_1}{\sigma_1} \,\frac{x_2-a_2}{\sigma_2} + \frac{(x_2 -
  a_2)^2}{\sigma_2^2\,(1 - \rho^2)} = 1 
\label{brandt}
\end{equation}
de donde vemos que 

\begin{eqnarray}
C_1 &=& \frac{1}{\sigma_1^2 \,(1 - \rho^2)} \nonumber\\
C_2 &=& \frac{1}{\sigma_2^2 \,(1 - \rho^2)} \nonumber\\
C_3 &=&  -2 \frac{\rho}{\sqrt{C_1\,C_2}}
\end{eqnarray}

Por lo tanto,

\begin{equation}
\rho= - \frac{C_3}{2\sqrt{C_1\,C_2}}
\end{equation}

Luego, con el valor de $C_1$ obtenemos $\sigma_1$, y con el valor de
$C_2$ obtenemos $\sigma_2$. Con estos valores y los valores medios
$a_1$ y $a_2$ del ajuste anterior, de la Ec.~(\ref{regresion_lineal})
encontramos la recta de regresi\'on
\begin{equation}
\frac{\Delta m_e}{(m_e)_0} = -2.57 \frac{\Delta \alpha}{\alpha_0}  -0.01. 
\label{relac_fenomenologica_metodo2}
\end{equation}
Esta relaci\'on es diferente de la relaci\'on encontrada en
Ec.~(\ref{relac_fenomenologica_metodo1}). Sin embargo, el signo de la
pendiente es negativo en ambos casos, debido a que $\Delta m_e
/(m_e)_0$ y $\Delta \alpha / \alpha_0$ est\'an anticorrelacionadas.
La relaci\'on propuesta por \citet{Ichi06}, y con la cual hicieron los
ajustes es
\begin{equation}
\frac{\Delta m_e}{(m_e)_0} = p \frac{\Delta \alpha}{\alpha_0}
\end{equation}
con $p$ fijo. Encuentran buenos ajustes para $p=0, 1/2, 2, 4$. Todos
estos valores son positivos, y aqu{\'\i} se ha mostrado que, para el
mejor ajuste de los datos (cuando se deja $\Delta \alpha / \alpha_0$ y
$\Delta m_e / (m_e)_0$ como par\'ametros libres), la correlaci\'on
entre las variaciones es negativa, es decir, $p$ deber{\'\i}a ser
negativo. Por lo tanto, conclu{\'\i}mos que la propuesta
de \citet{Ichi06} deber{\'\i}a ser descartada.

\cleardoublepage
\chapter{Modelo de Barrow \& Magueijo}
\label{chap:BM}

Como hemos discutimos en la Introducci\'on, la variaci\'on de las
constantes fundamentales es una predicci\'on de teor{\'\i}as que
intentan unificar las fuerzas de la naturaleza. En el l{\'\i}mite de
bajas energ{\'\i}as, \'estas pueden ser interpretadas como
teor{\'\i}as efectivas.  Por otra parte, existen varios modelos
fenomenol\'ogicos que, a partir de unas pocas hip\'otesis generales,
describen la variaci\'on de ciertas constantes fundamentales. A su
vez, estos modelos pueden ser l{\'\i}mites a bajas energ{\'\i}as de
teor{\'\i}as m\'as generales.  Entre ellos se encuentra el modelo de
 \citet{BM05}, que describe la evoluci\'on de la masa
del electr\'on.  Si bien, en algunos modelos, la variaci\'on de la masa
del electr\'on aparece vinculada a la variaci\'on de la constante de
estructura fina, no es necesario que sus variaciones
est\'en relacionadas, y dichos autores consideraron oportuno estudiar
las consecuencias cosmol\'ogicas de un modelo simple de variaci\'on de
la masa del electr\'on, independientemente del comportamiento de
$\alpha$.

Aqu{\'\i} deduciremos soluciones anal{\'\i}ticas aproximadas m\'as precisas
que las de Barrow \& Magueijo (2005). En la
Secci\'on \ref{sec:formalismo} resolvemos en forma m\'as rigurosa la
ecuaci\'on de Friedmann para el factor de escala, y luego mantenemos
todas las constantes de integraci\'on en la resoluci\'on de la
ecuaci\'on diferencial para el campo escalar. Se obtienen as{\'\i}
soluciones detalladas para la variaci\'on de la masa del electr\'on
como funci\'on del tiempo, que se muestran en la Secci\'on
\ref{soluciones}. Finalmente, los par\'ametros del modelo ser\'an
acotados utilizando datos.

Las diferentes observaciones astr\'onomicas y geof{\'\i}sicas que se
pueden utilizar para poner cotas a la variaci\'on de la masa del
electr\'on, corresponden a distintas \'epocas en la historia del
Universo. Esto las hace realmente \'utiles para testear modelos que
predigan un comportamiento temporal determinado. En esta tesis se
consideran distintas observaciones, las cuales se describen en la
Secci\'on \ref{sec:datos}, y se encuentran cotas a los par\'ametros
del modelo de Barrow \& Magueijo utilizando estas observaciones a
distintos tiempos, en la Secci\'on \ref{sec:ajuste_parametros}.

Luego, en la Secci\'on \ref{sec:wep} obtenemos l{\'\i}mites
adicionales para estos par\'ametros, a partir de las cotas existentes
para la validez del Principio de Equivalencia D\'ebil, que se obtienen
de experimentos tipo E\"otv\"os. Comparando con las cotas obtenidas a
partir de resultados cosmol\'ogicos, se llega a la conclusi\'on de que
el modelo debe ser descartado.  El tratamiento desarrollado en este
Cap{\'\i}tulo fue publicado en \citet{Scoccola08}.

\section{El formalismo}
\label{sec:formalismo}

\def\beq{\begin{equation}}
\def\eeq{\end{equation}}

\def\bea{\begin{eqnarray}}
\def\eea{\end{eqnarray}}

\def\bdm{\begin{displaymath}}
\def\edm{\end{displaymath}}

\newcommand{\ii}{\'\i}

\newcommand{\OmegaM}{\Omega_{M_0}}
\newcommand{\OmegaR}{\Omega_{R_0}}
\newcommand{\OmegaL}{\Omega_{\Lambda_0}}

La densidad lagrangiana del campo libre de Dirac, a partir de la cual se
puede derivar la ecuaci\'on de movimiento para el electr\'on, es \citep{Sakurai}:

\begin{equation}
\mathscr{L}= -c \hbar \bar \psi \gamma_\mu (\partial/\partial x_\mu)
\psi - m_e c^2\bar\psi \psi.
\label{eq_Dirac}
\end{equation}

Consideremos que la masa del electr\'on est\'a controlada por un campo
escalar ``dilat\'onico'' $\phi$ definido por $m_e=m_{e0} \exp[\phi]$,
donde $m_{e0}$ es el valor actual de la masa, que corresponde a $\phi=0$
hoy. Para variaciones peque\~nas de $m_e$, desarrollamos $\exp[\phi]=
1+ \phi$. La densidad lagrangiana total es

\begin{equation}
\mathscr{L}= -c \hbar \bar \psi \gamma_\mu (\partial/\partial x_\mu)
\psi - m_{e0} c^2\bar\psi \psi - m_{e0} c^2\bar\psi \psi \phi + \frac{\omega}{2} \partial_\mu \phi  \partial^\mu \phi.
\label{eq_Dirac_completa}
\end{equation}
Por lo tanto, la ecuaci\'on de movimiento del campo $\phi$ se obtiene de la
siguiente densidad lagrangiana

\begin{equation}
\mathscr{L} = \mathscr{L}_{int} + \mathscr{L}_\phi
\end{equation}
donde el t\'ermino cin\'etico es 
\begin{equation}
\mathscr{L}_\phi= \frac{\omega}{2} \partial_\mu \phi  \partial^\mu \phi
\end{equation}
y el t\'ermino de interacci\'on entre el campo $\phi$ y el campo de
Dirac del electr\'on es
\begin{equation}
\mathscr{L}_{int} = - m_{e0} c^2 \bar \psi \psi \phi. 
\label{termino_interaccion}
\end{equation}
 Este modelo tiende al modelo can\'onico sin variaci\'on de la
masa del electr\'on cuando el valor del par\'ametro
$\omega \rightarrow \infty$.

La ecuaci\'on de Euler-Lagrange a resolver es la siguiente:

\begin{equation}
\frac{\partial}{\partial x_\mu} \left( \frac{
  \partial\mathscr{L}}{\partial \left( \partial \phi / \partial
  x_\mu\right)} \right) - \frac{\partial \mathscr{L}}{\partial \phi} = 0.
\end{equation}

Escribimos expl{\'\i}citamente $\mathscr{L}_\phi$, teniendo en cuenta
que la convenci\'on de signos para la m\'etrica es $(+\ -\ -\ -\ )$, y
que $\phi$ es funci\'on s\'olo del tiempo:

\begin{equation}
\mathscr{L}_\phi = \frac{\omega}{2}  \partial_\mu \phi
\partial^\mu\phi = \frac{\omega}{2}  \left( \partial_0 \phi  \partial_0\phi
- \partial_i \phi  \partial_i\phi \right) =  \frac{\omega}{2}
\left( \frac{d \phi}{dx^0}\right)^2  =  \frac{\omega}{2} \frac{1}{c^2}
\left( \frac{d \phi}{dt}\right)^2.
\end{equation}
De aqu{\'\i},

\begin{equation}
 \frac{\partial\mathscr{L}}{\partial  \left( \partial_0 \phi \right)}=
 \omega \frac{d \phi}{dx^0} =  \frac{\omega}{c} \frac{d \phi}{dt} 
\end{equation}
por lo tanto
\begin{equation}
\partial_0 \left( \frac{\partial\mathscr{L}}{\partial  \left( \partial_0 \phi \right)}\right)=
 \omega \frac{d^2 \phi}{d(x^0)^2} =  \frac{\omega}{c^2} \frac{d^2 \phi}{dt^2}.
\end{equation}
Por otra parte

\begin{equation}
\frac{\partial \mathscr{L}}{\partial \phi} = - m_{e0} c^2 \bar \psi \psi.
\end{equation}

Por lo tanto, la ecuaci\'on de Euler-Lagrange queda:

\begin{eqnarray}
\frac{\partial}{\partial x_\mu} \left( \frac{
  \partial\mathscr{L}}{\partial \left( \partial \phi / \partial
  x_\mu\right)} \right) - \frac{\partial \mathscr{L}}{\partial \phi} &=&
\partial_0 \left( \frac{\partial\mathscr{L}}{\partial
  \left( \partial_0 \phi \right)} \right) - \frac{\partial
  \mathscr{L}}{\partial \phi}   \nonumber\\
&=& \frac{\omega}{c^2} \frac{d^2 \phi}{dt^2} + m_{e0} c^2 \bar \psi \psi
  = 0
\end{eqnarray}
que se puede reescribir como:

\begin{equation}
 \frac{d^2 \phi}{dt^2} = - \frac{m_{e0}}{\omega} c^4 \bar \psi \psi.
\end{equation}

Reemplazamos $\bar \psi \psi$ por su promedio macrosc\'opico, que en
el l{\'\i}mite no relativista es igual a $n_e + n_p$, siendo $n_e$ el
n\'umero de electrones y $n_p$ el n\'umero de positrones. Despreciando
$n_p$ frente a $n_e$, obtenemos

\begin{equation}
 \frac{d^2 \phi}{dt^2} = - \frac{m_{e0}}{\omega} c^4 n_e.
\end{equation}

Pasando a derivadas covariantes, $\frac{d^2\phi}{dt^2} \to
\ddot \phi + 3\frac{\dot{a}}{a}\dot{\phi}$,
por lo tanto, la ecuaci\'on queda:

\begin{equation}
\ddot \phi  +  3 \frac{\dot a}{a} \dot \phi = a^{-3} \left(\dot \phi
a^3 \right)^{\dot{}} = - \frac{m_{e0}}{\omega} c^4 n_e 
\end{equation}
o equivalentemente
\begin{equation}
\left(\dot \phi a^3 \right)^{\dot{}} =  - \frac{m_{e0}}{\omega} c^4
a^3 n_e = - \frac{m_{e0}}{\omega} c^4 n_{e0}.
\label{eq_phi}
\end{equation}
donde hemos tenido en cuenta que $a^3 n_e = a_0^3 n_{e0}$ y $a_0=1$.

Integrando la Ec.~(\ref{eq_phi}) obtenemos:

\begin{equation}
\dot \phi a^3 = -M t + A
\label{eq_phi_once_integrated}
\end{equation}
con 
\begin{equation}
M = \frac{m_{e0}n_{e_0}}{\omega} c^4 = \frac{\rho_{e0}}{\omega} c^4  
\label{relacion_M_omega}
\end{equation}
y $A$ una constante de integraci\'on. Esta es la ecuaci\'on que
debemos resolver para el campo $\phi$, teniendo en cuenta que el
factor de expansi\'on $a$ es una funci\'on del tiempo que debemos
encontrar para el modelo cosmol\'ogico considerado.

\subsection{Contribuci\'on de $\phi$ a la Ecuaci\'on de Friedmann}
\label{sub:efectos}

En esta tesis se deriva una ecuaci\'on m\'as precisa para el campo
$\phi$ de este modelo. Para ello, comencemos considerando la
ecuaci\'on de Friedmann

\begin{equation} \left(\frac{\dot a}{a}\right)^2 + \frac{c^2 k}{a^2} =
\frac{8 \pi G}{3 c^2} \epsilon
\label{friedmann}
\end{equation}
donde $\epsilon = \epsilon_M + \epsilon_R + \epsilon_\Lambda$ es la
densidad de energ\ii a total y $\dot a = da/ dt$.  Escribimos las
densidades de materia, radiaci\'on y energ{\'\i}a oscura (o constante
cosmol\'ogica), respectivamente, en t\'erminos de sus valores
actuales:
\beq
\epsilon_M = \epsilon_{M_0} \left(\frac{a_0}{a} \right)^3, \qquad
\epsilon_R = \epsilon_{R_0} \left(\frac{a_0}{a} \right)^4, \qquad
\epsilon_\Lambda = \epsilon_{\Lambda_0} \left(\frac{a_0}{a} \right)^0.
\eeq

Definimos $\Omega =\rho / \rho_{\mathrm{crit}}$ donde la densidad
cr\ii tica es:
\beq
\rho_{\mathrm{crit}} = \frac{3 c^2 H^2}{8 \pi G}
\eeq
y $H= \dot a / a$ es la tasa de expansi\'on de Hubble al tiempo $t$.

Sea un universo espacialmente plano ($k=0$). Multiplicando
Ec.~(\ref{friedmann}) por $H_0^{-2}$, obtenemos:

\beq
\left(\frac{\dot a}{H_0 a} \right)^2 = \frac{8 \pi G}{3 c^2 H_0^2}\left[
  \epsilon_{M_0} \left(\frac{a_0}{a} \right)^3 + \epsilon_{R_0}
  \left(\frac{a_0}{a} \right)^4 + \epsilon_{\Lambda_0} \right].
\eeq
Esta ecuaci\'on puede escribirse como 
\beq
\frac{1}{a^2}\left(\frac{da}{d\tau}\right)^2=
\frac{\Omega_{M_0}}{a^3} + \frac{\Omega_{R_0}}{a^4} + \Omega_{\Lambda_0} 
\eeq
donde hemos realizado el cambio de variables: $\tau= H_0 t$ y
considerado $a_0=1$. Notar que $\tau$ es adimensional.

El campo $\phi$ tiene una densidad de energ{\'\i}a dada por 
\begin{equation}
\epsilon_\phi = \frac{1}{2} \omega \partial_\mu \phi \partial^\mu \phi
= \frac{1}{2} \omega \left(\partial_0 \phi\right)^2  = \frac{1}{c^2}
\frac{\omega}{2} \dot \phi^2.
\end{equation}
Usando la ecuaci\'on (\ref{eq_phi_once_integrated}), la podemos
escribir como:
\begin{equation}
\epsilon_\phi = \frac{1}{c^2} \frac{\omega}{2} \frac{\left(-M t + A\right)^2}{a^6}.
\end{equation}

Esta densidad de energ{\'\i}a contribuye como un t\'ermino m\'as en la
 ecuaci\'on de Friedmann, que queda
\begin{equation}
\frac{1}{a^2}\left(\frac{da}{d\tau}\right)^2 = 
 \frac{\Omega_{M_0}}{a^3} + \frac{\Omega_{R_0}}{a^4} +
  \Omega_{\Lambda_0} + \frac{4 \pi}{3} \frac{G \omega}{c^4}\frac{\left(-\frac{M}{H_0^2} \tau + \frac{A}{H_0}\right)^2}{a^6}.
\label{eqexpansion}
\end{equation}

Para tiempos muy remotos, los t\'erminos m\'as importantes en la
Ec.~(\ref{eqexpansion}) son los correspondientes a la radiaci\'on y al
campo $\phi$. Para poder despreciar los efectos del campo $\phi$ en la
ecuaci\'on de Friedmann, pedimos que la contribuci\'on debida a \'este
sea mucho menor que la contribuci\'on de la radiaci\'on, es decir:

\begin{equation}
\frac{4 \pi}{3} \frac{G \omega}{c^4}\frac{\left(-\frac{M}{H_0^2} \tau_{el}
  + \frac{A}{H_0}\right)^2}{a^2_{el}} \ll \Omega_{R_0} 
\label{vinculo}
\end{equation}
donde $\tau_{el}$ es el tiempo $\tau$ a partir del cual vale este
modelo, cuando la temperatura del universo corresponde a la
energ{\'\i}a de producci\'on de pares electr\'on-positr\'on. 
\footnote{Lo que habria que pedir es:
\begin{equation}
\frac{4 \pi}{3} \frac{G \omega}{c^4}\frac{\left(-\frac{M}{H_0^2} \tau
  + \frac{A}{H_0}\right)^2}{a(\tau)^2} \ll \Omega_{R_0} 
\label{ecuacion_posta}
\end{equation}
para todo $\tau > \tau_{el}$.

Sin embargo, el miembro izquierdo de la desigualdad es varios
\'ordenes de magnitud menor que el miembro izquierdo de la
Ec.(\ref{vinculo}) para los datos con los que contamos en el
r\'egimen de dominio de materia y radiaci\'on, es decir BBN y CMB.
}

En lo que sigue, damos una estimaci\'on del valor de $\tau_{el}$ y del
factor de expansi\'on correspondiente, $a_{el}$. Ambos aparecen en la
condici\'on que impondremos al hacer el ajuste de los par\'ametros $A$
y $M$ del modelo.

\subsubsection{R\'egimen de validez del modelo}
\label{sub:threshold}

  El modelo de Barrow \& Magueijo tiene validez a partir del momento
en que la temperatura del Universo tiene un valor correspondiente a la
energ{\'\i}a m{\'\i}nima requerida para la formaci\'on de pares
electr\'on-positr\'on. En esta \'epoca, radiaci\'on y materia est\'an
en equilibrio, formando un s\'olo fluido, de manera tal que

\begin{equation}
E_\gamma = k T = 2 m_e c^2
\end{equation}
donde $k$ es la constante de Boltzmann. Entonces, la temperatura a la
cual se cruza el umbral de creaci\'on de pares es
\begin{equation}
T_{el} = \frac{2 m_e c^2}{k}.
\end{equation}
Por otro lado, sabemos que la temperatura decae con la expansi\'on de
manera inversamente proporcional
\begin{equation}
aT = \mathrm{cte} \Rightarrow T_{el} = \frac{a_0 T_0}{a_{el}}.
\end{equation}
Podemos estimar $\tau_{el}$ a partir de la ecuaci\'on de Friedmann para
la \'epoca de dominio de la radiaci\'on:
\begin{equation}
\frac{1}{a^2}\left(\frac{da}{d\tau}\right)^2 = \frac{\Omega_{R_0}}{a^4}
\Rightarrow a da =\sqrt{\Omega_{R_0}}d\tau \Rightarrow \frac{1}{2}
\left. a^2 \right|_{a_1}^{a_{el}} = \sqrt{\Omega_{R_0}} \left. \tau \right|_{\tau_1}^{\tau_{el}}.
\end{equation}
Suponemos que $\tau_1=0$ y $a_1=0$. Entonces
\begin{equation}
a_{el}= \sqrt{2}\Omega_{R_0}^{1/4}\tau_{el}^{1/2} = \frac{a_0
  T_0}{T_{el}}= \frac{a_0 k T_0}{2 m_e c^2}= \frac{ k T_0}{2 m_e c^2}
\end{equation}
donde hemos considerado $a_0=1$. De esta manera
\begin{equation}
\tau_{el}= H_o t_{el} = \frac{1}{2}\Omega_{R_0}^{-1/2}\left(\frac{k T_0}{2m_e c^2} \right)^2.
\end{equation}
Para obtener el valor num\'erico de este tiempo, tomamos los
siguientes valores:

\begin{eqnarray}
kT_0 &=& 2.35 \times 10^{-4} \,\rm{eV} \,(\mathrm{temperatura\ de\ CMB})
\nonumber \\
2m_e c^2  &=& 1.02 \times 10^6 \,\rm{eV} \,(\mathrm{energ\acute{\i}a\ para\ la\ producci\acute{o}n\
de\ pares\ e^{-} - e^{+}})\nonumber \\
\Omega_{R_0}h^2 &=& 2.47 \times 10^{-5} \nonumber \\
h &=& H_0/100\ {\rm km\ s^{-1}\ Mpc^{-1}} = 0.73  \nonumber \\
H_0^{-1} &=& 4.22 \times 10^{17} \,\rm{s}
\end{eqnarray}
Esto da un valor para el tiempo cuando cesa la producci\'on de pares
$e^{-}-e^{+}$ igual a
\begin{equation}
t_{el}= 1.52\times 10^{-4} \,\rm{s}.
\end{equation}
El valor num\'erico de $a_{el}$ es 
\begin{equation}
a_{el}= 2.304 \times 10^{-10}.
\end{equation}

\section{Soluciones}
\label{soluciones}

\def\beq{\begin{equation}}
\def\eeq{\end{equation}}

\def\bea{\begin{eqnarray}}
\def\eea{\end{eqnarray}}

\def\bdm{\begin{displaymath}}
\def\edm{\end{displaymath}}

Para encontrar la soluci\'on de la ecuaci\'on del campo $\phi$,
primero es necesario resolver la ecuaci\'on diferencial para el factor
de expansi\'on $a(t)$.

\subsection{Soluci\'on de la Ecuaci\'on de Friedmann}

Buscamos la expresi\'on para $a(t)$ resolviendo la ecuaci\'on de
Friedmann. Estudiamos dos etapas del universo:
\begin{enumerate}
\item r\'egimen de dominio de materia y radiaci\'on 
\item r\'egimen de dominio de materia y constante cosmol\'ogica.
\end{enumerate}

La transici\'on ocurre cuando la contribuci\'on a la ecuaci\'on de
Friedmann, tanto de la radiaci\'on como del t\'ermino de constante
cosmol\'ogica, son iguales. Esto sucede cuando el factor de
expansi\'on toma el valor:
\beq a= \sqrt[4]{{\Omega_{R_0}\over \Omega_{\Lambda_0}}}. \eeq

\subsubsection{Primer r\'egimen}
Consideremos un universo dominado por materia y radiaci\'on. 
La ecuaci\'on para el factor de expansi\'on queda:

\beq
{1\over a^2}\left({da\over d\tau}\right)^2 = {\Omega_{M_0}\over a^3}.
+ {\Omega_{R_0}\over a^4}  
\eeq

Haciendo el cambio de variables $d\eta = d\tau / a$, integrando, y tomando como
condiciones iniciales  que a $t=0$, $a=0$ y $\eta=0$, llegamos a que 
\beq
\eta = {2 \over \Omega_{M_0}} \left[\sqrt{\Omega_{M_0}a +
    \Omega_{R_0}} - \sqrt{\Omega_{R_0}}\right].
\eeq
Despejamos $a(\eta)$:

\beq
a^R(\eta) = {\Omega_{M_0}\over 4}\eta^2 + \sqrt{\Omega_{R_0}} \eta
\eeq
donde el supra\ii ndice $R$ denota que estamos en la \'epoca dominada
por materia y radiaci\'on.
Para encontrar $\tau$ en t\'erminos de $\eta$, seguimos los siguientes
pasos:
\beq
d\tau=a d\eta \Rightarrow \tau = \int_0^\tau d\tau' = \int_0^\eta
a(\eta') d\eta'= {\Omega_{M_0}\over 12} \eta^3 +
{\sqrt{\Omega_{R_0}}\over 2} \eta^2
\eeq
es decir que la relaci\'on entre $\tau$ y $\eta$ es:

\beq
\tau = {\Omega_{M_0}\over 12} \eta^3 +
{\sqrt{\Omega_{R_0}}\over 2} \eta^2.
\eeq

\subsubsection{Segundo r\'egimen}
Consideremos ahora un universo dominado por materia y constante
cosmol\'ogica.  La ecuaci\'on para el factor de expansi\'on queda:
\beq
{1\over a^2}\left({da\over d\tau}\right)^2 = {\Omega_{M_0}\over a^3}
+ \Omega_{\Lambda_0}.
\eeq
Cambiando a la variable $y= \sqrt{{\OmegaL / \OmegaM}}a^{3/2}$, llegamos a esta ecuaci\'on:
\beq
d\tau = {2\over 3} {1\over \sqrt{\OmegaL}}{dy\over \sqrt{1+y^2}} \quad
\Rightarrow \quad \tau ={2\over 3} {1\over \sqrt{\OmegaL}} \sinh^{-1} y
+ C.
\eeq

Reemplazando la expresi\'on de $y$ y despejando $a$ obtenemos:
\beq
a^{\Lambda} (\tau) = \sqrt[3]{{\OmegaM \over
    \OmegaL}}\sinh^{2/3}\left[{3\over 2} \sqrt{\OmegaL} \tau + C' \right].
\eeq

Las constantes de integraci\'on se eligen de manera tal de asegurar la
continuidad de $a(\eta)$ y de su derivada en el punto de ajuste
$\eta_A$, donde ambas soluciones se superponen.  Por lo tanto, el punto
de ajuste es aquel donde se cumple:
\beq
a^R(\eta_A) = {\Omega_{M_0}\over 4}\eta_A^2 + \sqrt{\Omega_{R_0}} \eta_A= \sqrt[4]{{\OmegaR \over \OmegaL}}.
\eeq

Resolviendo la ecuaci\'on cuadr\'atica, y teniendo en cuenta que
$\eta$ es siempre positivo, obtenemos el $\eta_A$:

\beq
\eta_A = {2 \sqrt{\OmegaR}\over \OmegaM} \left[ \sqrt{1 + {\OmegaM
      \over \OmegaR}{\sqrt[4]{\OmegaR \over \OmegaL}}} -1 \right].
\eeq

Si llamamos
\beq
\xi = {\OmegaM \over \OmegaR}{\sqrt[4]{\OmegaR \over \OmegaL}}=\OmegaM\ \OmegaR^{-3/4}\ \OmegaL^{-1/4}
\eeq
podemos escribir
\beq
\eta_A = {2 \sqrt{\OmegaR}\over \OmegaM} \left[\sqrt{1 + \xi} - 1 \right].
\eeq
En la variable $\tau$, el tiempo de ajuste es:
\beq
\tau_A = {\Omega_{M_0}\over 12} \eta_A^3 +
{\sqrt{\Omega_{R_0}}\over 2} \eta_A^2 \quad \Rightarrow \quad \tau_A =
{2\over 3}\ {\OmegaR^{3/2}\over \OmegaM^2} \left[ 2 -
  (2-\xi)\sqrt{1+\xi}\ \right].
\eeq
Considerando $a^R(\eta_A) = a^\Lambda (\tau_A)$, ajustamos la
constante $C'$: 

\beq
C' = \sinh^{-1}{(\xi^{-1/2})} - {2 + (\xi -2)\sqrt{1+\xi}\over \xi^2}.
\eeq

\noindent Resumiendo: 

En $\tau_A$, $a^R$ se pega con $a^\Lambda$ de manera
continua, con derivada primera continua. Las expresiones para el
factor de expansi\'on en cada uno de los reg\ii menes son:

\bea
a^R(\eta) &=& {\OmegaM \over 4}\eta^2 + \sqrt{\OmegaR} \eta \qquad {\rm
donde} \quad \tau = \frac{1}{2} \eta^2 \left[ \frac{\OmegaM}{6} \eta +
\sqrt{\OmegaR} \right] \nonumber\\
a^\Lambda(\tau) &=& \sqrt[3]{{\OmegaM \over \OmegaL}}
\sinh^{2/3}\left[{3\over 2} \sqrt{\OmegaL} \tau + C' \right]. 
\eea

Podemos calcular cu\'anto vale $\tau$ hoy, $\tau_0$, sabiendo que
$a^{\Lambda}(\tau_0)=1$. As\ii\ llegamos a que:

\beq
\tau_0 = {2\over 3} {1 \over \sqrt{\OmegaL}}\left[ \sinh^{-1}\left(
  \sqrt{{\OmegaL\over \OmegaM}}
  \right) - C' \right].
\eeq

\subsection{Soluciones para el  campo $\phi$}

\subsubsection{Dominio de materia y constante cosmol\'ogica}

La ecuaci\'on para el campo $\phi$ es:

\beq
\dot \phi a^3 = - M t + A.
\label{eqM}\eeq

Pasando a la variable $\tau$, y considerando $a=a^\Lambda(\tau)$, llegamos a esta ecuaci\'on 
\beq
\int_{\phi(\tau)}^{\phi(\tau_0)} d \phi = - \int_{\tau}^{\tau_0}   \ { M\ \OmegaL \over H_0^2\ \OmegaM}\ {\left( \tau - {H_0\ A
    \over M} \right)\over \sinh^2\left[ {3\over 2}\sqrt{\OmegaL} \tau
    + C' \right]}\ d\tau.
\eeq

Integrando (ver Ap\'endice \ref{apendice_2}), y teniendo en cuenta que
$m_e = m_{e0}\ \exp{[\phi]}$ y por lo tanto $\phi_0 = 0$, y adem\'as 
\beq
\phi = \ln \left( {m_e \over m_{e0}}\right) = \ln \left( {m_{e0} + m_e - m_{e0}
  \over m_{e0}}\right) = \ln \left( 1 + {\Delta m_e \over m_{e0}}\right) \sim {\Delta m_e \over m_{e0}}
\eeq
llegamos a:

\begin{multline}
\frac{\Delta m_e}{m_{e0}}(\tau)  = \\
  \frac{M}{H_0^2}\, \frac{2}{3\OmegaM}\left[ \sqrt{\OmegaL} \tau \coth
    \left( C' +  \frac{3}{2}\sqrt{\OmegaL}\tau  \right)- \frac{2}{3} \ln \left(\sinh \left( C' +
  \frac{3}{2}\sqrt{\OmegaL}\tau \right) \right) \right.
    \nonumber\\ 
\left.   +  \frac{2}{3}  \left(
  \ln\gamma - \frac{\sqrt{1 + \gamma^2}}{\gamma} \left( C' + \ln
  \left(\gamma + \sqrt{1+\gamma^2}\right) \right)\right)   \right] 
  \nonumber\\ 
+ \frac{A}{H_0}\,\frac{2\sqrt{\OmegaL}}{3\OmegaM}\left[-\coth\left(C'
  + \frac{3}{2}\sqrt{\OmegaL}\tau \right) + \frac{\sqrt{1
  + \gamma^2}}{\gamma} \right]
\end{multline}
donde hemos llamado 
\beq
\gamma = \sqrt{{\OmegaL \over \OmegaM}}.
\eeq

\subsubsection{Dominio de materia y radiaci\'on}

Partimos nuevamente de la ecuaci\'on para el campo $\phi$ 
\beq
\dot \phi a^3 = - M t + A
\eeq
y tenemos en cuenta que:
\beq
t = H_0^{-1} \tau = H_0^{-1} {1 \over 2} \eta^2  \left[ {\OmegaM \over 6} \eta + \sqrt{\OmegaR} \right].
\eeq 

\noindent Pasando a la variable $\eta$, y usando $a=a^R(\eta)$, tendremos que integrar

\beq
\int_{\phi_A}^{\phi} d\phi'  = \int_{\eta_A}^{\eta} {-{ M \over 2
      H_0} \eta'^2 \left({\OmegaM \over 6} \eta' + \sqrt{\OmegaR}
    \right) + A \over H_0 \eta'^2 \left( {\OmegaM \over 4} \eta' +
    \sqrt{\OmegaR}\right)^2} d\eta'.
\label{integral_phi_MR}
\eeq

\noindent La expresi\'on para $\phi_A$ est\'a dada en la Ec.~(\ref{phi_A}).
La soluci\'on de esta ecuaci\'on es (ver Ap\'endice \ref{apendice_1}): 

\begin{multline}
\frac{\Delta m_e}{m_{e0}}(\eta) = \frac{2}{3} \frac{M}{H_0^2}
  \frac{1}{\OmegaM}\left[ -2 \ln \left(\frac{2(\beta \eta + 1)}{1+\sqrt{1+\xi}} \right)
+\frac{1}{\beta\eta +1} -\frac{2}{1+\sqrt{1+\xi}} \right. \nonumber \\ 
\left.  + \frac{2}{3}f(\xi)\sqrt{1+\xi} +\frac{1}{4}\ln \left(\frac{\OmegaL}{\OmegaR}\right)
  -\frac{2}{3} \left( \sinh^{-1}\gamma - C' \right)
  \frac{\sqrt{1+\gamma^2}}{\gamma} \right]  \nonumber \\
+ \frac{A}{H_0}
  \frac{\OmegaM}{\OmegaR^{3/2}} \left[ \frac{1}{2}\ln
  \left(\frac{\beta \eta +1}{\beta \eta}\right) 
+ \frac{1}{2}\ln\left(\frac{\sqrt{1+\xi} -1}{\sqrt{1+\xi} +1}\right)
  -\frac{1}{4\beta \eta}   \right. \nonumber \\
\left.  -\frac{1}{4(\beta \eta +1)}+  \frac{\left( \xi - \frac{2}{3}\right)\sqrt{1+\xi} +
  \frac{2}{3}\frac{\sqrt{1+\gamma^2}}{\gamma}}{\xi^2} \right]   
\end{multline}
donde por conveniencia hemos definido $\beta= \frac{1}{4} \OmegaM
\OmegaR^{-1/2}$.

Finalmente, damos la expresi\'on para la tasa de variaci\'on de la
masa del electr\'on, que luego ser\'a \'util para comparar con las
cotas observacionales obtenidas a partir de relojes at\'omicos, en el
laboratorio. Dado que 
\beq
\phi= \ln \left(\frac{m_e}{m_{e0}}\right) \Rightarrow
\frac{m_e}{m_{e0}} = e^\phi \Rightarrow \frac{\dot m_e}{m_{e0}} =
e^\phi \dot \phi \simeq \dot \phi. 
\eeq
Por lo tanto, considerando la  Ec.~(\ref{eqM}), expresamos
\beq
\frac{\dot m_e}{m_e}= \frac{-M\, t + A}{a^3}.
\eeq

\section{Datos}
\label{sec:datos}


En esta secci\'on resumimos los l{\'\i}mites existentes para la
variaci\'on de la masa del electr\'on, obtenidos a partir de datos
observacionales correspondientes a distintas etapas cosmol\'ogicas.

\subsection{Nucleos{\'\i}ntesis primordial}

La nucleos{\'\i}ntesis primordial es una de las herramientas m\'as
importantes para estudiar el universo temprano. El cociente
bari\'on-fot\'on, o equivalentemente la asimetr{\'\i}a en los bariones
$\eta_B\equiv (n_B -n_{\bar B})/n_\gamma$ se puede determinar tanto a partir
de c\'alculos te\'oricos como de  observaciones de las abundancias de los
elementos qu{\'\i}micos livianos. Un cambio en el valor de $m_e$ en la
\'epoca de la formaci\'on de estos elementos
respecto de su valor actual afecta cantidades f{\'\i}sicas tales como
la suma de densidades de energ{\'\i}a de electrones y positrones, la
suma de sus presiones, y la diferencia entre las densidades en
n\'umero de electrones y positrones.  Los cambios m\'as importantes en
las abundancias primordiales debido a variaci\'on de $m_e$ resultan del
cambio en las tasas de decaimiento d\'ebil, m\'as que del cambio en la
tasa de expansi\'on. En \citet{Scoccola08} se consideraron los datos
disponibles para las abundancias de D, ${}^4$He y ${}^7$Li.  El
an\'alisis estad{\'\i}stico indic\'o que no hay buen ajuste para el
conjunto total de datos. Sin
embargo, se logran ajustes razonables si se excluye alguno de los datos.
 Se discuten las cotas obtenidas utilizando
todas las abundancias y s\'olo un subgrupo, llegando a un ajuste de
$\eta_B$ consistente con el valor de WMAP s\'olo si las abundancias
consideradas son las de $\rm{D}$ y ${}^4\rm{He}$. La cota que se
obtiene en este caso es la siguiente:
\beq
\frac{\Delta m_e}{\left(m_e\right)_0} = -0.022 \pm 0.009.
\eeq

En el caso de considerar  otros pares de abundancias, el valor del mejor
ajuste para $\eta_B$ se aleja mucho del valor obtenido por WMAP.  La
abundancia del ${}^7$Li no se tuvo en cuenta porque a\'un no hay
consenso acerca de las cotas para la misma.  Se ha se\~nalado
\citep{richard05} que debe entenderse mejor el transporte turbulento en
las zonas radiativas de las estrellas para tener una estimaci\'on
confiable de la abundancia de ${}^7$Li. Por otra parte, \citet{MR04}
reanalizaron los datos disponibles para el ${}^7$Li y obtuvieron
estimaciones marginalmente consistentes con WMAP. M\'as a\'un,
\citet{PF07} destacaron que la discrepancia se acent\'ua si se
considera la contaminaci\'on con ${}^6$Li.  Es por eso que se adopt\'o
el criterio m\'as conservador de considerar la cota que se obtiene al
ajustar las abundancias de $\rm{D}$ y ${}^4\rm{He}$.

\subsection{Fondo c\'osmico de radiaci\'on}

El fondo c\'osmico de radiaci\'on provee una de las cotas para la
variaci\'on de la masa del electr\'on en el Universo temprano, durante
la \'epoca de formaci\'on del hidr\'ogeno neutro. En el
Cap{\'\i}tulo \ref{chap:CMB} de esta Tesis, se encontraron varias
cotas para la variaci\'on de la masa del electr\'on. Para este
an\'alisis utilizaremos la cota que se obtiene del ajuste con los
datos de WMAP3 y el espectro de potencias $P(k)$ del cat\'alogo
2dFGRS:
\begin{equation}
\frac{\Delta m_e}{\left(m_e\right)_0} = -0.029 \pm 0.034.
\end{equation}
La diferencia entre esta cota y la obtenida con el conjunto de datos
WMAP5 es m{\'\i}nima a los efectos del ajuste de los par\'ametros del modelo
de Barrow \& Magueijo, y no altera las conclusiones de este
Cap{\'\i}tulo.

\subsection{Sistemas de absorci\'on en quasares}
\label{quasars}

Los sistemas de absorci\'on en espectros de quasares constituyen
``laboratorios'' naturales donde buscar variaciones temporales de las
constantes fundamentales. En particular, un m\'etodo para poner
l{\'\i}mites a la variaci\'on de $\mu=\frac{m_p}{m_e}$ fue
desarrollado por
\citet{VL93}. Est\'a basado en el hecho de que las longitudes de onda
de las l{\'\i}neas vibro-rotacionales electr\'onicas dependen de la
masa reducida de las mol\'eculas, con distintas dependencias para
distintas transiciones. De esta manera, es posible distinguir el
corrimiento al rojo (redshift) cosmol\'ogico
de una l{\'\i}nea, del corrimiento provocado por una variaci\'on en
$\mu$.  Para una dada transici\'on del hidr\'ogeno molecular ${\rm
H}_2$, la longitud de onda en el laboratorio (reposo), $\lambda_i^0$,
se puede relacionar con la del sistema de absorci\'on, $\lambda_i$,
mediante $\frac{\lambda_i}{\lambda_i^0} = (1 + z_{abs}) (1 + K_i
\frac{\Delta \mu}{\mu})$, donde $z_{abs}$ es el redshift de
absorci\'on y $K_i$ es el coeficiente que determina la sensibilidad de
la longitud de onda $\lambda_i$ a la masa del electr\'on. Usando
observaciones de sistemas de absorci\'on en quasares a muy alto
redshift y medidas de laboratorio, varios autores han puesto cotas a
la variaci\'on de $\mu$
\citep{P98,LE02,Ivanchik03,Ivanchik05}. Los datos disponibles
actualizados hasta el a\~no 2007,
para cada redshift, se citan en la tabla \ref{molecular} y son los que
consideramos para testear el modelo de Barrow-Magueijo. Llamamos a este conjunto de datos QSO I.
\begin{table}[h!]
\begin{center}
\caption{QSO I: Redshift de absorci\'on, valor de $\frac{\Delta m_e}{\left(m_e\right)_0}$  con el correspondiente error (en unidades de $10^{-5}$), 
y la referencia.}
\label{molecular}
\begin{tabular}{|c|c|c|}
\hline
 Redshift  & $\frac{\Delta m_e}{\left(m_e\right)_0}\pm
 \sigma$  &  Referencia \\ 
\hline
3.025 & $-1.47 \pm 0.83 $  & \citet{Ivanchik05}\\\hline
2.811 & $-6.25 \pm 13.75 $ &  \citet{P98}\\\hline
2.5947 & $ -2.11 \pm 1.39$ &  \citet{Ivanchik05} \\\hline
\end{tabular}
\end{center}
\end{table}

En el a\~no 2008 se actualizaron las cotas para estos sistemas de
absorci\'on en quasares. Esto se logr\'o gracias a una mejora
substancial en el m\'etodo de calibraci\'on de las longitudes de onda,
y tambi\'en incorporando nuevas transiciones al ajuste, que antes no
se ten{\'\i}an en cuenta por estar situadas dentro de una l{\'\i}nea
Lyman $\alpha$ del quasar. Los nuevos datos del conjunto QSO I se muestran en la Tabla
\ref{molecular_2008}.  En la secci\'on \ref{sec:ajuste_parametros} se
repite el ajuste de los par\'ametros del modelo usando estos datos.
\begin{table}[h!]
\begin{center}
\caption{QSO I: Valores actualizados de la Tabla \ref{molecular}. La \'ultima
l{\'\i}nea corresponde al m\'etodo de transiciones inversas del NH$_3$
$\left(\frac{\Delta m_e}{\left(m_e\right)_0}\right.$ y errores en unidades de
$\left.10^{-6}\right)$.}
\label{molecular_2008}
\begin{tabular}{|c|c|c|}
\hline
 Redshift  & $\frac{\Delta m_e}{\left(m_e\right)_0}\pm
 \sigma$  &  Referencia \\ 
\hline
3.025 & $8.2 \pm 7.4$ & \citet{King08}  \\\hline
2.811 & $-1.4 \pm 3.9$ & \citet{King08}  \\\hline
2.5947  &   $0.1 \pm 6.2$  &  \citet{King08} \\\hline
0.86 & $0.74 \pm 1.23$ & \citet{Murphy08}  \\\hline
\end{tabular}
\end{center}
\end{table}

Un m\'etodo alternativo para medir $\mu$ a alto redshift, sugerido por
\citet{FK07}, es usar la sensibilidad a las variaciones en $\mu$  de las
transiciones inversas del amon{\'\i}aco a $24$ GHz. Un cambio en la
frecuencia debido a una variaci\'on de $\mu$ se puede discernir del
redshift cosmol\'ogico compar\'andolas con transiciones cuya
sensibilidad a esta cantidad f{\'\i}sica sea menor.  Para comparar, se
utilizan transiciones rotacionales de mol\'eculas tales como CO,
HCO$^{+}$ y HCN.  Por ahora se conoce un solo sistema de absorci\'on
en quasares con absorci\'on en NH$_3$. Agregamos la cota que se
obtiene a partir de este sistema al conjunto QSO I. La misma se
muestra en la \'ultima entrada de la Tabla
\ref{molecular_2008}. 

Otro m\'etodo para acotar la variaci\'on de las constantes
fundamentales est\'a basado en la comparaci\'on entre la transici\'on
hiperfina de 21 cm, en absorci\'on, para el hidr\'ogeno ($\nu _a)$ y
una transici\'on resonante \'optica $(\nu _b)$, tambi\'en del
hidr\'ogeno. El cociente $\frac{\nu_a}{\nu_b}$ es proporcional a
$x=\alpha ^2g_p\frac{m_e}{m_p}$ donde $g_p$ es el factor $g$ del
prot\'on
\citep{Tzana07}. Por lo tanto, un cambio en esta cantidad resultar\'a
en una diferencia en el redshift medido a partir de la l{\'\i}nea de
21 cm y el espectro medido a partir de l{\'\i}neas de absorci\'on en
el \'optico, dada por $\frac{\Delta
x}x=\frac{z_{opt}-z_{21}}{\left(1+z_{21}\right) }$.  Dado que estamos
trabajando en el contexto del modelo de \citet{BM05}, la \'unica
constante fundamental que puede variar es $m_e$. La Tabla
\ref{opticoradio} muestra las cotas presentadas por
\citet{Tzana07} combinando las medidas de los redshifts obtenidos del
\'optico y del radio. A este conjunto de datos lo llamaremos QSO II.
 Este m\'etodo tiene el inconveniente de que es dif{\'\i}cil
determinar si ambas l{\'\i}neas (radio y \'optico) fueron originadas
en el mismo sistema de absorci\'on.  Por lo tanto, una diferencia en
la velocidad de las nubes de absorci\'on puede esconder una
variaci\'on de $x$.
\begin{table}[h!]
\begin{center}
\caption{QSO II: Redshift de absorci\'on y valor de $\frac{\Delta m_e}{\left(m_e\right)_0}$  con su correspondiente error (en unidades de $10^{-5}$), obtenidos de comparar redshifts de radio y de l{\'\i}neas moleculares.}
\label{opticoradio}
\begin{tabular}{|c|c|c|c|c|}
\cline{1-2}\cline{4-5}
 Redshift & $\frac{\Delta m_e}{\left(m_e\right)_0}\pm \sigma$
 &&Redshift & $\frac{\Delta m_e}{\left(m_e\right)_0}\pm \sigma$ \\
\cline{1-2}\cline{4-5}
 0.24 &   $1.21 \pm   2.10$  &&  1.78 & $ -2.59 \pm  0.90$ \\ \cline{1-2}\cline{4-5}
  0.31 & $-0.61 \pm  4.27$  &&  1.94 & $ 3.30  \pm   0.44$ \\ \cline{1-2}\cline{4-5}
  0.40 & $ 3.22 \pm  3.15$ &&  2.04 & $ 5.20 \pm  2.76$    \\ \cline{1-2}\cline{4-5}
  0.52 & $-2.95 \pm  1.05 $ && 2.35 & $ -2.54 \pm  1.82$ \\ \cline{1-2}\cline{4-5} 
  0.52 & $  0.26 \pm 3.67 $ \\\cline{1-2}
 \end{tabular}
\end{center}
\end{table}

\subsection{Cotas a partir de datos geof{\'\i}sicos}
\label{geo}

La vida media de emisores de part{\'\i}culas $\beta$ de vida media
larga ha sido determinada tanto en medidas en laboratorio, como
comparando con la edad de meteoritos, la cual se encuentra del
an\'alisis del decaimiento de part{\'\i}culas $\alpha$.  Las cotas
m\'as fuertes para la variaci\'on de la vida media, $\lambda$, viene
de la comparaci\'on del decaimiento del $^{187}\rm{Re}$ durante la
formaci\'on del sistema solar y el presente, \citep{Olive04b}:
$\frac{\Delta \lambda}{\lambda} = (-0.016 \pm 0.016)$.  \citet{SV90}
derivaron una relaci\'on entre el corrimiento en la vida media y una
posible variaci\'on de las constantes fundamentales. Como aqu{\'\i} estamos
considerando s\'olo  la variaci\'on de $m_e$, la cota que consideramos
se obtiene de la siguiente relaci\'on: 
\beq
\frac{\Delta \lambda}{\lambda} = a \frac{\Delta m_e}{\left(m_e\right)_0}
\eeq
donde $a=-600$ para $^{187}\rm{Re}$.

\subsection{Cotas de laboratorio}
\label{clocks}

La comparaci\'on entre frecuencias de diferentes transiciones
at\'omicas a lo largo del tiempo es \'util para poner cotas fuertes a
la tasa de variaci\'on de las constantes fundamentales en la
actualidad.  En particular, la frecuencia hiperfina del Cesio se puede
aproximar por $\nu_{{\rm Cs}} \simeq g_{{\rm Cs}} \frac{m_e}{m_p}
\alpha^2 R_y F_{{\rm Cs}} (\alpha)$, donde $g_{{\rm Cs}}$ es el factor
nuclear $g$, $R_y$ es la constante de Rydberg expresada como
frecuencia y $F_{{\rm Cs}}(\alpha)$ es una funci\'on adimensional de
$\alpha$ y no depende de $m_e$ al menos a primer orden, mientras que
las frecuencias de transiciones \'opticas se pueden expresar como
$\nu_{opt} \simeq R_y F(\alpha)$.
Varios autores \citep{Bize03,Fischer04,Peik04} han
medido diferentes transiciones \'opticas y las han comparado con la
frecuencia de la separaci\'on hiperfina del nivel fundamental en
$^{133}{\rm Cs}$ neutro. Estas medidas se pueden usar para poner l{\'\i}mites
a la variaci\'on de $\dot m_e/\left(m_e\right)_0$. En la Tabla
\ref{table-clocks} se listan cotas obtenidas de diferentes
experimentos. Estos datos ser\'an comparados con la predicci\'on para
$\dot m_e/\left(m_e\right)_0$ que se deduce de la ecuaci\'on~(\ref{eqM}):
\beq
\dot \phi= \frac{\dot m_e}{\left(m_e\right)_0} =  \frac{-M t + A}{a^3} = H_0 \frac{-\frac{M}{H_0^2} t + \frac{A}{H_0}}{a^3}. 
\eeq

\begin{table}[h!]
\begin{center}
\caption{Relojes at\'omicos comparados, valor de $\frac{\dot m_e}{\left(m_e\right)_0}$ con su correspondiente error (en unidades de $10^{-15}$), intervalo de tiempo para el cual se midi\'o la diferencia, y la referencia.}
\label{table-clocks}
\begin{tabular}{|c|c|c|c|}
\hline
 Frecuencias  & $\frac{\dot m_e}{\left(m_e\right)_0} \pm \sigma
   \left[10^{-15} \rm{yr}^{-1}\right]$ &
 $\Delta t [\rm{yr}]$&  Referencia \\ \hline
Hg$^{+}$ and Cs & $ 0.2 \pm 7.0 $ & 5  &  \citet{Fischer04}\\\hline
Yb$^+$ and Cs & $1.2 \pm 4.4 $ & 2.8 & \citet{Peik04}\\\hline
Hg$^+$ and Cs & $ 0 \pm 7 $ & 2 & \citet{Bize03} \\\hline
\end{tabular}
\end{center}
\end{table}

\section{Ajuste de los par\'ametros del modelo}
\label{sec:ajuste_parametros}


\label{ajustes_para}

El modelo que estamos estudiando tiene dos par\'ametros libres, cuyos
valores se pueden estimar buscando el mejor ajuste para el conjunto de
datos descripto en la secci\'on anterior. Ajustaremos las cantidades
adimensionales $M/H_0^2$ y $A/H_0$, por el m\'etodo de m\'axima
verosimilitud.  Para ello, hacemos una grilla de valores para
$M/H_0^2$ y $A/H_0$ y evaluamos el $\chi^2$ en cada uno de los puntos
de la grilla. Para asegurar que el valor del campo $\phi$ sea
peque\~no para todo tiempo, lo cual debe ser cierto ya que hemos
aproximado $\exp[\phi]\simeq 1 + \phi$, agregamos a la expresi\'on del
$\chi^2$ un nuevo t\'ermino que tenga en cuenta el v{\'\i}nculo dado
por la ecuaci\'on (\ref{vinculo}). Adem\'as, tomamos una expresi\'on
elevada al cuadrado porque el $\chi^2$ siempre debe ser positivo. La
expresi\'on para el $\chi^2$ queda:

\begin{equation}
  \chi^2 = \sum_i \frac{\left( \phi(t_i) - \frac{\Delta
        m_e}{m_{e_0}}(t_i)  \right)^2}{\sigma_i^2} 
  + \left[\frac{1}{\Omega_{R_0}}\frac{4 \pi}{3} \frac{G \omega}{c^4}\frac{\left(-\frac{M}{H_0^2} \tau_{el}
        + \frac{A}{H_0}\right)^2}{a^2_{el}}\right]^2.
\end{equation}
Existe una relaci\'on entre $\omega$ y $M/H_0^2$. Recordemos la
Ec.~(\ref{relacion_M_omega}):
\begin{equation}
M= \frac{\rho_{e0}}{\omega} c^4.
\end{equation}
Reescribimos esta relaci\'on buscando una expresi\'on para $\rho_{e0}$. 
Sea $N_b$ el n\'umero total de bariones, $N_{\rm H}$ el n\'umero de n\'ucleos
de \'atomos de Hidr\'ogeno,  $N_{\rm He}$ el n\'umero de n\'ucleos de
\'atomos de Helio, y $N_e$ el n\'umero total de electrones. Se tiene
que, dado que el universo es neutro: 
\begin{equation}
N_e = N_{\rm H} + 2N_{{\rm He}}
\end{equation}
y dado que  cada n\'ucleo de hidr\'ogeno aporta un bari\'on y cada
n\'ucleo de helio aporta cuatro bariones,
\begin{equation}
N_b = N_{\rm H} + 4N_{{\rm He}}= N_e + 2N_{{\rm He}}.
\end{equation}
Ahora, sea $f_{{\rm He}}$ la fracci\'on de bariones en forma de n\'ucleos de
${\rm He}$ con respecto al n\'umero total de bariones, es decir: $f_{{\rm He}}=
4N_{{\rm He}}/N_b$ (ya que cada ${\rm He}$ tiene 4 bariones). Entonces
\begin{equation}
N_e = N_b - 2N_{{\rm He}} = N_b - 2 \frac{f_{{\rm He}}}{4}N_b =  N_b \left(1 - \frac{f_{{\rm He}}}{2}\right).
\end{equation}
Por otra parte, la densidad (en masa) de  electrones es la densidad
num\'erica multiplicada por la masa del electr\'on. Por ende, se tiene que 
\begin{equation}
\rho_{e0} = \frac{M_e}{V} = \frac{N_e m_e}{V} =
\frac{N_b \left(1 - \frac{f_{{\rm He}}}{2}\right)}{V} m_e =  \left(1 - \frac{f_{{\rm He}}}{2}\right) \rho_b \frac{m_e}{m_p}
\end{equation}
donde en la \'ultima igualdad se tuvo en cuenta que
\begin{equation}
\rho_b =   \frac{N_b m_b}{V} \simeq  \frac{N_b m_p}{V} \Rightarrow
\frac{N_b}{V} = \frac{\rho_b}{m_p}. 
\end{equation}

Escribimos la densidad (en masa) de bariones en funci\'on de la densidad
cr{\'\i}tica de energ{\'\i}a:
\begin{equation}
\rho_b = c^{-2} \Omega_b \rho_c.
\end{equation} 

Por lo tanto,

\begin{equation}
\rho_{e0} = \left(1 - \frac{f_{{\rm He}}}{2}\right) \rho_b \frac{m_e}{m_p} =
\left(1 - \frac{f_{{\rm He}}}{2}\right) \frac{m_e}{m_p} \Omega_b\rho_c c^{-2}
\end{equation}
y dado que 
$\rho_c=\frac{3 \ c^2}{8\pi}\frac{H_0^2}{G}$, se tiene que 
\begin{equation}
\rho_{e0} = \left(1 - \frac{f_{{\rm He}}}{2}\right) \frac{m_e}{m_p} \Omega_b \frac{3 }{8\pi}\frac{H_0^2}{G}
\end{equation}
por lo tanto,
\begin{equation}
\omega= \frac{\rho_{e0}}{M} c^4= \frac{\rho_{e0}}{H_0^2} c^4 \frac{1}{\left( \frac{M}{H_0^2}\right)} =  \left(1 - \frac{f_{{\rm He}}}{2}\right) \frac{m_e}{m_p} \Omega_b \frac{3 }{8\pi}\frac{c^4}{G}
\frac{1}{\left( \frac{M}{H_0^2}\right)} 
\end{equation}
o lo que es lo mismo:
\begin{equation}
\frac{\omega G}{c^4}= \frac{3 }{8\pi} \left(1 -
\frac{f_{{\rm He}}}{2}\right) \frac{m_e}{m_p} \Omega_b \frac{1}{\left(
\frac{M}{H_0^2}\right)}.
\label{omegaGc4}
\end{equation}

Por lo tanto, la expresi\'on para el $\chi^2$ que vamos a minimizar es
la siguiente:

\begin{equation}
\chi^2 = \sum_i \frac{\left( \phi(t_i) - \frac{\Delta
    m_e}{m_{e_0}}(t_i)  \right)^2}{\sigma_i^2} + \left[ \frac{1}{2}
    \frac{\Omega_b}{\Omega_R} \left( 1 - \frac{f_{{\rm He}}}{2}\right)
    \frac{m_e}{m_p}\frac{1}{\left(\frac{M}{H_0^2}\right)} \left(\frac{-\frac{M}{H_0^2}\tau_{el}+\frac{A}{H_0}}{a_{el}}\right)^2\right]^2.
\end{equation}
La suma es sobre la totalidad de los datos, y $t_i$ se refiere al
tiempo correspondiente a cada uno de ellos. Este primer an\'alisis
estad{\'\i}stico fue realizado antes del a\~no 2008 y por ende, se
utilizaron las cotas para $\Delta m_e / m_e$ obtenida con los quasares
de la Tabla \ref{molecular}. El resultado del an\'alisis
estad{\'\i}stico muestra que no se logra buen ajuste para el conjunto
completo de datos. Sin embargo, se puede encontrar un buen ajuste
excluyendo el quasar a $z=1.94$ de la Tabla~\ref{opticoradio}.  Con el
objetivo de estudiar c\'omo afecta cada conjunto de datos a los
valores obtenidos para los par\'ametros del modelo, repetimos el
an\'alisis estad{\'\i}stico, excluyendo de a un grupo de datos por
vez. Encontramos que la cota que proviene de nucleos{\'\i}ntesis es
crucial para determinar el valor de $A/H_0$. Por otra parte, el grupo
de datos de la Tabla~\ref{molecular} es importante para determinar el
valor de $M/H_0^2$. Los resultados se muestran en la
Tabla~\ref{table:results}, donde hemos llamado QSO I al conjunto de
datos de la Tabla \ref{molecular}, y QSO II al de la
Tabla \ref{opticoradio}.  Estas conclusiones no se modifican cuando se
utilizan las cotas dadas en la Tabla \ref{molecular_2008}.

\begin{table}[ht!]
\begin{center}
\renewcommand{\arraystretch}{1.3}
\begin{tabular}{|c|c|c|c|}
\hline
 datos & $M/H_0^2$ & $A/H_0$ & $\frac{\chi^2_{min}}{N-2}$ \\ \hline
todos los datos & $-7.30_{-2.02}^{+2.10}$ & $3.60_{-1.50}^{+1.44}$ &
1.14 \\ \hline todos $-$ nucleo & $-7.30_{-2.02}^{+2.11}$ &
$-0.10_{-13.3}^{+13.5}$ & 1.22 \\ \hline todos $-$ CMB &
$-5.60_{-2.54}^{+2.46}$ & $3.60_{-1.50}^{+1.44}$ & 1.16 \\ \hline todos
$-$ decaim $\beta$ & $-7.40_{-2.02}^{+2.11}$ & $3.60_{-1.50}^{+1.44}$ &
1.13 \\ \hline todos $-$ QSO I & $ -10.40_{-3.70}^{+3.61}$ &
$3.60_{-1.59}^{+1.45}$ & 1.29 \\ \hline todos $-$ QSO II &
$-5.70_{-2.44}^{+2.53}$ & $3.60_{-1.49}^{+1.44}$ & 0.30 \\ \hline
todos $-$ relojes & $-7.30_{-2.02}^{+2.11}$ & $3.60_{-1.50}^{+1.44}$ &
1.44 \\ \hline
\end{tabular}
\caption{Par\'ametros para el modelo que da el mejor ajuste de los datos.
    El valor para el par\'ametro $M/H_0^2$ est\'a en unidades de $10^{-6}$,
    y el valor para el par\'ametro $A/H_0$ est\'a en unidades de  $10^{-13}$.}
\label{table:results}
\end{center}
\end{table}

Dado que el modelo est\'a descripto en t\'erminos de la constante de
acoplamiento $\omega$, derivamos su valor para el mejor ajuste, a
partir de los resultados anteriores.  Partimos de la siguiente
relaci\'on:

\begin{equation}
\frac{G \omega}{c^4}= \frac{3}{8\pi} \ \left(1 - \frac{f_{{\rm He}}}{2}\right)\ \Omega_b\ \frac{m_e}{m_p}
\ \left(\frac{M}{H_0^2}\right)^{-1}
\end{equation}
y usamos los siguientes valores num\'ericos:

\begin{eqnarray}
G &=& 6.67 \times 10^{-8} {\rm cm}^3 {\rm g}^{-1} {\rm s}^{-2} \nonumber\\
\Omega_b &=& \frac{\Omega_b\ h^2}{h^2}= \frac{0.0223}{(0.73)^2}= 0.0418
\nonumber\\
\frac{m_p}{m_e} &=& 1836.15.\nonumber\\
\end{eqnarray}

Falta estimar el valor de $f_{\rm He}$. Para ello, hacemos las siguientes
consideraciones:

\begin{eqnarray}
f_{{\rm He}} &=& \frac{4 N_{{\rm He}}}{N_b} \Rightarrow N_b= \frac{4 N_{{\rm He}}}{f_{{\rm He}}}
= N_{\rm H} + 4N_{{\rm He}} \nonumber\\
f_{{\rm He}} &=& \frac{4N_{{\rm He}}}{N_{\rm H} + 4 N_{{\rm He}}} =
\frac{4}{\frac{N_{\rm H}}{N_{{\rm He}}} + 4} = \frac{1}{\frac{1}{4\frac{N_{{\rm He}}}{N_{\rm H}}} + 1}.
\end{eqnarray}

Por otra parte, observacionalmente se determina que el cociente entre
la masa en ${\rm He}$ y la masa en ${\rm H}$ es 

\begin{equation}
\frac{M_{{\rm He}}}{M_{\rm H}} = 0.24 = \frac{N_{{\rm He}}\ m_{{\rm He}}}{N_{\rm H}\ m_{\rm H}}
\end{equation}
y de la tabla peri\'odica, se determina que
\begin{equation}
\frac{m_{{\rm He}}}{m_{\rm H}} = \frac{4.0026}{1.0079} = 3.9712.
\end{equation}

Entonces, $f_{\rm He}= 0.1946788$, y por ende: 

\begin{equation}
1 - \frac{f_{{\rm He}}}{2} = 0.9027.
\end{equation}

Por lo tanto, dado que del ajuste de los par\'ametros del modelo se
encuentra que 

\begin{equation}
\frac{M}{H_0^2} =  -7.30_{-2.02}^{+2.10}  \times 10^{-6}
\label{primer_valor_M}
\end{equation}
el valor para $\omega G / c^4$ es

\begin{equation}
\frac{\omega G}{c^4} =  -0.336_{-0.093}^{+0.097} 
\end{equation}
donde hemos encontrado los errores de $\omega G /c^4$ propagando los
errores de $M/H_0^2$, de la siguiente manera:

\begin{equation}
\frac{\omega G}{c^4} = \mathrm{cte}\times  \frac{1}{\left(
\frac{M}{H_0^2}\right)} \Rightarrow \Delta \left( \frac{\omega G}{c^4}\right) =  \left(\frac{\omega G}{c^4}
\right) \frac{\Delta \left(\frac{M}{H_0^2} \right)}{ \left(\frac{M}{H_0^2} \right)}.
\end{equation}
Esto lo hemos podido hacer dado que el intervalo a 1$\sigma$
alrededor del mejor ajuste para el par\'ametro $M/H_0^2$ no contiene
al cero.

La cantidad $\omega G /c^4$ es adimensional. Para comparar con el
resultado de Barrow \& Magueijo, debemos usar cantidades
adimensionales ya que \'estas no dependen de las unidades. Barrow \&
Magueijo utilizan unidades naturales, por eso presentan sus resultados
como $G |\omega| > 0.2$. En unidades usuales, este resultado se
traduce en $G |\omega|/c^4 > 0.2$, que es el que usamos para comparar
con nuestros resultados.

\begin{figure}[ht!]
\begin{center}
\includegraphics[scale=1.,angle=0]{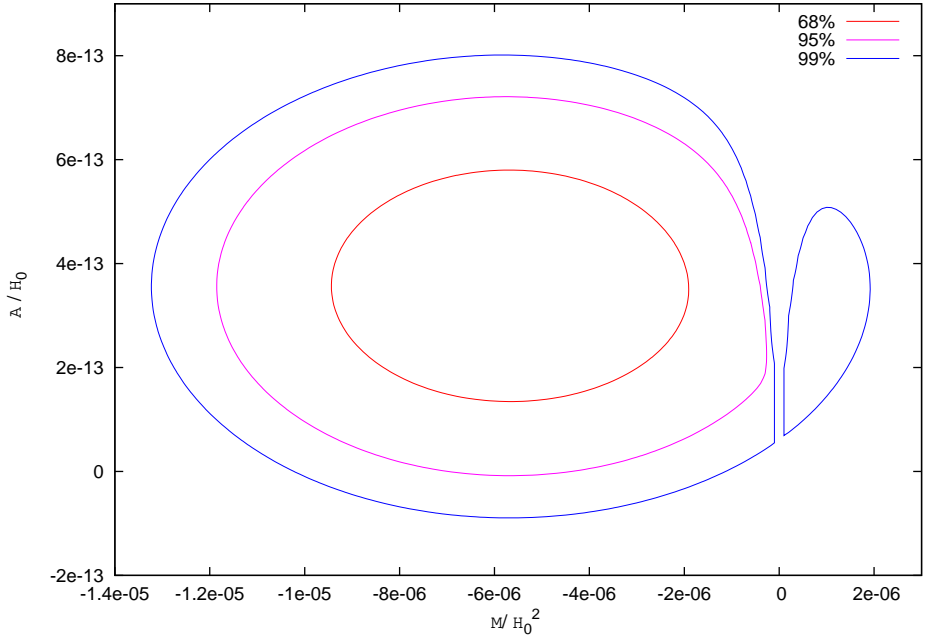}
\end{center}
\caption{Contorno de confiabilidad para los par\'ametros del modelo de Barrow \& Magueijo.}
\label{plot2D}
\end{figure}

El contorno de confianza que resulta del an\'alisis estad{\'\i}stico
puede verse en la figura \ref{plot2D}. El corte que se observa en el
mismo se debe a que la condici\'on impuesta en el $\chi^2$ diverge
para $M=0$.

\begin{figure}[ht!]  \begin{center}
\includegraphics[scale=1.2,angle=0]{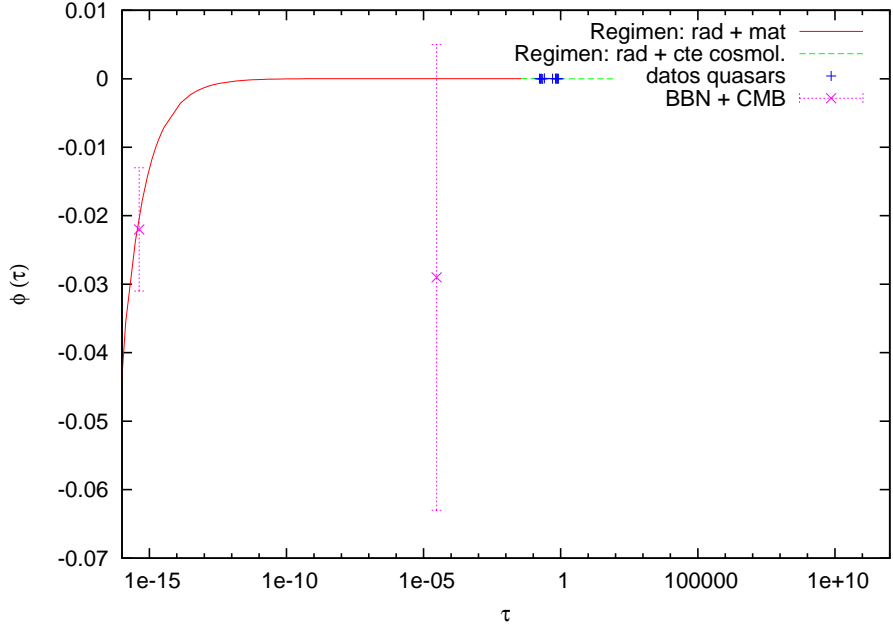}
\end{center} \caption[Soluci\'on del modelo de Barrow \& Magueijo para el mejor ajuste de los par\'ametros.]{Soluci\'on del modelo de Barrow \& Magueijo correspondiente a los par\'ametros del mejor ajuste. Se presentan tambi\'en los datos utilizados para el ajuste, con sus correspondientes barras de error.}
\label{grafico_phi} \end{figure}

En el a\~no 2008, fueron actualizados algunos de los datos de la secci\'on
\ref{quasars}, que se muestran en la Tabla \ref{molecular_2008}.  
Adicionalmente, se ha publicado una cota para la variaci\'on de la
masa de electr\'on obtenida con un nuevo m\'etodo que utiliza las
l{\'\i}neas de las transiciones inversas del NH$_3$.  Hemos repetido
el ajuste teniendo en cuenta dichas actualizaciones. El an\'alisis
muestra que no se obtiene un buen ajuste al considerar el conjunto
completo de datos. Sin embargo, se logra un buen ajuste cuando se
descartan tres sistemas de absorci\'on en quasares del grupo
de \citet{Tzana07} (el que se encuentra a $z=1.94$, que habia sido
descartado tambi\'en en el ajuste anterior, el que est\'a a $z=1.78$,
y la entrada 4 de la Tabla \ref{opticoradio} de la izquierda, 
con $z=0.52$). La nueva cota para la variaci\'on de la masa del
electr\'on obtenida con los datos de WMAP5 no va a cambiar el ajuste
de $M/H_0^2$ debido a que las barras de error para las cotas que
provienen de CMB (y BBN) son mucho mayores que las correspondientes a
los dem\'as datos, como puede verse en la Fig.~(\ref{grafico_phi}).

El resultado del nuevo ajuste, realizado con 16 datos, se presenta en
la Tabla \ref{table:results_2008}.  Observar el cambio en el valor de
la cota para el par\'ametro $M/H_0^2$. Este valor es m\'as consistente
con cero que el valor obtenido en el ajuste anterior (ver
Ec.~(\ref{primer_valor_M})). Lo que cambi\'o entre un ajuste y el otro
es el grupo de datos QSO I. El nuevo conjunto de datos es menos
favorable para la variaci\'on de $m_e$. Esto se traduce en un valor
m\'as consistente con cero para uno de los par\'ametros que controla
la amplitud de la funci\'on $\phi(t)$, $M/H_0^2$, responsable de la
variaci\'on de $m_e$.

\begin{table}[ht!]
\renewcommand{\arraystretch}{1.3}
\begin{center}
\begin{tabular}{|c|c|c|}
\hline
 $M/H_0^2$ & $A/H_0$ & $\frac{\chi^2_{min}}{N-2}$ \\ \hline
 $1.00_{-0.58}^{+0.77}$  & $3.30_{-1.32}^{+1.39}$  & 0.87 \\ \hline
\end{tabular}
\caption{Par\'ametros del mejor ajuste considerando los datos
 actualizados al 2008. Las unidades de $M/H_0^2$ son $10^{-6}$, y de
    $A/H_0$ son $10^{-13}$.}
\label{table:results_2008}
\end{center}
\end{table}

\begin{figure}[ht!]  \begin{center}
\includegraphics[scale=1.,angle=0]{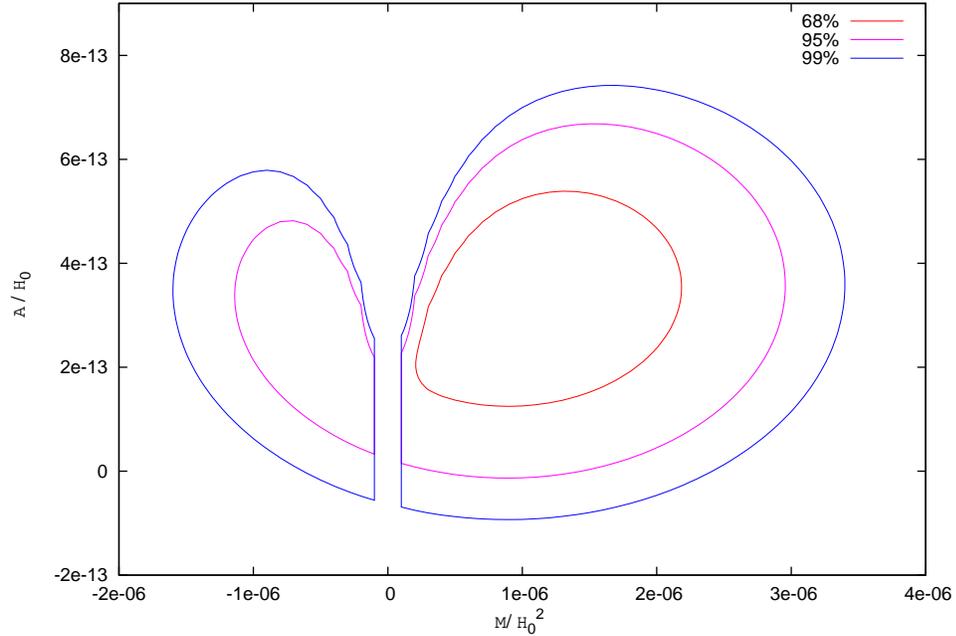}
\end{center} \caption{Contorno de confiabilidad para los par\'ametros del modelo de Barrow \& Magueijo,
seg\'un los datos actualizados.}
\label{plot2D_2008} \end{figure}

A partir de la cota para $M/H_0^2$, encontramos la cota para $G \omega/c^4$
\beq
\frac{G \omega}{c^4} =  2.456_{-1.433}^{+1.912}.
\eeq
El contorno de confianza que resulta de este an\'alisis
estad{\'\i}stico puede verse en la figura
\ref{plot2D_2008}. Nuevamente, el corte que se observa es debido a que
la condici\'on impuesta en el $\chi^2$ diverge para $M=0$.

\section{Principio de Equivalencia D\'ebil}
\label{sec:wep}

De manera complementaria, se pueden obtener cotas para el par\'ametro
$\omega$ del modelo a partir de los l{\'\i}mites impuestos por los
experimentos que testean la validez del Principio de Equivalencia
D\'ebil.  Para obtener una expresi\'on que podamos comparar con dichos
datos, consideremos las variaciones espaciales del campo escalar
$\phi$. Dado que estudiaremos las implicancias de dichas variaciones a
nivel local, despreciaremos las variaciones temporales de $\phi$.  A
partir de la ecuaci\'on de Euler-Lagrange:

\begin{equation}
\frac{\partial}{\partial x_\mu} \left( \frac{
  \partial\mathscr{L}}{\partial \left( \partial \phi / \partial
  x_\mu\right)} \right) - \frac{\partial \mathscr{L}}{\partial \phi} = 0
\end{equation}
y teniendo en cuenta que 
\begin{equation}
\mathscr{L}= \mathscr{L}_\phi + \mathscr{L}_{int} = \frac{1}{2}\omega
\partial_\mu\phi \partial^\mu \phi - m_e c^2 \bar \psi \psi \phi 
\end{equation}
llegamos a la siguiente ecuaci\'on:
\begin{equation}
\Box \phi = - \frac{m_e c^2}{\omega} n_e
\end{equation}
Cuando despreciamos las variaciones temporales, $\Box \phi = - \nabla^2
\phi$, y por lo tanto, la ecuaci\'on se transforma en:
\begin{equation}
\nabla^2 \phi = \frac{m_e c^2}{\omega} n_e = \frac{c^2}{\omega} \rho_e
= \frac{c^2}{\omega} \zeta \rho
\label{eq:phi_espacial}
\end{equation}
donde $\zeta= \rho_e / \rho$.

Para encontrar la soluci\'on de esta ecuaci\'on, usaremos el m\'etodo
de la Funci\'on de Green, y consideraremos que toda la masa ($M$) est\'a
concentrada en el centro de masa del cuerpo, de manera tal que $\rho=
M \delta^3(\vec{r})$. Dado que 

\begin{equation}
\nabla^2 G(x,x') = \delta(x-x') \Rightarrow G(x,x')= -\frac{1}{4\pi} \frac{1}{|x-x'|}
\end{equation}
la soluci\'on de la ecuaci\'on (\ref{eq:phi_espacial}) es
\begin{equation}
\phi = - \frac{1}{4 \pi} \frac{c^2}{\omega} \zeta M \frac{1}{r}
\label{eq:sol_phi_espacial}
\end{equation}

Consideremos el caso de una part{\'\i}cula de prueba en presencia de
un cuerpo m\'as masivo como, por ejemplo, el sol o la tierra.  En este caso,
adem\'as de la fuerza gravitatoria, la part{\'\i}cula experimenta una
fuerza debida a las variaciones espaciales del campo escalar $\phi$
generado por los electrones presentes en el sol (o en la tierra).

Supongamos que la part{\'\i}cula de prueba tiene una masa total $m_t$
y que la masa en electrones es $m_t^e$.  Planteamos la segunda Ley de
Newton:

\begin{equation}
\sum \vec{F} = - m_t \vec{\nabla} \Phi - m_t^e c^2 \vec{\nabla} \phi =
m_t \vec{a}
\end{equation}
de donde se deduce la siguiente expresi\'on para la aceleraci\'on que
sufre la part{\'\i}cula de prueba:
\begin{equation}
\vec{a} = - \vec{\nabla} \Phi - \frac{m_t^e}{m_t} c^2 \vec{\nabla}\phi
= - \vec{\nabla} \Phi - \zeta_t c^2 \vec{\nabla}\phi
\end{equation}
donde $\Phi$ es el potencial newtoniano.\footnote{Hemos considerado
que la fuerza inducida por $\phi$ sobre la materia viene dada por
$m_t^e c^2 \vec{\nabla} \phi$ ya que la densidad lagrangiana de
interacci\'on es
\begin{eqnarray}
\mathscr{L}_{int} &=& - m_e c^2 \bar \psi \psi \phi =  - m_e c^2
n(\vec{r})\phi(\vec{r}) =  - m_e c^2
\delta(\vec{r}-\vec{r}_0)\phi(\vec{r}) \nonumber\\
L_{int} &=& \int d^3x \mathscr{L}_{int} =  - m_e c^2  \phi(\vec{r}_0) 
\end{eqnarray}
y por lo tanto, el potencial de interacci\'on en $\vec{r}_0$, inducido por $\phi$, es $-
m_e c^2 \phi(\vec{r}_0)$.}

El campo gravitatorio se deriva de la funci\'on potencial $\Phi = - G
M /r$. De esta manera:
\begin{equation}
\vec{\nabla} \Phi= \frac{GM}{r^2} \hat{r}
\end{equation}

Por otra parte, el campo escalar $\phi$ que generan los electrones en
el sol (ec.~\ref{eq:sol_phi_espacial}), lleva a la siguiente
expresi\'on:

\begin{equation}
\vec{\nabla} \phi= \frac{1}{4\pi} \frac{c^2}{\omega} \zeta_s M \frac{1}{r^2} \hat{r}
\end{equation}

En conclusi\'on, la expresi\'on para la aceleraci\'on que sufre una
part{\'\i}cula de prueba en el campo gravitatorio del sol, y en el
caso de que el campo $\phi$ tenga variaciones espaciales, es:

\begin{equation}
\vec{a} = - \frac{GM}{r^2} \left[ 1 + \frac{1}{4\pi} \frac{c^4}{G\omega}
  \zeta_s \zeta_t \right] \hat{r}
\label{eq:aceleracion}
\end{equation}
Vemos que la aceleraci\'on depende de la relaci\'on entre la cantidad
de electrones y la cantidad de bariones, es decir, depende de la
composici\'on qu{\'\i}mica de los cuerpos.

\subsection{Experimentos de Eötvös}

Dados dos cuerpos $A$ y $B$, que caen con aceleraciones $\vec{a}_A$ y $\vec{a}_B$
respectivamente, se define el par\'ametro de Eötvös de la siguiente
manera:

\begin{equation}
\eta(A,B) \equiv \frac{2 (a_A - a_B)}{a_A + a_B}
\end{equation}

En nuestro caso de inter\'es, cuando la aceleraci\'on tiene la
expresi\'on de la Ec.~(\ref{eq:aceleracion}), el par\'ametro $\eta$
toma la forma:

\begin{equation}
\eta = \frac{1}{4\pi} \frac{c^4}{G \omega} \zeta_s \frac{\zeta_A -
  \zeta_B}{1 + \frac{1}{4\pi}\ \frac{c^4}{G\omega} \zeta_s
  \left(\zeta_A - \zeta_B \right)} \simeq  \frac{1}{4\pi} \frac{c^4}{G
  \omega} \zeta_s \left(\zeta_A - \zeta_B \right)
\end{equation}

Este par\'ametro se ha medido en distintos experimentos cuyos
resultados y referencias se detallan en la
Tabla \ref{table:eotvos}. Por lo tanto, se puede obtener una cota sobre
el par\'ametro $ c^4/G\omega$. Sin embargo, necesitamos saber el
valor del par\'ametro $\zeta$ de los elementos con los cuales se
realiza el experimento.

Sea $n$ la densidad num\'erica de \'atomos de la substancia
monoat\'omica analizada, de n\'umero at\'omico $Z$, y peso at\'omico
$P$. Recordamos que $Z$ representa el n\'umero de protones en el
n\'ucleo (dado que el \'atomo es el\'ectricamente neutro, esto
equivale al n\'umero de electrones), y $P$ es el promedio de la masa
de los is\'otopos de dicho elemento, pesados seg\'un la abundancia con
la cual aparecen en la tierra, y en unidades de masa at\'omica (uma),
que se define como la doceava parte de la masa de un \'atomo de
$\mathrm{{}^{12}C}$. Por lo tanto, la expresi\'on para el par\'ametro
$\zeta$ es:

\begin{equation}
\zeta= \frac{\rho_e}{\rho} = \frac{n\ Z\ m_e}{n\ P}
\end{equation}

Dado que $1\ \mathrm{uma} = 1.661 \times 10^{-24}\, \mathrm{g} = 931.57\,
\mathrm{MeV}$, y que la masa del electr\'on es igual a $0.511\,
\mathrm{MeV}$, el par\'ametro $\zeta$ se escribe como:

\begin{equation}
\zeta = \frac{ Z\ \times 0.511}{ P \times 931.57}
\end{equation}
donde $P$ se escribe en umas, como aparece en la tabla peri\'odica de
los elementos.

Si en lugar de ser monoat\'omica, la substancia est\'a compuesta de
una mezcla de distintos elementos, el par\'ametro $\zeta$ es:

\begin{equation}
\zeta = \frac{\sum_i N_i Z_i m_e}{\sum_i N_i P_i} 
\end{equation}
donde $N_i$ es el n\'umero de \'atomos de cada elemento. Si tenemos
una cierta cantidad de masa $M_i$ del elemento $i$, y esta cantidad es
un porcentaje $p_i$ de la masa total $M$ de la substancia, tenemos que:

\begin{equation}
N_i = \frac{M_i}{P_i} = \frac{p_i M}{P_i} 
\end{equation}
Por lo tanto, el par\'ametro $\zeta$ queda:

\begin{equation}
\zeta = \frac{\sum_i \frac{p_i}{P_i}  Z_i m_e}{\sum_i p_i} = \sum_i \frac{p_i}{P_i}  Z_i m_e
\end{equation}
o lo que es lo mismo:
\begin{equation}
\zeta = \sum_i \frac{p_i \ Z_i}{P_i}  \frac{0.511}{931.57}
\end{equation}
donde $P_i$ se expresa en umas.

Haremos un ajuste del par\'ametro $c^4/G\omega$ con los datos de
la Tabla \ref{table:eotvos}.

\begin{table}[h!]
\centering
\begin{tabular}{|l|c|c|c|}
\hline\hline
Materiales & $\eta(A,B)\times10^{11}$ & Referencia & en $\Phi$ del:\\
\hline
Al-Au & $1.3\pm1.0$ & \citet{RKD64}          & sol  \\
Al-Pt & $0.0\pm0.1$ & \citet{Braginski72}    & sol  \\
Cu-W  & $0.0\pm4.0$ & \citet{KeiFall82}      & sol  \\
Be-Al & $-0.02\pm0.28$ & \citet{Su94}        & tierra  \\
Be-Cu & $-0.19\pm0.25$ & \citet{Su94}        & tierra  \\
Si/Al-Cu & $0.51\pm0.67$ &  \citet{Su94}     & sol  \\
\hline
\end{tabular}
\caption{Resultados de experimentos de  E{\"o}tv{\"o}s.}
\label{table:eotvos}

\end{table}

Como mencionamos anteriormente, es necesario calcular el par\'ametro
$\zeta$ para los elementos utilizados en dichos experimentos. Los
resultados aparecen en la Tabla \ref{table:zeta}. En el experimento
realizado con Si/Al-Cu, la muestra de Si/Al tiene una masa de $10.0432
\,\mathrm{g}$ ($62.7\%$ Si - $37.3\%$ Al). Para el caso del Sol y de
la Tierra, necesitamos la composici\'on qu{\'\i}mica detallada, la
cual se muestra en la Tabla \ref{table:comp_sun_earth}\footnote{La composici\'on qu{\'\i}mica del Sol y de la Tierra, fueron tomadas respectivamente de
\noindent{\emph{ http://astronomy.swin.edu.au/cosmos/C/Chemical+Composition}}\\
\noindent{\emph{ http://seds.lpl.arizona.edu/nineplanets/nineplanets/earth.html}
}.}.

\begin{table}[h!]
\centering
\begin{tabular}{|c|c|c|c|}
\hline\hline
elemento & $Z$ & $P$ & $\zeta \times 10^4$\\
\hline
Al &  13 & 26.98  & 2.643 \\
Au &  79 & 196.97 & 2.200 \\
Pt &  78 & 195.08 & 2.193 \\
Cu &  29 &  63.55 & 2.503 \\
W  &  74 & 183.84 & 2.208 \\ 
Be &   4 &   9.01 & 2.435 \\
Si &  14 &  28.09 & 2.734 \\
Si-Al & --- & --- & 2.700 \\
Sol  & --- & --- &  4.704 \\
Tierra & --- & --- & 2.568 \\
\hline
\end{tabular}
\caption{Valores del par\'ametro $\zeta$.}
\label{table:zeta}
\end{table}

\begin{table}[h!]
\centering
\begin{tabular}{|l|c|c|}
\hline\hline
Elemento & $p_i$ (solar) & $p_i$ (terrestre) \\
\hline
      Hidr\'ogeno &	73$\%$    &      ---  \\
      Helio       &	25$\%$    &      ---  \\
      Ox{\'\i}geno   &	0.80$\%$  &    29.5$\%$\\
      Carbono   &	0.36$\%$  &      ---  \\
      Hierro     &	0.16$\%$  &    34.6$\%$\\
      Ne\'on     & 	0.12$\%$  &      ---  \\
      Niquel      &         ---   &    2.4$\%$\\ 
      Nitr\'ogeno &	0.09$\%$  &      ---  \\
      Silicio  &	0.07$\%$  &    15.2$\%$\\
      Magnesio &	0.05$\%$  &    12.7$\%$\\
      Sulfuro  & 	0.04$\%$  &     1.9$\%$\\
      Titanio &          ---      &     0.05$\%$\\
      Otros combinados &0.04$\%$  &      ---  \\
\hline
\end{tabular}
\caption{Composici\'on qu{\'\i}mica del Sol y de la Tierra. Los valores son porcentajes en masa.}
\label{table:comp_sun_earth}
\end{table}

Para encontrar una estimaci\'on del par\'ametro $c^4/G \omega$,
hacemos una grilla de valores de este par\'ametro y calculamos el
$\chi^2$, cuya expresi\'on es
\begin{equation}
\chi^2= \sum_i  \frac{\left(\frac{1}{4\pi} \frac{c^4}{G
  \omega} \zeta_s \left(\zeta_A - \zeta_B \right) - \eta_i\right)^2}{\sigma_i^2}
\end{equation}
donde $\eta_i$ y $\sigma_i$ son los datos y errores de la
Tabla \ref{table:eotvos}.  Al minimizar esta cantidad, encontramos el
mejor ajuste para el par\'ametro $c^4/G \omega$. A 1$\sigma$, el resultado es
\begin{equation}
\frac{c^4}{G \omega} = (7.5 \pm 66.6) \times 10^{-5}.
\end{equation}
Como este intervalo contiene al cero, no podemos hacer una
propagaci\'on de errores para encontrar una cota para el par\'ametro
que nos interesa, que es $G \omega / c^4$.  Sin embargo, podemos dar
una cota inferior para el mismo. Efectivamente, dado que
\begin{equation}
\left|\frac{c^4}{G \omega}\right| < 74.1 \times 10^{-5}
\end{equation}
llegamos a que el l{\'\i}mite impuesto por el Principio de Equivalencia D\'ebil es
\begin{equation}
\left|\frac{G \omega}{c^4}\right| > 1349.
\end{equation}

\section{Validez del modelo}

En el paper original \citep{BM05} se hacen algunas aproximaciones
acerca de la evoluci\'on del factor de escala, por lo que la
soluci\'on del modelo se simplifica. En esta tesis se presentan
soluciones m\'as precisas de la evoluci\'on del factor de escala, y en
consecuencia, se obtienen soluciones m\'as exactas para el campo
escalar $\phi$ que induce variaciones en la masa del electr\'on. Otra
mejora en las soluciones proviene de no despreciar la primera
constante de integraci\'on, la cual es de suma importancia en el
Universo temprano. Efectivamente, una vez que se integra la
ecuaci\'on~(\ref{eqM}) se puede escribir
\begin{equation}
\dot{\phi}a^{3}=-M\left( t - \frac{A}{M}\right) 
=-M\left( t + 8.47\times 10^{2} {\rm yr} \right) \label{eqM2}
\end{equation}
donde se han usado los valores del mejor ajuste de los par\'ametros
$M/H_0^2$ y $A/H_0$, y el valor $h=0.73$. Notar que el segundo
t\'ermino del miembro derecho domina en el Universo temprano, en
particular durante la nucleos{\'\i}ntesis primordial.

En \citet{BM05} se presenta una cota de $ G|\omega|/c^4 > 0.2$. En
dicho trabajo, este resultado se obtiene a partir de la precisi\'on de las
cotas para la variaci\'on de $m_e$ provenientes de los quasares,
pero no se consideran las medidas individuales, con sus valores medios y
errores. Adem\'as la expresi\'on que relaciona $\Delta m_e / (m_e)_0$
y el par\'ametro $ G|\omega|/c^4 $ es muy simplificada.  En cambio, en
esta tesis se han utilizado todos los datos disponibles que ponen
l{\'\i}mites a la variaci\'on de la masa del electr\'on, en distintos
tiempos cosmol\'ogicos. Hay una diferencia entre los an\'alisis, dado
que en
\citet{BM05} se considera como cota el valor $\Delta m_e/ (m_e)_0 <
10^{-5}$, mientras que en esta tesis se ha considerado cada dato con
su correspondiente error. Al obtener cotas a partir de experimentos
que testean la validez del Principio de Equivalencia D\'ebil, ellos
obtienen una cota $G|\omega|/c^4 > 10^3$, comparable con la obtenida
en esta tesis a partir de la misma consideraci\'on. Sin embargo, al
comparar con la cota obtenida al restringir la variaci\'on temporal de
la masa del electr\'on a lo largo de la historia del Universo con los
datos existentes hasta el momento, se obtiene que $1.023 < G\omega/c^4
< 4.368$ (intervalo de confianza a 1 $\sigma$). (Antes de la
actualizaci\'on de datos del 2008, hab{\'\i}amos obtenido $-0.615 <
G\omega/c^4 < -0.045 $ a 3 $\sigma$). De esta comparaci\'on llegamos a
la conclusi\'on de que el modelo propuesto en \citet{BM05} para la
variaci\'on temporal de la masa del electr\'on es inconsistente y debe
ser descartado.

Un \'ultimo comentario acerca del signo de $\omega$. En el ajuste con
datos anteriores a 2008, a 2$\sigma$, el valor de $\omega$ es
negativo. Esto no debe resultar sorprendente. De hecho, t\'erminos
cin\'eticos negativos en el Lagrangiano ya han sido considerados por
otros autores, por ejemplo, en modelos de k-essence con una componente
de energ{\'\i}a \emph{fantasma} (phantom)
\citep{caldwell02}. De todas maneras, con los datos actualizados esto
no sucede.

\cleardoublepage
\chapter{Radios Planetarios}
\label{chap:radios}

\def\beq{\begin{equation}}
\def\eeq{\end{equation}}

\def\bea{\begin{eqnarray}}
\def\eea{\end{eqnarray}}

\def\bdm{\begin{displaymath}}
\def\edm{\end{displaymath}}

Consideremos un sistema f{\'\i}sico en el laboratorio, libre de la
acci\'on de la gravedad. Desde el punto de vista de la
termodin\'amica, el estado del sistema se puede describir con una
funci\'on potencial \emph{ energ{\'\i}a interna} $U$ que depende de la
entrop{\'\i}a $S$, el volumen $V$, y el n\'umero de part{\'\i}culas
$N$. \'Estas son variables extensivas. Las derivadas primeras del
potencial definen las variables intensivas \emph{temperatura} $T$,
\emph{presi\'on} $P$, y \emph{potencial qu{\'\i}mico}
$\mu$\footnote{Una variable intensiva es una cantidad f{\'\i}sica cuyo
valor no depende de la  masa de la substancia.}.
Quedan establecidas, entonces, las llamadas \emph{ecuaciones de
estado}. En particular
\beq
P= -\left.\frac{\partial U}{\partial V}\right)_{S,N}.
\eeq

Cuando sometemos el sistema a la acci\'on de la gravedad, cada
part{\'\i}cula del ``gas'' experimenta la fuerza gravitatoria, que se
suma a las fuerzas internas de interacci\'on con las otras
part{\'\i}culas. Sin embargo, la dependencia de la energ{\'\i}a
interna $U$ con las otras variables termodin\'amicas no cambia cuando
el sistema est\'a en un campo gravitacional y las ecuaciones de estado
son las mismas.

Estudiemos ahora un gas autogravitante como sistema
termodin\'amico. Supongamos que el sistema tiene simetr{\'\i}a
esf\'erica, est\'a en equilibrio hidrost\'atico, y adem\'as no hay
fuentes ni sumideros de masa, es decir, la masa se conserva. Para
encontrar su estructura (cantidades termodin\'amicas en funci\'on de
la coordenada radial) debemos resolver el siguiente sistema de
ecuaciones:
\begin{eqnarray}
\frac{dP}{dr} &=& -\frac{G m(r)\rho(r)}{r^2} \\
\frac{dm(r)}{dr} &=& 4 \pi \rho(r) r^2.
\end{eqnarray}
Es necesario, adem\'as, utilizar la ecuaci\'on de estado que relaciona
la presi\'on con la densidad. \'Esta se obtiene para el sistema
particular de estudio, en el laboratorio, cuando est\'a libre de la
acci\'on de la gravedad.

El objetivo de este cap{\'\i}tulo es encontrar cu\'al es la
dependencia de las cantidades termodin\'amicas con las constantes
fundamentales, y c\'omo se ve afectada la estructura de un planeta si
las constantes fundamentales toman un valor diferente del actual. En
particular, analizaremos c\'omo cambia el radio del planeta.
Estudiaremos luego la estructura interna de tres cuerpos del sistema
solar: La Tierra, el planeta Mercurio, y la Luna, y buscaremos una
relaci\'on te\'orica entre la variaci\'on del radio y la variaci\'on
de las constantes fundamentales. Para estos sistemas contamos con
cotas observacionales para posibles variaciones de sus radios entre la
\'epoca de formaci\'on del sistema solar y la actualidad.
 Con los l{\'\i}mites observacionales, y las expresiones te\'oricas
obtenidas, se dar\'a una cota para la variaci\'on ``local'' de las
constantes fundamentales en los \'ultimos 4.000 millones de a\~nos.

\section{Deducci\'on te\'orica}

Partimos del hamiltoniano de un sistema general de electrones
e iones, donde despreciamos la gravedad pero consideramos la
interacci\'on electromagn\'etica entre las
part{\'\i}culas. Identificando el hamiltoniano con la energ{\'\i}a
interna, deducimos la dependencia con las constantes fundamentales de
todas las cantidades termodin\'amicas.

Luego, consideramos un planeta como sistema autogravitante, y
estudiamos c\'omo la gravedad influye en que la estructura del mismo
no sea invariante frente a un cambio de escala. Asimismo, analizamos
c\'omo la variaci\'on de las constantes fundamentales afecta la
estructura del planeta, actuando a trav\'es de una modificaci\'on
efectiva de la constante de gravitaci\'on de Newton.

Finalmente, deducimos una expresi\'on para la variaci\'on del radio del
planeta cuando las constantes fundamentales var{\'\i}an en el tiempo.

\subsection{Hamiltoniano y cambios de escala}

\subsubsection{Sistema sin gravedad}

Consideremos el hamiltoniano (en unidades cgs) de un sistema
cualquiera de electrones y n\'ucleos, tal como mol\'eculas o
s\'olidos, en ausencia de gravedad 

\beq H= \sum_i \frac{p^2_i}{2m_e} + \frac{1}{2} \sum_{i \neq j}
\frac{e^2}{R_{ij}} - \frac{1}{2} \sum_{a,i} \frac{Z_a e^2}{R_{ai}}
\eeq en donde $R_{ij} = |{\bf r}_i - {\bf r}_j|$. El segundo t\'ermino
corresponde a la repulsi\'on electr\'on-electr\'on, y el tercero a la
interacci\'on electr\'on-n\'ucleo. Hemos despreciado la energ{\'\i}a
cin\'etica de los n\'ucleos, que es muy peque\~na comparada con la de
los electrones.\footnote{Es la aproximaci\'on de Born-Oppenheimer. Las
correcciones por movimiento de los n\'ucleos son del orden
$O(m_e/M)$.}

Si en el hamiltoniano introducimos un factor de escala para las
longitudes

\beq
{\bf r}= \beta {\bf x}  \qquad \qquad {\bf p}= {\bf y}/\beta
\label{escaleo_r}
\eeq
el hamiltoniano toma la forma

\beq
H= \frac{1}{2} \left[ \frac{1}{m_e \beta^2} \sum_i y_i^2 +
  \frac{e^2}{\beta} \left( \sum_{i \neq j} \frac{1}{X_{ij}} - \sum_{a,i} \frac{Z_a}{X_{ai}}\right)\right].
\eeq

Si introducimos la representaci\'on de coordenadas ${\bf y}=-i\hbar
{\bf \nabla}$, podemos elegir $\beta$  de modo tal que los coeficientes de
la energ{\'\i}a cin\'etica y la energ{\'\i}a potencial sean iguales

\beq
\frac{\hbar^2}{m_e\beta^2}=\frac{e^2}{\beta} \Rightarrow \beta
=\frac{\hbar^2}{ e^2 m_e} = \frac{\hbar}{ \alpha m_e c}
=a_B
\eeq
donde $a_B$ es el radio de Bohr. Con este factor de escala, el
hamiltoniano toma la forma 

\beq
H= \frac{e^2}{a_B}H_A
\label{escaleo_H}
\eeq
en donde el hamiltoniano adimensional, independiente de las constantes
fundamentales, es:
\beq
H_A= \frac{1}{2} \left[ - \sum_i \nabla_i^2 + \sum_{i\neq j}
  \frac{1}{X_{ij}} - \sum_{a,i} \frac{Z_a}{X_{ai}} \right].
\eeq

Las ecuaciones (\ref{escaleo_r}) y (\ref{escaleo_H}) muestran que en
condiciones adiab\'aticas y en ausencia de gravedad la variaci\'on de
las constantes cambia las longitudes y las energ{\'\i}as con los
factores de escala
\beq
r=a_Bx \qquad \qquad E= \frac{e^2}{a_B}E_A.
\eeq
Como consecuencia, el volumen $V$ cambiar\'a como 
\beq
V= a_B^3 V_A.
\label{cambio_vol}
\eeq

\subsubsection{Reescaleo de variables termodin\'amicas}

Para obtener c\'omo cambian otras cantidades termodin\'amicas,
recordemos la definici\'on de la Entalp{\'\i}a ${\mathcal H}$:
\beq
{\mathcal H}= U + PV = {\mathcal H}(S,P,N).
\eeq
Por tener dimensiones de energ{\'\i}a, su cambio frente al reescaleo es
\beq
 {\mathcal H}= 
\frac{e^2}{a_B} {\mathcal H_A}.
\label{cambio_entalpia}
\eeq
Por otra parte,
\beq
P= \frac{\partial  {\mathcal H}}{\partial V} = \frac{e^2}{a_B}
\frac{\partial {\mathcal H_A}}{\partial V_A} \frac{\partial V_A}{\partial V} =
\frac{e^2}{a_B^4}\frac{\partial {\mathcal H_A}}{\partial V_A} =
\frac{e^2}{a_B^4} P_A.
\eeq
La compresibilidad adiab\'atica variar\'a como
\beq
\kappa_S = - \left. \frac{1}{V} \frac{\partial V}{\partial P}
\right)_{S,N} = - \frac{1}{V} {\mathcal H}_{pp} = \frac{a_B^4}{e^2} \kappa_A.
\eeq

Supongamos que las constantes fundamentales dependen del tiempo. En
consecuencia, el factor de escala $a_B$ tambi\'en tendr\'a una
dependencia temporal
\bea
r(t) &=& a_B(t) x \nonumber\\
r(t_0) &=& a_{B}(t_0) x 
\eea
donde $t_0$ es el valor del tiempo hoy. Sin embargo, $x$ no depende del
tiempo, porque la dependencia temporal est\'a en las constantes
fundamentales. Por lo tanto, al hacer el cociente queda
\beq 
r = \beta r_0 
\eeq 
donde hemos redefinido $\beta = a_B(t)/a_{B0}$, con $a_{B0}= a_{B}(t_0)$. Tambi\'en definimos
por conveniencia $\gamma = \frac{e^2}{e^2_0}/\beta$. Los
sub{\'\i}ndices $0$ representan las cantidades termodin\'amicas cuando el
valor de las constantes fundamentales es el actual.
De este modo, obtenemos las relaciones de escala

\beq
\begin{tabular}{ccc}
$r= \beta r_0$ & \qquad $V= \beta^3V_0$ &  \qquad $\rho = \beta^{-3}\rho_0$ \\
${\mathcal H}= \gamma {\mathcal H}_0$  & \qquad  $P= \gamma / \beta^3 P_0$ &  \qquad  $\kappa=\beta^3/ \gamma\ \kappa_0$. 
\end{tabular}
\label{rel-escalas}
\eeq

Las relaciones termodin\'amicas se deducen a partir del hamiltoniano
del sistema en ausencia de gravedad, por lo tanto, estas relaciones de
escala siguen valiendo a\'un cuando estudiemos el sistema incluyendo
la fuerza gravitatoria.

\subsubsection{Sistema autogravitante}

Supongamos ahora que tenemos en cuenta los efectos de la gravedad en
el sistema. El hamiltoniano ser\'a

\beq
 H= \sum_i \frac{p^2_i}{2m_e} + \frac{1}{2} \sum_{i \neq j}
\frac{e^2}{R_{ij}} - \frac{1}{2} \sum_{a,i} \frac{Z_a e^2}{R_{ai}} - G
 \sum_{\alpha \neq \beta} \frac{m_\alpha m_\beta}{R_{\alpha \beta}}
\eeq
donde los {\'\i}ndices $i,j$ hacen referencia a los electrones, $a$
 hace referencia a los n\'ucleos, y $\alpha, \beta$ hacen referencia a
 cualquier part{\'\i}cula. En la \'ultima suma, los t\'erminos que
 m\'as contribuyen son los que tienen el doble producto de las masas
 de los n\'ucleos, debido a que la masa de cualquier n\'ucleo es mucho
 mayor que la masa del electr\'on. Por ello, el hamiltoniano se puede
 aproximar como
\beq
 H= \sum_i \frac{p^2_i}{2m_e} + \frac{1}{2} \sum_{i \neq j}
\frac{e^2}{R_{ij}} - \frac{1}{2} \sum_{a,i} \frac{Z_a e^2}{R_{ai}} - G
 \sum_{a \neq b} \frac{m_a m_b}{R_{ab}}
\eeq
donde nuevamente $a$ y $b$ hacen referencia a los n\'ucleos del sistema.

En este caso, el cambio de variables 
\beq
{\bf r}= \beta {\bf x}  \qquad \qquad {\bf p}= {\bf y}/\beta
\eeq no nos permitir\'a expresar al hamiltoniano como un factor de
escala por un hamiltoniano adimensional, independiente de las
constantes fundamentales.

Sin embargo, las relaciones de escala para las cantidades
termodin\'amicas se mantienen, y las usaremos en las secciones
siguientes, dado que nos interesan las ecuaciones de estado, que se
calculan a partir del hamiltoniano libre de gravedad. Por \'ultimo
destacamos que el sub{\'\i}ndice $0$ no har\'a referencia a las
cantidades ``hoy'', sino a las cantidades reescaleadas con un factor
que tiene en cuenta el valor de las constantes al tiempo $t$, y el
valor actual. Esto ser\'a explicado mejor en las secciones subsiguientes.

\subsection{Ecuaci\'on de equilibrio hidrost\'atico}

Retomemos el estudio de un sistema autogravitante, con simetr{\'\i}a
esf\'erica, en equilibrio hidrost\'atico, en el cual la masa se
conserva.  Veamos c\'omo se modifica la ecuaci\'on de equilibrio
hidrost\'atico cuando reescaleamos las cantidades
termodin\'amicas. Supongamos que las constantes fundamentales $\alpha$
y $m_e$ pueden tener valores arbitrarios, diferentes de sus valores
actuales. Usaremos la \emph{hip\'otesis adiab\'atica}, que establece
que en presencia de una variaci\'on adiab\'atica de las constantes
fundamentales, la forma de las ecuaciones no cambia.

Con esta condici\'on, la ecuaci\'on de equilibrio hidrost\'atico toma
la forma habitual:
\beq
\frac{dP}{dr} = -G \frac{m(r)\rho(r)}{r^2}
\eeq
en donde
\beq
m(r) =  4\pi \int_0^r \rho(r') r'^2 dr'
\eeq 
En estas ecuaciones tanto $\alpha$ como $m_e$ pueden variar lentamente
con el tiempo.\footnote{``Lentamente'' significa que la escala de
tiempo de variaci\'on es mucho m\'as grande que las escalas
t{\'\i}picas en el sistema solar.}

El sistema est\'a en equilibrio, tiene un radio total $R$ y una
estructura radial en las cantidades termodin\'amicas dada por las
funciones

\bea
P(r) &&\qquad \rm{Presi\acute{o}n}  \nonumber\\
\rho(r) && \qquad \rm{Densidad}  \nonumber\\
d(r) = r &&\qquad \rm{Distancia\, desde\, el\, centro\, hasta\, cierta\, capa} \nonumber\\
 \kappa_S(r)&& \qquad \rm{Compresibilidad\, adiab\acute{a}tica}
\eea
donde $r$ es la coordenada radial que va desde $0$ a $R$.

Supongamos que $r= a_B x$, es decir, medimos $r$ en unidades de radios de
Bohr $a_B$. Si hacemos un cambio de escala en el
sistema\footnote{Similar a la expansi\'on del Universo.}, y ahora
medimos en unidades de $a_{B0}$, estamos cambiando la
parametrizaci\'on, y $r_0=a_{B0}x$. El cambio de escala queda definido
por $r= \frac{a_B}{a_{B0}} r_0 = \beta r_0$. Las variables termodin\'amicas
cambian seg\'un
\beq
P(r)= \frac{\gamma}{\beta^3}P_0(r_0) \qquad \rho(r)= \beta^{-3}
\rho_0(r_0) \qquad \kappa= \frac{\beta^{3}}{\gamma} \kappa_0(r_0).
\eeq

Cuando reescaleamos las longitudes, $d(R)=R_0$, tal que $R=\beta
R_0$. Sin embargo, esto no significa que el nuevo radio del planeta
sea $R_0$. Esto se debe a que el hamiltoniano con gravedad no es
invariante de escala, y por lo tanto, la configuraci\'on de equilibrio
del planeta puede ser diferente si las constantes fundamentales tienen
un valor distinto del actual. Esto implica que la nueva
configuraci\'on de equilibrio es tal que el nuevo radio es
$\tilde{R}_0$, y la estructura del planeta es

\bea
\tilde{P}(r_0) &&\qquad \rm{Presi\acute{o}n}  \nonumber\\
\tilde{\rho}(r_0) && \qquad \rm{Densidad}  \nonumber\\
\tilde{d}(r_0) \neq r_0 &&\qquad \rm{Distancia\, desde\, el\, centro\, hasta\, cierta\, capa} \nonumber\\
\tilde{\kappa}_S(r_0)&& \qquad \rm{Compresibilidad\, adiab\acute{a}tica}
\eea
donde $r_0$ es la coordenada radial que va desde $0$ a $R_0$.

Veamos c\'omo cambia la ecuaci\'on de equilibrio hidrost\'atico cuando
hacemos este cambio de escala.  Vamos a hacer la aproximaci\'on de que
la masa se conserva\footnote{Dado que la masa del electr\'on es mucho
menor que la masa del prot\'on y de bariones m\'as pesados, el cambio
en la masa por el reescaleo es despreciable.}, es decir la masa $m$ en
un punto $r$ medida en el sistema de coordenadas $S$ es la misma que
la masa $m_0$ en el punto $r_0$ medida en el sistema de coordenadas
$S_0$, donde $r=\beta r_0$. Es decir,
\beq
m(r)=m_0(r_0).
\eeq
Asimismo, relacionamos la densidad en uno y otro sistema de
coordenadas
\beq
\rho(r)= \frac{m(r)}{4/3\pi r^3} = \frac{1}{\beta^3}
\frac{m(r)}{4/3\pi r_0^3} =
\frac{1}{\beta^3} \frac{m_0(r_0)}{4/3\pi r_0^3} = \frac{1}{\beta^3}
\rho_0(r_0).
\eeq

Reemplazando en la ecuaci\'on de equilibrio hidrost\'atico

\beq
\frac{d \left( \gamma/\beta^3 P_0\right)}{d\left(\beta
  r_0\right)} = -G \frac{m_0(r_0) \beta^{-3}\rho_0(r_0)}{\beta^2 r_0^2}
\eeq
y simplificando queda

\beq
\frac{d P_0(r_0)}{dr_0}= - \frac{G}{\gamma \beta}
\frac{m_0(r_0)\rho_0(r_0)}{r_0^2} = - G_{ef}
\frac{m_0(r_0)\rho_0(r_0)}{r_0^2}.
\label{equil_hidro}
\eeq

Vemos que la ecuaci\'on de equilibrio hidrost\'atico mantiene su forma
si reescaleamos la constante de gravitaci\'on de Newton de la
siguiente manera

\beq G_{ef}= \frac{G}{\gamma \beta} = \frac{G}{e^2/e^2_0} = G
\left(\frac{\alpha_0}{\alpha} \right).
\eeq

Esta $G_{ef}$ depende del valor de la constante de estructura fina,
pero no depende del valor de la masa del electr\'on. As{\'\i}, la
 estructura de equilibrio del sistema, soluci\'on de la
Ec.~(\ref{equil_hidro}), ser\'a diferente si $\alpha$ var{\'\i}a,
mientras que si $m_e$ var{\'\i}a s\'olo se ver\'a un cambio en el
tama\~no del sistema debido al reescaleo, pero ning\'un cambio
\emph{intr{\'\i}nseco}. Interpretemos esto f{\'\i}sicamente:

\subsubsection{Cambio de $m_e$}

El cambio en la masa del planeta por un cambio en $m_e$ es
despreciable. Bajo esta consideraci\'on, $G$ no cambia, y tampoco lo hace
el radio, m\'as all\'a de un reescaleo. El radio del planeta es
independiente de su masa.

\subsubsection{Cambio de $\alpha$}

Si $\alpha$ var{\'\i}a, cambia la fuerza electromagn\'etica. El sistema
est\'a menos ligado si $\alpha$ es m\'as peque\~na, y dado que la
repulsi\'on entre los n\'ucleos da la compresibilidad del material,
\'este es m\'as compresible. El radio del
planeta es menor, debido a que la gravedad se ve resaltada; es el
mismo efecto que se producir{\'\i}a si $G$
fuese mayor. Es decir, se obtiene una $G$ efectiva, tal que $G_{ef} > G$.

\subsection{Cambios en el radio}

Queremos determinar c\'omo se modifica el radio de un planeta si las
constantes fundamentales var{\'\i}an. Es decir, queremos comparar
$\tilde{R}_0$ (el radio del planeta si las constantes fundamentales
tienen el valor actual), con $R_0$, que es el radio del planeta si las
constantes fundamentales tienen otro valor, pero medido en las mismas
unidades que $\tilde{R}_0$\footnote{Para entender esto, hacemos la
siguiente analog{\'\i}a con la expansi\'on del Universo y el tama\~no
de los c\'umulos de galaxias:

El tama\~no f{\'\i}sico (no com\'ovil) de un c\'umulo de galaxias
crece con la expansi\'on del Universo. Supongamos  que
queremos comparar el tama\~no t{\'\i}pico de un c\'umulo actual
con el tama\~no de los primeros c\'umulos del Universo, para
determinar si hay alg\'un tipo de evoluci\'on intr{\'\i}nseca. Al
mirar los c\'umulos lejanos, los estamos viendo c\'omo eran en el
pasado, cuando el factor de expansi\'on era $a(t)$, menor que el
actual $a_0$. Para comparar el tama\~no en aquella \'epoca con el
tama\~no hoy, tenemos que reescalear el tama\~no del c\'umulo
lejano, de manera tal de simular que el Universo ya se expandi\'o, y
luego comparar con el tama\~no hoy. En otras palabras, se debe medir
en las mismas unidades para poder comparar.}.  Por lo tanto, la
cantidad de inter\'es ser\'a $\xi(M) = \tilde{R}_0 - R_0$ (ambas en
unidades de $a_{B0}$) y no $ \tilde{R}_0 - R$ (que est\'an en
distintas unidades).

Para resolver la ecuaci\'on (\ref{equil_hidro}) conviene usar la masa
como variable independiente. Esto tiene la ventaja adicional que la
masa es invariante en el tiempo
\beq
d m_0(r_0) = \rho_0(r_0) 4 \pi r_0^2dr_0. 
\eeq
Sustituyendo en (\ref{equil_hidro}) e integrando entre una capa (donde
la masa acumulada desde el centro sea $m$) y la superficie
\beq
P_0(m) = -\int_m^0 dP_0(m_0) =  \frac{G_{ef}}{4\pi} \int_m^{M_T}
\frac{m_0 dm_0}{r_0^4}
\eeq
donde hemos usado la condici\'on de contorno $P=0$ en la
superficie. Esta es una soluci\'on formal, dado que $r_0(m_0)$ es desconocido.
La funci\'on \emph{Presi\'on  reescaleada} es $P_0(m)$.

Definamos la siguiente funci\'on
\beq
f(m)= \frac{1}{4\pi} \int_m^{M_T} \frac{m_0 dm_0}{r_0^4}
\eeq
es decir
\beq
P_0(m)= G_{ef} f(m).
\eeq


\subsubsection{>Qu\'e representan las cantidades con sub{\'\i}ndice $0$?}

Si hubiesemos partido de un planeta de masa $M_T$, con distribuci\'on
de presi\'on $P_0(r_0)$ y de densidad $\rho_0(r_0)$, y hubiesemos
reescaleado con $\beta$ tal que s\'olo var{\'\i}e $m_e$, hubieramos
llegado a la distribuci\'on $P(r)$ y $\rho(r)$ en equilibrio
hidrost\'atico, y al radio $R$, tal que $R=\beta R_0$. En este caso,
las cantidades con sub{\'\i}ndice $0$ representan la estructura del
planeta {\it hoy}.

En cambio, si var{\'\i}a $\alpha$ no podemos partir de $(P,\rho,R)$,
hacer el mismo reescaleo y llegar a $(P_0,\rho_0,R_0)$ en equilibrio
hidrost\'atico. Las cantidades con sub{\'\i}ndice $0$ s\'olo son un
reescaleo, pero no representan la estructura del planeta  {\it hoy}.

Al resolver la ecuaci\'on de equilibrio hidrost\'atico reescaleada,
tenemos la presi\'on reescaleada:
\beq
P_0(m)= G_{ef} f(m).
\eeq
Si $\alpha$ var{\'\i}a, $G_{ef} \neq G$, mientras que si s\'olo $m_e$
var{\'\i}a, $G_{ef}=G$. Adem\'as, si $G_{ef}=G$, $P_0$ coincide con la
presi\'on actual $\tilde{P}_0$. En cambio, si $G_{ef} \neq G$, $P_0$
es la presi\'on reescaleada, pero no coincide con la distribuci\'on de
presi\'on hoy.

Supongamos que tenemos un cambio en $P_0$ dado por un cambio en
$G_{ef}$ (punto a punto en la coordenada masa). En este caso, 
\beq
\frac{\Delta P_0}{P_0} \simeq \frac{\Delta G_{ef}}{G} \quad
{\rm significa} \quad \frac{\tilde{P}_0 - P_0}{\tilde{P}_0}\simeq \frac{G-G_{ef}}{G}.
\eeq
donde hemos despreciado el cambio en $f(m)$ por un cambio en $G_{ef}$,
de manera an\'aloga al tratamiento de \citet{will}.

Vale aclarar que en cada punto las relaciones termodin\'amicas para las
distribuciones $P_0$ y $\tilde{P}_0$ son las mismas, por lo que en
ambos casos vale

\beq
\kappa_0 = -\frac{1}{V_0} \frac{\partial V_0}{\partial P_0}.
\eeq

Si la presi\'on reescaleada cambia por variaci\'on de las constantes
fundamentales, produciendo un $\Delta P_0$, el cambio en el volumen
ser\'a $\Delta V_0$, que ser\'a la diferencia entre el elemento de
volumen hoy y el elemento del volumen en coordenadas reescaleadas.

Sea la transformaci\'on 

\beq
\vec{\tilde{r}}_0(m) = \vec{r}_0(m) + \vec{\xi}(m)
\label{transformacion_r}
\eeq
que mide el desplazamiento de una capa de masa por variaci\'on de las
constantes fundamentales. Entonces $\xi(M_T)$ es la diferencia entre
el radio real hoy, y el radio reescaleado en el pasado, si las
constantes toman otro valor.

Sea la transformaci\'on $\rm{T}: W \subset \mathbf{R}^3
 \rightarrow \mathbf{R}^3$ dada por la Ec.~(\ref{transformacion_r}).
 Escrita en componentes es
\bea
\tilde{x}_0 = x_0 + \xi_1 \nonumber \\
\tilde{y}_0 = y_0 + \xi_2 \nonumber \\
\tilde{z}_0 = z_0 + \xi_3. 
\eea

El Jacobiano de $\rm{T}$ mide c\'omo se distorsionan los vol\'umenes
por la transformaci\'on. As{\'\i}, $V_0$ se transforma en
$\tilde{V}_0 = J V_0$, con
\beq
J = \left| \frac{\partial
  (\tilde{x}_0,\tilde{y}_0,\tilde{z}_0)}{(x_0,y_0,z_0)}\right| = det
\left( \begin{array}{ccc} 1+\frac{\partial \xi_1}{\partial x_0}
  &\frac{\partial \xi_2}{\partial x_0}  &\frac{\partial
    \xi_3}{\partial x_0} \\ \frac{\partial \xi_1}{\partial y_0} & 1+
  \frac{\partial \xi_2}{\partial y_0}  &\frac{\partial \xi_3}{\partial
    y_0} \\ \frac{\partial \xi_1}{\partial z_0} &  \frac{\partial
    \xi_2}{\partial z_0}  & 1+\frac{\partial \xi_3}{\partial
    z_0}\end{array} \right) \simeq 1 + \vec{\nabla}_0\cdot \vec{\xi} + O(2)
\eeq
donde hemos considerado que la transformaci\'on es diferencial. De
esta manera,
\beq
\frac{\Delta V_0}{V_0} = \frac{\tilde{V}_0 - V_0}{V_0} \simeq  \vec{\nabla}_0\cdot \vec{\xi}
\eeq
donde $\Delta V_0$ significa la diferencia entre el volumen real y el
reescaleado (a una dada masa acumulada $m$) y $V_0$ significa el
volumen reescaleado. 
Suponiendo que $\vec{\xi}$ es radial, y tomando la divergencia en
coordenadas esf\'ericas,
\beq
\vec{\nabla}_0 \cdot \vec{\xi} = \frac{1}{r_0^2} \frac{d(r_0^2 \xi)}{dr_0}.
\eeq

De 
\beq
\kappa_0 = -\frac{1}{V_0} \frac{\partial V_0}{\partial P_0}
\eeq
llegamos a que 
\beq
\Delta P_0 = -\frac{1}{\kappa_0} \frac{\Delta V_0}{V_0}
=-\frac{1}{\kappa_0} \vec{\nabla}_0 \cdot \vec{\xi} =
-\frac{1}{\kappa_0} \frac{1}{r_0^2} \frac{d(r_0^2 \xi)}{dr_0},
\eeq
por lo que 
\beq
\Delta P_0 = \frac{\Delta G_{ef}}{G} P_0= -\frac{1}{\kappa_0} \frac{1}{r_0^2} \frac{d(r_0^2 \xi)}{dr_0}.
\eeq
De aqu{\'\i}
\beq
\frac{d(r_0^2 \xi)}{dr_0} = - \frac{\Delta G_{ef}}{G} P_0
\kappa_0(r_0) r_0^2.
\eeq
Entonces
\beq
\xi(r_0) = - \frac{\Delta G_{ef}}{G} \frac{1}{r_0^2} \int_0^{r_0} P_0(r_0')
\kappa_0(r_0') r_0'^2 dr_0'.
\label{eq:xi_r0}
\eeq
En su estudio acerca de consecuencias observables de posibles violaciones al
Principio de Equivalencia Fuerte, \citet{will} presenta una expresi\'on
equivalente a \'esta, aunque en otro contexto y deducida a partir de
 consideraciones diferentes.

Escribiendo (\ref{eq:xi_r0}) en funci\'on de la masa, y evaluando en $M_T$, llegamos a 

\beq
\xi(M_T) = - \frac{\Delta G_{ef}}{G} \frac{1}{R_0(M_T)^2}
\int_0^{R_0} P_0(r_0') \kappa_0(r_0') r_0'^2 dr_0'
\label{eq:xi}
\eeq
donde $R_0$ es el radio reescaleado, que lo aproximamos por el radio
actual $\tilde{R}_0$, $P_0(r')$ es la distribuci\'on real de presi\'on
hoy, y $\kappa_0(r_0')$ es la reescaleada, pero que la aproximamos por
la real hoy.
La justificaci\'on para aproximar el radio es la siguiente

\beq
R_0(M_T) = \tilde{R}_0(M_T) - \xi(M_T) = \tilde{R}_0(M_T) \left( 1 -
\frac{\xi(M_T)}{\tilde{R}_0(M_T) } \right) \simeq  \tilde{R}_0(M_T)
\eeq
porque en la Ec.~(\ref{eq:xi}) ya estamos a orden $O(\frac{\Delta
G_{ef}}{G})$.
Asimismo
\beq
\frac{1}{R_0(M)^2} = \frac{1}{ \tilde{R}_0(M_T) ^2} \frac{1}{ \left( 1 -
\frac{\xi(M_T)}{\tilde{R}_0(M_T) } \right)^2 } \simeq \frac{1}{ \tilde{R}_0(M_T) ^2} 
\eeq
al orden en el que estamos trabajando.

Por otra parte, justificamos que $\kappa_0(r_0')$ es aproximadamente
la distribuci\'on real por lo siguiente. Las relaciones
termodin\'amicas se deducen a partir del hamiltoniano sin gravedad. Al
reescalear de manera de expresar las cantidades termodin\'amicas en
funci\'on de los que valen con las constantes fundamentales actuales,
encontramos las mismas relaciones termodin\'amicas que
encontrar{\'\i}amos hoy. Para ver que la funci\'on es la misma (porque
$R_0 \neq \tilde{R}_0$)
\beq
\kappa_0(r_0') = \kappa_0(r_{verd}) + \left.\frac{\partial
  \kappa_0}{\partial r}\right)_{verd} \cdot \left( r_{verd} -
  r'\right)
\eeq
y $\left( r_{verd} - r'\right) \sim \xi$, por lo cual lo despreciamos.

Por lo tanto, el cambio en el radio del planeta en unidades de $a_{B0}$ ser\'a

\beq
\Delta R =\xi(M_T)= - \frac{\Delta G_{ef}}{G} \frac{1}{\tilde{R}_0^2}
\int_0^{\tilde{R}_0} P_0(r_0') \kappa_0(r_0') r_0'^2 dr_0'.
\label{Delta_R}
\eeq

\def\beq{\begin{equation}}
\def\eeq{\end{equation}}

\def\bea{\begin{eqnarray}}
\def\eea{\end{eqnarray}}

\def\bdm{\begin{displaymath}}
\def\edm{\end{displaymath}}

\section{Modelado de los planetas}

La variaci\'on temporal de las constantes fundamentales produce una
alteraci\'on en el valor de la constante de Newton efectiva
$G_{ef}$. Esto, a su vez, provoca cambios en la estructura interna de
los planetas.

Ajustaremos el valor de $\Delta G_{ef}/G$ utilizando las cotas
observacionales para la variaci\'on de los radios de la Tierra,
Mercurio y la Luna, presentados por \citet{McElhinny}. Luego,
haremos una estimaci\'on para $\Delta \alpha / \alpha_0$ entre la
\'epoca de formaci\'on del sistema solar y la actualidad.

Sin embargo, para poder relacionar $\Delta G_{ef}/G$ con $\Delta R$ es
necesario modelar la estructura interna de los planetas mencionados
anteriormente.

\subsection{Estructura de la Tierra}

Las ondas s{\'\i}smicas proveen una manera de explorar el interior de
la Tierra, y su comportamiento predictivo hace posible obtener modelos
de alta resoluci\'on de las propiedades internas de nuestro planeta.

Consideremos que la Tierra tiene simetr{\'\i}a esf\'erica, y que sus
propiedades termodin\'amicas en funci\'on del radio est\'an bien descriptas
por el \emph{Preliminary Reference Earth Model}
(PREM)\citep{lay_wallace} (ver Tabla \ref{tabla_PREM}). Los
par\'ametros de dicho modelo son la densidad $\rho$, las velocidades
de ondas s{\'\i}smicas $P$ y $S$, el coeficiente de atenuaci\'on de
\emph{shear} $Q_\mu$, el m\'odulo de bulk adiab\'atico o
incompresibilidad $K_S$, la rigidez $\mu$, la presi\'on y la gravedad.

El m\'odulo de bulk o incompresibilidad K se define como la
resistencia de un material al cambio de volumen, cuando est\'a sujeto
a presi\'on. Si $\Delta V$ es el cambio de volumen por un cambio de
presi\'on $\Delta P$, tenemos

\beq
\frac{\Delta V}{V}= \frac{-\Delta P}{K}
\eeq
es decir que $K= 1/\kappa$, donde $\kappa$ es la compresibilidad.

Para estimar el valor de la integral involucrada en la expresi\'on de
$\Delta R$ para el caso de la Tierra, dividimos al planeta en capas y,
llamando $r_i$ al radio de cada capa $i$, hacemos la siguiente
aproximaci\'on:

\beq
\int_0^R P(r') \kappa_S(r')  r'^2 dr' \simeq \sum_{i=1}^N P(r_i) \kappa_S(r_i)
 r_i^2 (r_i - r_{i-1}). 
\eeq

Usando esta aproximaci\'on en la Ec.~(\ref{Delta_R}), obtenemos la
relaci\'on:
\beq
\frac{\Delta R}{\left( \frac{\Delta G_{ef}}{G}\right)} = -288.09\,
\mathrm{km}. 
\eeq
Considerando que el radio actual de la Tierra es $\tilde{R}_0= 6371\, {\rm km}$
(seg\'un el modelo PREM), se obtiene 
\beq
\frac{\Delta R}{\tilde{R}_0}= -0.04522 \left( \frac{\Delta G_{ef}}{G}\right). 
\eeq

\subsection{Estructura de Mercurio}

Para calcular la cota sobre el radio de Mercurio, necesitamos conocer
la presi\'on y la compresibilidad en funci\'on de la coordenada
radial. Usamos los modelos calculados por \citet{harder_schubert}, de
los cuales, elegimos el presentado en \citet{spohn}.  \'Este es un
modelo con simetr{\'\i}a esf\'erica, de dos capas: un n\'ucleo de
hierro, de densidad $\rho_c=8\, \mathrm{gcm^{-3}}$ y un manto de
silicatos, con una densidad de $\rho_m= 3.35\, \mathrm{gcm^{-3}}$. El
radio de Mercurio es $R= 2439\, \mathrm{km}$, y el radio del n\'ucleo
es $R_c= 1860\,\mathrm{km}$.

\subsubsection{C\'alculos para la presi\'on}

Para encontrar la presi\'on en funci\'on del radio, suponemos que el
planeta est\'a en equilibrio termodin\'amico, es decir que se
satisface

\beq
\frac{dP}{dr}= -G \frac{m(r)\rho(r)}{r^2}
\eeq
entonces
\beq
\int_R^r dP = P(r) - P(R) = P(r) = - \int_R^r \frac{G m(r') \rho(r')}{r'^2}dr'
\eeq
donde hemos considerado que $P(R)=0$. Teniendo en cuenta que 
\beq
m(r)= 4 \pi \int_0^r \rho(x) x^2 dx
\eeq
y llamando
\beq
g(r)= \frac{Gm(r)}{r^2}
\eeq
escribimos
\beq
P(r)= \int_r^R g(r') \rho(r') dr'.
\label{expresion_presion}
\eeq

Buscamos la expresi\'on para la presi\'on dentro y fuera del n\'ucleo.

\subsubsection{Dentro del n\'ucleo}

Para $0 \leq r < R_c$ separamos la integral (\ref{expresion_presion}) en
dos partes
\beq
P(r)= \int_r^{R_c} g(r') \rho(r') dr' + \int_{R_c}^{R} g(r') \rho(r') dr'. 
\eeq
En la primera integral, $r < R_c$, es decir $\rho(r')=\rho_c$, por
lo tanto, la resolvemos de la siguiente manera
\beq
\int_r^{R_c} g(r') \rho(r') dr' = 4\pi G \rho_c \int_r^{R_c}
\frac{1}{r'^2} \int_0^{r'} \rho(x)x^2 dx dr'.
\eeq
Ahora $0\leq x \leq r' < R_c$,  y por ende $\rho(x)= \rho_c$. De
esta manera,
\beq
\int_r^{R_c} g(r') \rho(r') dr' = 4\pi G \rho_c^2 \int_r^{R_c}
\frac{1}{r'^2} \int_0^{r'} x^2 dx dr' = 4\pi G \rho_c^2 \frac{1}{6}(R_c^2 -r^2).
\eeq

Resolvemos ahora la segunda integral, donde $R_c < r' < R$, y por lo
tanto, $\rho(r')= \rho_m$
\beq
\int_{R_c}^{R} g(r') \rho(r') dr' = 4 \pi G \rho_m \int_{R_c}^{R}
\frac{1}{r'^2} \int_0^{r'} \rho(x) x^2 dx dr'.
\label{segundaintegral}
\eeq
Separamos la integral en la variable $x$ de la siguiente manera
\beq
\int_0^{r'} \rho(x) x^2 dx = \rho_c\int_0^{R_c} x^2 dx + \rho_m
\int_{R_c}^{r'}  x^2 dx = \rho_c \frac{R_c^3}{3} + \rho_m \frac{(r'^3
  - R_c^3)}{3}.
\eeq
Reemplazando en la ecuaci\'on (\ref{segundaintegral}),
\beq
\int_{R_c}^{R} g(r') \rho(r') dr' = 4 \pi G \left[\rho_m\left(\rho_c
- \rho_m\right) \frac{R_c^3}{3} \left(-\frac{1}{R} + \frac{1}{R_c}\right) +
\frac{\rho_m^2}{6} \left( R^2 - R_c^2\right) \right]. 
\eeq
Por lo tanto, la presi\'on dentro del n\'ucleo, en funci\'on del radio, es
\beq
P^{(c)}(r)= 4 \pi G \left[ \frac{\rho_c^2}{6}\left(R_c^2 - r^2\right) + \rho_m\left(\rho_c
- \rho_m\right) \frac{R_c^3}{3} \left(-\frac{1}{R} + \frac{1}{R_c}\right) +
\frac{\rho_m^2}{6} \left( R^2 - R_c^2\right) \right].
\eeq

\subsubsection{Fuera del n\'ucleo}

En el manto, $R_c < r < R$, por lo tanto
\beq
P(r)= \int_r^R g(r') \rho(r') dr' = 4 \pi G \rho_m \int_r^R
\frac{1}{r'^2} \int_0^{r'} \rho(x)x^2 dx dr'.
\eeq
Aqu{\'\i}, $R_c < r < r'<R$, por lo tanto separamos
\beq
\int_0^{r'} \rho(x)x^2 dx = \rho_c\int_0^{R_c} x^2 dx +
\rho_m\int_{R_c}^{r'} x^2 dx = \rho_c \frac{R_c^3}{3} + \rho_m \frac{\left(
 r'^3-R_c^3\right)}{3}.
\eeq
En consecuencia, la presi\'on en el manto, en funci\'on del radio es
\beq
P^{(m)}(r)= 4 \pi G \left[ \rho_m \left(\rho_c - \rho_m \right)
  \frac{R_c^3}{3} \left(-\frac{1}{R} + \frac{1}{r} \right) +
  \frac{\rho_m^2}{6} \left( R^2 - r^2 \right)\right].
\eeq

\subsubsection{C\'alculos para la compresibilidad}

Para la incompresibilidad, usamos los valores del m\'odulo de bulk 
de la Tabla I del trabajo de \citet{harder_schubert} y consideramos que 
\beq
K(P) = K_0 + \frac{\partial K}{\partial P} P.
\eeq
Para el manto, $K_0^{(manto)}=120 \mathrm{ GPa}$, y $ \frac{\partial
K}{\partial P}^{(manto)}= 4.25$. Para el n\'ucleo,
$K_0^{(core)}=127 \mathrm{ GPa}$, y $ \frac{\partial K}{\partial
P}^{(core)}= 2.20$.
Es decir que 
\bea
K^{(m)}(P)= 120\, \mathrm{GPa} + 4.25 P^{(m)} \,(\mathrm{GPa}) \\
K^{(c)}(P)= 127\, \mathrm{GPa} + 2.20 P^{(c)} \,(\mathrm{GPa}) \\
\eea
Recordamos que la incompresibilidad $K$ es la inversa de la
compresibilidad $\kappa$.

Finalmente, buscamos la expresi\'on para $\frac{\Delta R}{\left(\Delta
G_ {ef}/G_{ef}\right)}$:

\bea
\frac{\Delta R}{\left(\frac{\Delta G_{ef}}{G_{ef}}\right)}&=& -
\frac{1}{R^2} \int_0^R K^{-1}_s(r')P(r') r'^2 dr' \nonumber\\ &=& -
\frac{1}{R^2} \left[\int_0^{R_c} K^{-1 \,(c)}_s(r')P^{(c)}(r') r'^2
  dr' + \int_{R_c}^R K^{-1 \,(m)}_s(r')P^{(m)}(r') r'^2 dr'\right]\nonumber\\{}
\label{Delta_R_mercurio}
\eea
la cual se puede integrar semi-anal{\'\i}ticamente usando el programa
\emph{Mathematica}. Para determinar en que unidades de presi\'on quedar\'a
el resultado de la integral, analizamos las unidades de las cantidades
que la componen
\bea
G &=&  \left(6.67428 \pm 0.00067 \right) \times 10^{-11} \ {\rm m}^3
\ {\rm kg}^{-1} \ {\rm s}^{-2} \nonumber\\
\left[r\right] &=& {\rm km} \nonumber\\
\left[\rho\right] &=& {\rm g\, cm^{-3}}.
\eea
Teniendo en cuenta que $1 {\rm g}\, {\rm cm}^{-3} = 10^{12}\, {\rm kg}\,  {\rm
km}^{-3}$, y que $ 1 {\rm Pa} = 1 {\rm kg} \, m^{-1}\, {\rm s}^{-2}$,
llegamos a que las unidades de $P(r)$ ser\'an:
\beq
\left[ P(r) \right] =  10^{12}\, {\rm kg}\, {\rm km}^{-1}\,  {\rm s}^{-2} =
10^{9} \mathrm{Pa} = 1 \mathrm{GPa}.
\eeq
Por lo tanto, si medimos las distancias en ${\rm km}$, el resultado de
la Ec.~(\ref{Delta_R_mercurio}) estar\'a tambi\'en en ${\rm km}$.

\beq
\frac{\Delta R}{\left(\frac{\Delta G_{ef}}{G}\right)}= -48.91 \,{\rm km}. 
\eeq

Considerando que el radio actual de Mercurio es $\tilde{R}_0= 2439\,
{\rm km}$, se obtiene
\beq
\frac{\Delta R}{\tilde{R}_0}= -0.02005 \left( \frac{\Delta
G_{ef}}{G}\right).  
\eeq

\subsection{Estructura de la Luna}

El modelo de la Luna que vamos a considerar est\'a descripto en
\citet{Kuskov_kronrod}. Se trata de un modelo con simetr{\'\i}a
esf\'erica, que consta de zonas diferenciadas: corteza, manto y un
posible  n\'ucleo.  En este estudio despreciaremos el n\'ucleo,
dado que a\'un no se ha podido confirmar su existencia, radio y
composici\'on. Por otra parte, hay evidencias de que si existiera,
deber{\'\i}a ser peque\~no ($< 460\, {\rm km}$)
\citep{Wieczorek}.  Por lo tanto, consideraremos que 
las caracter{\'\i}sticas del manto inferior se extienden hasta el
centro de la luna.

Las expresiones para la densidad y compresibilidad en funci\'on de la
coordenada radial se obtendr\'an a partir de las caracter{\'\i}sticas
de la estructura interna.  Por otra parte, la presi\'on en el interior
de la Luna se puede describir con la f\'ormula
\beq
P(r)= P_0 \left[ 1- \left( \frac{r}{ R}\right)^2 \right]
\label{presion_luna}
\eeq donde la presi\'on central es $P_0=
47.1\,\mathrm{kbar}$\footnote{Recordemos las unidades: \beq 1
\mathrm{kbar} = 10^3 \mathrm{bar} = 10^9 \mathrm{dy/cm^2} = 10^9
\mathrm{g\ cm^{-1} seg^{-2}} \eeq}, el radio
medio  es $R=1738\,\mathrm{km}$, y $r$ es la coordenada radial.

\subsubsection{Corteza}

La corteza est\'a compuesta de una mezcla de Al-Ca, con una densidad
promedio de $3\, \mathrm{g\, cm^{-3}}$ y se extiende desde la
superficie hasta una profundidad de $58\,\mathrm{km}$. En la interfaz
corteza-manto, la densidad est\'a en el siguiente rango:
$3.24<\rho_{m-cr}<3.32 \,\mathrm{g\, cm^{-3}}$.  En
\citet{Kuskov_kronrod} se consideran dos modelos, seg\'un el valor de
la densidad en la interfaz:
\beq
\begin{array}{ccc}
&{\rm Modelo \, I} & \qquad \rho_{m-cr}= 3.24 \,\mathrm{g\, cm^{-3}}\nonumber\\
&{\rm Modelo\, II} & \qquad \rho_{m-cr}= 3.32 \,\mathrm{g\, cm^{-3}}.\nonumber
\end{array}
\eeq

Para hallar la compresibilidad en la corteza, suponemos que la
densidad tiene un crecimiento lineal con la profundidad $h$, es decir
\beq
\rho(h)= \rho_{m-cr} + \left. \frac{d\rho}{dh}\right)_{h_{m-cr}}\cdot
\left( h - h_{m-cr}\right)
\eeq
donde $h$ es la variable que caracteriza la profundidad y $h_{m-cr}=58\,\mathrm{km}$
es la profundidad de la interfaz corteza-manto. La densidad promedio
es
\bea
\bar\rho  &=& \frac{1}{h_{m-cr}} \int_0^{h_{m-cr}}\left[\rho_{m-cr} +
\left. \frac{d\rho}{dh}\right)_{h_{m-cr}}\cdot
\left( h - h_{m-cr}\right) \right] dh \nonumber \\
 &=& \rho_{m-cr} - \frac{1}{2}
 \left. \frac{d\rho}{dh}\right)_{h_{m-cr}}\cdot h_{m-cr}
\eea
de aqu{\'\i} deducimos que
\beq
 \left. \frac{d\rho}{dh}\right)_{h_{m-cr}} = \frac{2}{h_{m-cr}} \left(
 \rho_{m-cr} - \bar \rho \right),
\eeq
por lo que podemos escribir la expresi\'on para la densidad como
\beq
\rho(h) = \rho_{m-cr} + \frac{2}{h_{m-cr}} \left( \rho_{m-cr} -
  \bar \rho \right)\cdot \left( h - h_{m-cr} \right). 
\label{dens_luna_corteza}
\eeq
Por otra parte,
\beq
\kappa_S = - \frac{1}{V} \left.\frac{\partial V}{\partial P}\right)_S =
\frac{1}{\rho} \left.\frac{\partial \rho}{\partial P}\right)_S =
\frac{1}{\rho} \frac{d\rho}{dh} \frac{\partial h}{\partial P}= \frac{1}{\rho}
\frac{d\rho}{dh} \frac{1}{\left( \frac{\partial P}{\partial h}\right)}
\eeq
y derivando la expresi\'on~(\ref{presion_luna}) obtenemos la
compresibilidad en funci\'on de la profundidad
\beq
\kappa_S (h) = \frac{1}{\rho(h)} \cdot
\left. \frac{d\rho}{dh}\right)_{h_{m-cr}} \cdot \frac{R}{2P_0 \left( 1
  -\frac{h}{R}\right)}.
\label{compr_luna_corteza}
\eeq
En funci\'on de la coordenada radial, la compresibilidad queda
expresada por 
\beq
\kappa_S (r) = \frac{1}{\rho(r)} \cdot
\left. \frac{d\rho}{dh}\right)_{h_{m-cr}} \cdot \frac{R^2}{2P_0 r}.
\label{compr_luna_corteza_radial}
\eeq

Reemplazando los valores num\'ericos queda (para $1680 < r < 1738 \,
{\left[\rm km\right]}$):

\bea
\kappa_S(r) &=&
\frac{265.38}{\left(3.24+\frac{0.24}{29}(1680-r)\right)r} \,
\frac{1}{\mathrm{kbar}} \qquad \mathrm{para\,el\,Modelo\,I}\nonumber\\
\kappa_S(r) &=&
\frac{353.83}{\left(3.32+\frac{0.32}{29}(1680-r)\right)r} \,
\frac{1}{\mathrm{kbar}} \qquad \mathrm{para\,el\,Modelo\,II}.
\eea

\subsubsection{Manto}

El manto lunar consiste en varias regiones de velocidades
s{\'\i}smicas constantes, separadas por fronteras
discont{\'\i}nuas. En las Tablas \ref{velocidades_moon_1} (para el
Modelo I) y \ref{velocidades_moon_2} (para el Modelo II), se muestra
la estructura interna de la luna, dando las propiedades de las
distintas capas que conforman el manto, incluyendo los valores de las
velocidades medias compresional ($V_P$) y de shear ($V_S$). Las
distintas capas del manto se indican con: MS I (manto superior I), MS
II (manto superior II), M int. (manto intermedio), y M inf. (manto
inferior).

Usaremos la siguiente relaci\'on entre las velocidades s{\'\i}smicas
$V_P$ y $V_S$, el m\'odulo de bulk  adiab\'atico $K_S$
y la densidad $\rho$ para sacar valores de la incompresibilidad para
las distintas capas del manto:
\beq
\frac{K_S}{\rho}= V_P^2 - \frac{4}{3} V_S^2.
\eeq
Si $\rho$ se mide en $\mathrm{g \, cm^{-3}}$, mientras que $V_P$, y
$V_S$ en $\mathrm{km \, s^{-1}}$, las unidades de $K_S$ son $10\,
\mathrm{kbar}$. Recordar que la compresibilidad $\kappa_S$ es la inversa del
m\'odulo de bulk $K_S$. 

Finalmente calculamos
\beq
\frac{\Delta R}{\left( \frac{\Delta G_{ef}}{G} \right)} =  -
\frac{1}{R^2} \int_0^{R} \kappa_S(r') P(r') r'^2 dr'.
\eeq
Separamos la integral en 5 integrales, con $\kappa_S$ constante en
cada una de ellas
\bea
\frac{\Delta R}{\left( \frac{\Delta G_{ef}}{G} \right)} &=&  -
\frac{1}{R^2}  \sum_{i=1}^{4} \kappa_s(r_i) P_0 \int_{r_{i-1}}^{r_i} \left[ 1 -
  \left(\frac{r'}{R}\right)^2\right] r'^2 dr'  
\nonumber\\
&&- \frac{1}{R^2}  P_0\int_{R_{m-cr}}^{R} \kappa_s(r') \left[ 1 -
  \left(\frac{r'}{R}\right)^2\right] r'^2 dr'\nonumber\\
&=&  - \frac{1}{R^2}  \sum_{i=1}^{4} \kappa_s(r_i) P_0 \left( \frac{r_i^3}{3} -
\frac{r_i^5}{5 R^2} - \frac{r_{i-1}^3}{3} + \frac{r_{i-1}^5}{5 R^2} \right)\nonumber\\
&&- \frac{1}{R^2}  P_0\int_{R_{m-cr}}^{R} \kappa_s(r') \left[ 1 -
  \left(\frac{r'}{R}\right)^2\right] r'^2 dr'.
\eea

Por lo tanto, el resultado para el Modelo I es:
\beq
\frac{\Delta R}{\left( \frac{\Delta G_{ef}}{G} \right)} =   -83.622\,\mathrm{km}
\eeq
mientras que para el Modelo II, el resultado es:
\beq
\frac{\Delta R}{\left( \frac{\Delta G_{ef}}{G} \right)} =  -78.498 \,\mathrm{km}.
\eeq

Considerando que el radio actual de la Luna es $\tilde{R}_0= 1738\,
{\rm km}$, se obtiene para el Modelo I
\beq
\frac{\Delta R}{\tilde{R}_0}= -0.04811 \left( \frac{\Delta
G_{ef}}{G}\right)  
\eeq
y para el Modelo II
\beq
\frac{\Delta R}{\tilde{R}_0}= -0.04517 \left( \frac{\Delta
G_{ef}}{G}\right). 
\eeq

\begin{table}[!ht]
\renewcommand{\arraystretch}{1.3}
\begin{center}
\begin{tabular}{|c|c|c|c|c|c|c|}
\hline
regi\'on & $V_S$ &  $V_P$ & $\rho$ & $\kappa_S$ & Prof. inicial  & Prof. final  \\
{ } & $\left[\mathrm{km \,s^{-1}}\right]$ &  $\left[\mathrm{km \,s^{-1}}\right]$ & $\left[\mathrm{g\,cm^{-3}}\right]$ & $\left[\mathrm{kbar^{-1}}\right]$ &   $\left[\mathrm{km}\right]$ &  $\left[\mathrm{km}\right]$  \\
\hline
corteza &  ---  & --- & eq.(\ref{dens_luna_corteza}) & eq.(\ref{compr_luna_corteza})  & 0  & 58  \\
\hline
MS I & 4.493  &  7.674 & 3.239  &  0.00097 & 58 & 270\\
\hline
MS II & 4.464  &  7.674 & 3.260  & 0.00095 & 270 & 400\\
\hline
M int. & 4.305  &  7.568 & 3.367  & 0.00091 & 400  & 800\\
\hline
M inf. &   4.549  &  8.223 & 3.389  & 0.00074 &  800  & n\'ucleo\\
 \hline 
\end{tabular}
\end{center}
\caption{Estructura para el Modelo I.}
\label{velocidades_moon_1}
\end{table}

\begin{table}[!ht]
\renewcommand{\arraystretch}{1.3}
\begin{center}
\begin{tabular}{|c|c|c|c|c|c|c|}
\hline
regi\'on & $V_S$ &  $V_P$ & $\rho$ & $\kappa_S$ & Prof. inicial  & Prof. final  \\
{ } & $\left[\mathrm{km \,s^{-1}}\right]$ &  $\left[\mathrm{km \,s^{-1}}\right]$ & $\left[\mathrm{g\,cm^{-3}}\right]$ & $\left[\mathrm{kbar^{-1}}\right]$ &   $\left[\mathrm{km}\right]$ &  $\left[\mathrm{km}\right]$  \\
\hline
corteza &  ---  & --- & eq.(\ref{dens_luna_corteza}) & eq.(\ref{compr_luna_corteza})  & 0  & 58  \\
\hline
MS I & 4.511  &  7.748 & 3.319  & 0.00092 & 58 & 270\\
\hline
MS  II & 4.461  &  7.772 & 3.319  & 0.00089 & 270 & 400\\
\hline
M int. & 4.294  &  7.526 & 3.375  & 0.00092 & 400 & 800\\
\hline
M inf. &   4.532  &  8.202 & 3.376  & 0.00074 & 800 & n\'ucleo\\
 \hline 
\end{tabular}
\end{center}
\caption{Estructura  para el Modelo II.}
\label{velocidades_moon_2}
\end{table}

\section{Ajuste y resultados}

Para realizar el ajuste, utilizaremos los l{\'\i}mites para la
variaci\'on de los radios planetarios dados en  \citet{McElhinny}.
Para la Tierra, estas cotas fueron obtenidas a partir de datos
paleomagn\'eticos con un m\'etodo que compara las posiciones de
regiones que presentan evidencias del campo magn\'etico terrestre
correspondientes al mismo tiempo geol\'ogico, y estima cual ser{\'\i}a
el radio de la Tierra en esa \'epoca teniendo en cuenta el posible
movimiento de placas tect\'onicas. 
La cota para la Luna se obtiene estudiando su superficie y determinando
la edad de sus ``mares'' y ``tierras altas''. As{\'\i} se estim\'o
que la superficie de la Luna adquiri\'o su forma actual hace  
$3.9\times 10^9$ a\~nos, cuando tuvieron lugar los \'ultimos eventos
de mayores cambios de estructura.  Finalmente, en las escarpas de la
superficie de Mercurio hay evidencias de una contracci\'on del manto
que ocurri\'o hace m\'as de $4 \times 10^9$ a\~nos.  En la Tabla
\ref{datos_Delta_R} se detallan los l{\'\i}mites obtenidos para la 
variaci\'on del radio de estos planetas.
\begin{table}[!ht]
\renewcommand{\arraystretch}{1.3}
\begin{center}
\begin{tabular}{|c|c|c|}
\hline
Planeta & $R_a/\tilde{R}_0$ &  edad  $\left( 10^9\, {\rm a\tilde{n}os} \right)$  \\
\hline
Tierra & $1.020 \pm 0.028$ & $0.4$   \\
\hline
Mercurio & $1.0000 \pm 0.0004 $ & $3.5$ \\
\hline
Luna & $1.0000 \pm 0.0006$ & $3.9$ \\
\hline 
\end{tabular}
\end{center}
\caption{Paleoradios $R_a$ dados en funci\'on del radio actual de la
 Tierra, Mercurio, y la Luna.}
\label{datos_Delta_R}
\end{table}

Usaremos estos datos para realizar un ajuste de $\Delta G_{ef}/G$ y
buscar una cota a la variaci\'on de la constante de estructura fina
$\alpha$, entre el tiempo de formaci\'on de los planetas y
hoy. Supondremos que $\alpha$ tuvo un valor constante en el pasado,
aunque no necesariamente igual a su valor actual.
Recordemos 
\beq
\Delta R= \xi(M_T)= \tilde{R}_0 - R_0 \qquad \Rightarrow \frac{\Delta R}{\tilde{R}_0} = 1 - \frac{R_0}{\tilde{R}_0}.
\label{expresion_final_Delta_R}
\eeq
siendo ${R_0}/{\tilde{R}_0}$ el cociente entre el paleoradio y el
radio actual del planeta.  En \citet{McElhinny} se explica que un
cambio hom\'ologo en el planeta no ser{\'\i}a detectado por las
observaciones, dado que escalea todas las dimensiones lineales de
igual manera. Es decir, {\bf las cotas para la variaci\'on del radio
est\'an dadas para diferencias que excedan un simple cambio de
escala. Es por eso que las cotas dadas en la Tabla \ref{datos_Delta_R}
se comparan con $R_0/\tilde{R}_0$ de la expresi\'on
(\ref{expresion_final_Delta_R})}.

Realizamos el ajuste para $\Delta G_{ef}/G$. En las secciones
anteriores obtuvimos valores $N_i$ para el cociente entre ${\Delta
R}/{\tilde{R}_0}$ y $\Delta G_{ef}/G$ en el caso de la Tierra,
Mercurio y la Luna. Estos valores se resumen en la Tabla
\ref{tabla:Ni}. De esta manera
\beq
\left(\frac{R_0}{\tilde{R}_0}\right)_i = 1 - \left(\frac{\Delta R}{\tilde{R}_0}\right)_i = 1 - N_i
\, \frac{\Delta G_{ef}}{G} \qquad i=1,2,3.
\eeq

\begin{table}[!ht]
\renewcommand{\arraystretch}{1.3}
\begin{center}
\begin{tabular}{|c|c|c|c|c|}
\hline
Planeta & Tierra & Mercurio & Luna (modelo I) & Luna (modelo II)  \\
\hline
$N_i$ &  $-0.04522$   &  $-0.02005$   &  $-0.04811$  &  $-0.04517$   \\
\hline
\end{tabular}
\end{center}
\caption{Valores del cociente entre  ${\Delta
R}/{\tilde{R}_0}$ y $\Delta G_{ef}/G$ para la Tierra, Mercurio, y
la Luna.}
\label{tabla:Ni}
\end{table}

Minimizamos el $\chi^2$ definido por 

\bea
\chi^2 = \sum_{i=1}^3 \frac{\left[\left( 1 - N_i \cdot \frac{\Delta G_{ef}}{G}\right)  -
\left(\frac{R_0}{\tilde{R}_0}\right)_i  \right]^2}{\sigma_i^2}.
\eea

El valor de la cota que se obtiene para $\Delta G_{ef}/G$, ya sea que
se utilice el modelo I o II para la Luna, es
\beq
\frac{\Delta G_{ef}}{G} =  0.000 \pm 0.011.
\eeq

Buscamos la cota para la variaci\'on de $\alpha$. Para ello,
consideramos que 
\beq
\frac{\Delta G_{ef}}{G} = \frac{G - G_{ef}}{G} = 1 - \frac{G_{ef}}{G}
= 1 - \frac{\alpha_0}{\alpha}.
\eeq
Obtenemos el siguiente ajuste
\beq
\frac{\Delta \alpha}{\alpha_0}= \frac{\alpha}{\alpha_0} - 1 =
0.000\pm 0.011.
\eeq

 El l{\'\i}mite encontrado resulta relevante por tratarse de una
cota local, e independiente de otros resultados a tiempos similares. Si
bien esta cota es m\'as d\'ebil que la obtenida a partir del reactor
nuclear natural que oper\'o en Oklo hace $1.8 \times 10^{9}$ a\~nos,
la cual tiene una precisi\'on de $10^{-8}$, o de los decaimientos
$\beta$, cuya precisi\'on es de $10^{-7}$, esta cota resulta
interesante por ser obtenida a partir de consideraciones muy
generales, mientras que las otras aqu{\'\i} mencionadas se obtuvieron
haciendo fuertes suposiciones en los modelos utilizados.

\cleardoublepage
\chapter{Resumen y Conclusiones}
\label{chap:conclusiones}

Esta tesis constituye un an\'alisis de la variaci\'on de la masa del
electr\'on $m_e$, y de la constante de estructura fina $\alpha$, en
distintas etapas de la evoluci\'on del Universo.

Una parte importante de este trabajo se focaliza en la \'epoca de
formaci\'on del hidr\'ogeno neutro, que se conoce como recombinaci\'on
del Universo. Hemos estudiado el escenario est\'andar de la
recombinaci\'on, y modificado las ecuaciones para las fracciones de
ionizaci\'on, para hacer posible que las constantes en estudio tomen
valores arbitrarios. Sin embargo, mantenemos la forma de las
ecuaciones, como propone la hip\'otesis adiab\'atica cuando las
variaciones son peque\~nas.  Analizamos las dependencias de las
cantidades f{\'\i}sicas con las constantes fundamentales, incluyendo
las cantidades relevantes en la recombinaci\'on del helio, y otros
fen\'omenos que modifican la historia de recombinaci\'on, tales como
las transiciones desde los estados tripletes del nivel $n=2$ al $n=1$
en el hidr\'ogeno, o el efecto del
hidr\'ogeno neutro sobre la recombinaci\'on del helio. Destacamos que
las dependencias en este nuevo escenario, que se detallan en el
Cap{\'\i}tulo \ref{ch:fisica_recomb}, es una de las contribuciones
originales de esta tesis \citep{Scoccola08b}. Modificamos las versiones
del c\'odigo R{\sc ecfast} que resuelven las ecuaciones de ionizaci\'on en
el escenario est\'andar y en el escenario actualizado, para poder
utilizarlas luego para hacer los ajustes con los datos.

Utilizando datos del fondo c\'osmico de radiaci\'on, informaci\'on
acerca de $H_0$ obtenida a partir del HST Key Project, y el espectro
de potencias de las galaxias del cat\'alogo 2dFGRS, hemos encontrado
l{\'\i}mites observacionales a la variaci\'on de $\alpha$ y $m_e$
entre la recombinaci\'on y la actualidad.  Realizamos an\'alisis
estad{\'\i}sticos, donde ajustamos el valor de un conjunto de
par\'ametros cosmol\'ogicos, junto con la variaci\'on de las
constantes. Analizamos la variaci\'on de cada una de ellas por
separado, y el caso en que ambas constantes var{\'\i}an
simult\'aneamente. Realizamos distintos ajustes utilizando diferentes
conjuntos de datos, y comparamos nuestros resultados con los
existentes en la literatura. En esta tesis se presentan resultados
usando un prior de $H_0$ que fue calculado a partir de objetos
astron\'omicos cercanos. Estos an\'alisis son los \'unicos en la
literatura donde se considera la posibilidad de que el resultado final
para el valor de $H_0$ presentado por el HST Key Project est\'e
afectado por la variaci\'on de las constantes, debido a que algunos de
los objetos utilizados para medir $H_0$ se encuentran a alto redshift,
donde las constantes pudieron tener un valor diferente del actual.

Se presentan cotas actualizadas con los datos de fondo c\'osmico de
radiaci\'on incluyendo los datos de WMAP5 y el espectro de potencias
del 2dFGRS, para la variaci\'on de $\alpha$, la variaci\'on de $m_e$ y
la variaci\'on conjunta.  Adem\'as, se estim\'o una relaci\'on
fenomenol\'ogica entre la variaci\'on de las constantes, que fue
encontrada a posteriori, en el an\'alisis donde se variaron ambas
constantes de manera independiente. Comparamos nuestros an\'alisis con
los realizados en \citet{Ichi06}, donde se propone que las variaciones
est\'an relacionadas a partir de una ley de potencias, motivada en el
contexto de una teor{\'\i}a de cuerdas heter\'oticas, y realizan el
ajuste presuponiendo dicha relaci\'on. Dado que los valores del
exponente de la ley de potencias utilizada por estos autores son
positivos y que del ajuste independiente que realizamos en esta tesis
estimamos una correlaci\'on negativa entre las variaciones de $\alpha$
y $m_e$, conclu{\'\i}mos que la propuesta de \citet{Ichi06}
deber{\'\i}a ser descartada.  Los resultados est\'an publicados
en \citet{landau08}.

Se presenta la \'unica cota existente para la variaci\'on de $m_e$
entre el desacople de la materia y la radiaci\'on y la \'epoca actual,
y \'esta fue obtenida a partir de los datos m\'as actualizados del
fondo c\'osmico de radiaci\'on junto con el espectro de potencias del
cat\'alogo 2dFGRS. La cota encontrada con los datos de WMAP3 y el
espectro del 2dFGRS fue publicada en \citet{Scoccola08}, antes de que
salieran publicados los datos de WMAP5.

Los datos actuales comprenden mediciones muy precisas de las
anisotrop\'{\i}as del fondo c\'osmico de radiaci\'on. Sin embargo, no
se ha podido medir con precisi\'{o}n la polarizaci\'{o}n. En los
pr\'{o}ximos a\~nos, una gran cantidad de experimentos (algunos en
marcha, otros previstos) permitir\'{a}n obtener mediciones precisas de
la polarizaci\'{o}n del fondo c\'osmico de radiaci\'on. Algunos de
ellos son CMBPol \citep{cmbpol}, QUIET \citep{quiet},
Clover \citep{clover}, BICEP \citep{bicep}, y Planck \citep{planck}.
Pero a pesar de que las misiones futuras tendr\'an mayor precisi\'on,
a\'un as{\'\i} no permitir\'an ajustar la variaci\'on de las
constantes fundamentales con una precisi\'on mejor que $10^{-3}$.

La cotas obtenidas observacionalmente para la variaci\'on de las
constantes fundamentales constituyen una de las pocas herramientas con
las cuales testear teor{\'\i}as que intentan unificar las cuatro
interacciones de la naturaleza, como hemos mencionado en la
Introducci\'on. En esta tesis, analizamos el modelo fenomenol\'ogico
de Barrow-Magueijo, que propone una variaci\'on de la masa del
electr\'on inducida por cambios de un campo escalar en el
espacio-tiempo. Mejoramos las soluciones originalmente presentadas por
dichos autores, teniendo en cuenta de manera m\'as detallada la
evoluci\'on del factor de expansi\'on del Universo, y haciendo menos
aproximaciones que en el trabajo original. Estimamos los par\'ametros
del modelo usando las cotas para la variaci\'on de la masa del
electr\'on que se obtienen en esta tesis a partir de los datos del
fondo c\'osmico de radiaci\'on, y otros datos astron\'omicos y de
laboratorio correspondientes a distintos tiempos cosmol\'ogicos. Por
otra parte, utilizando cotas observacionales obtenidas en experimentos
de E\"otv\"os que testean la validez del Principio de Equivalencia
D\'ebil, encontramos un l{\'\i}mite inferior para el par\'ametro que
aparece en el Lagrangiano propuesto en este modelo, que viene de la
variaci\'on en la masa del electr\'on inducida por cambios espaciales
del campo escalar. Comparando esta \'ultima con la cota obtenida con
los datos a distintos tiempos, cuando estudiamos evoluci\'on temporal
del campo, se llega a la conclusi\'on de que el modelo es
inconsistente y debe ser descartado. Estos resultados fueron
publicados en \citet{Scoccola08}.

Por \'ultimo, estudiamos la variaci\'on de las constantes
fundamentales entre la actualidad y la \'epoca de formaci\'on de los
planetas del Sistema Solar. A partir de consideraci\'on muy generales,
analizamos c\'omo se modificar{\'\i}a el radio de un planeta por
variaci\'on de las constantes $\alpha$ y $m_e$. Explicamos por qu\'e
la variaci\'on de la masa del electr\'on no induce cambios en el radio
que sean factibles de ser observados.  Obtenemos una cota para la
variaci\'on de la constante de estructura fina a partir de
l{\'\i}mites sobre la variaci\'on del radio de la Tierra, la Luna y
Mercurio, desde la \'epoca de los \'ultimos cambios estructurales
mayores en la superficie de los mismos. 
Si bien es una cota d\'ebil, es importante notar que se trata de un
resultado local, e independiente de otras medidas. Adem\'as, fue
obtenido a partir de consideraciones muy generales.

\cleardoublepage
\appendix


\chapter{Modelo \emph{PREM}}

\begin{table}[h!]
\begin{center}
\begin{tabular}{|c|c|c|c|c|c|c|c|c|c|}
\hline
Radio	&	Prof.	&	Densidad	&	Vp	&	Vs	&	$Q_\mu$	&	$K_{adiab}$&	$\mu$	&	Presi\'on	&	Gravedad	\\
(km)	&	(km)	&  ($\mathrm{g/cm^3}$)	&	(km/s)	&	(km/s)	&		&	(kbar)	&	(kbar)	&	(kbar)	&	($\mathrm{cm/s^2)}$\\
\hline
0	&	6371	&	13.08	&	11.26	&	3.66	&	85	&	14253	&	1761	&	3638.5	&	0	\\
200	&	6171	&	13.07	&	11.25	&	3.66	&	85	&	14231	&	1755	&	3628.9	&	73.1	\\
400	&	5971	&	13.05	&	11.23	&	3.65	&	85	&	14164	&	1739	&	3600.3	&	146	\\
600	&	5771	&	13.01	&	11.2	&	3.62	&	85	&	14053	&	1713	&	3552.7	&	218.6	\\
800	&	5571	&	12.94	&	11.16	&	3.59	&	85	&	13898	&	1676	&	3486.6	&	290.6	\\
1000	&	5371	&	12.87	&	11.1	&	3.55	&	85	&	13701	&	1630	&	3402.3	&	362	\\
1200	&	5171	&	12.77	&	11.03	&	3.51	&	85	&	13462	&	1574	&	3300.4	&	432.5	\\
1221.5	&	5149.5	&	12.76	&	11.02	&	3.5	&	85	&	13434	&	1567	&	3288.5	&	440	\\
1221.5	&	5149.5	&	12.16	&	10.35	&	0	&	0	&	13047	&	0	&	3288.5	&	440	\\
1400	&	4971	&	12.06	&	10.24	&	0	&	0	&	12679	&	0	&	3187.4	&	494.1	\\
1600	&	4771	&	11.94	&	10.12	&	0	&	0	&	12242	&	0	&	3061.4	&	555.4	\\
1800	&	4571	&	11.8	&	9.98	&	0	&	0	&	11775	&	0	&	2922.2	&	616.6	\\
2000	&	4371	&	11.65	&	9.83	&	0	&	0	&	11273	&	0	&	2770.4	&	677.1	\\
2200	&	4171	&	11.48	&	9.66	&	0	&	0	&	10735	&	0	&	2606.8	&	736.4	\\
2400	&	3971	&	11.29	&	9.48	&	0	&	0	&	10158	&	0	&	2432.4	&	794.2	\\
2600	&	3771	&	11.08	&	9.27	&	0	&	0	&	9542	&	0	&	2248.4	&	850.2	\\
2800	&	3571	&	10.85	&	9.05	&	0	&	0	&	8889	&	0	&	2055.9	&	904.1	\\
3000	&	3371	&	10.6	&	8.79	&	0	&	0	&	8202	&	0	&	1856.4	&	955.7	\\
3200	&	3171	&	10.32	&	8.51	&	0	&	0	&	7484	&	0	&	1651.2	&	1004.6	\\
 \hline 
\end{tabular}
\end{center}
\caption{Par\'ametros del modelo PREM}
\label{tabla_PREM}
\end{table}

\begin{table}[h!]
\begin{center}
\begin{tabular}{|c|c|c|c|c|c|c|c|c|c|}
\hline
Radio	&	Prof.	&	Densidad	&	Vp	&	Vs	&	$Q_\mu$	&	$K_{adiab}$&	$\mu$	&	Presi\'on	&	Gravedad	\\
(km)	&	(km)	&  ($\mathrm{g/cm^3}$)	&	(km/s)	&	(km/s)	&		&	(kbar)	&	(kbar)	&	(kbar)	&	($\mathrm{cm/s^2)}$\\
\hline
3400	&	2971	&	10.02	&	8.19	&	0	&	0	&	6743	&	0	&	1441.9	&	1050.6	\\
3480	&	2891	&	9.9	&	8.06	&	0	&	0	&	6441	&	0	&	1357.5	&	1068.2	\\
3480	&	2891	&	5.56	&	13.71	&	7.26	&	312	&	6556	&	2938	&	1357.5	&	1068.2	\\
3600	&	2771	&	5.5	&	13.68	&	7.26	&	312	&	6440	&	2907	&	1287	&	1052	\\
3800	&	2571	&	5.4	&	13.47	&	7.18	&	312	&	6095	&	2794	&	1173.4	&	1030.9	\\
4000	&	2371	&	5.3	&	13.24	&	7.09	&	312	&	5744	&	2675	&	1063.8	&	1015.8	\\
4200	&	2171	&	5.2	&	13.01	&	7.01	&	312	&	5409	&	2559	&	957.6	&	1005.3	\\
4400	&	1971	&	5.1	&	12.78	&	6.91	&	312	&	5085	&	2445	&	854.3	&	998.5	\\
4600	&	1771	&	5	&	12.54	&	6.82	&	312	&	4766	&	2331	&	753.5	&	994.7	\\
4800	&	1571	&	4.89	&	12.29	&	6.72	&	312	&	4448	&	2215	&	655.2	&	993.1	\\
5000	&	1371	&	4.78	&	12.02	&	6.61	&	312	&	4128	&	2098	&	558.9	&	993.2	\\
5200	&	1171	&	4.67	&	11.73	&	6.5	&	312	&	3803	&	1979	&	464.8	&	994.6	\\
5400	&	971	&	4.56	&	11.41	&	6.37	&	312	&	3471	&	1856	&	372.8	&	996.9	\\
5600	&	771	&	4.44	&	11.06	&	6.24	&	312	&	3133	&	1730	&	282.9	&	998.8	\\
5650	&	721	&	4.41	&	10.91	&	6.09	&	312	&	3067	&	1639	&	260.7	&	1000.6	\\
5701	&	670	&	4.38	&	10.75	&	5.94	&	312	&	2999	&	1548	&	238.3	&	1001.4	\\
5701	&	670	&	3.99	&	10.26	&	5.57	&	143	&	2556	&	1239	&	238.3	&	1001.4	\\
5771	&	600	&	3.97	&	10.15	&	5.51	&	143	&	2489	&	1210	&	210.4	&	1000.3	\\
5871	&	500	&	3.84	&	9.64	&	5.22	&	143	&	2181	&	1051	&	171.3	&	998.8	\\
5921	&	450	&	3.78	&	9.38	&	5.07	&	143	&	2037	&	977	&	152.2	&	997.9	\\
5971	&	400	&	3.72	&	9.13	&	4.93	&	143	&	1899	&	906	&	133.5	&	996.8	\\
5971	&	400	&	3.54	&	8.9	&	4.76	&	143	&	1735	&	8.6	&	133.5	&	996.8	\\
6061	&	310	&	3.48	&	8.73	&	4.7	&	143	&	1630	&	773	&	102	&	993.6	\\
 \hline 
\end{tabular}
\end{center}
\caption{Par\'ametros del modelo PREM (cont.)}
\label{tabla_PREM_cont}
\end{table}

\begin{table}[h!]
\begin{center}
\begin{tabular}{|c|c|c|c|c|c|c|c|c|c|}
\hline
Radio	&	Prof.	&	Densidad	&	Vp	&	Vs	&	$Q_\mu$	&	$K_{adiab}$&	$\mu$	&	Presi\'on	&	Gravedad	\\
(km)	&	(km)	&  ($\mathrm{g/cm^3}$)	&	(km/s)	&	(km/s)	&		&	(kbar)	&	(kbar)	&	(kbar)	&	($\mathrm{cm/s^2)}$\\
\hline
6106	&	265	&	3.46	&	8.64	&	4.67	&	143	&	1579	&	757	&	86.4	&	992	\\
6151	&	220	&	3.43	&	8.55	&	4.64	&	143	&	1529	&	741	&	71.1	&	990.4	\\
6151	&	220	&	3.35	&	7.98	&	4.41	&	80	&	1270	&	656	&	71.1	&	990.4	\\
6186	&	185	&	3.36	&	8.01	&	4.43	&	80	&	1278	&	660	&	59.4	&	989.1	\\
6221	&	150	&	3.36	&	8.03	&	4.44	&	80	&	1287	&	665	&	47.8	&	987.8	\\
6256	&	115	&	3.37	&	8.05	&	4.45	&	80	&	1295	&	669	&	36.1	&	986.6	\\
6291	&	80	&	3.37	&	8.07	&	4.46	&	80	&	1303	&	674	&	24.5	&	985.5	\\
6291	&	80	&	3.37	&	8.07	&	4.46	&	600	&	1303	&	674	&	24.5	&	985.5	\\
6311	&	60	&	3.37	&	8.08	&	4.47	&	600	&	1307	&	677	&	17.8	&	984.9	\\
6331	&	40	&	3.37	&	8.1	&	4.48	&	600	&	1311	&	680	&	11.2	&	984.3	\\
6346.6	&	24.4	&	3.38	&	8.11	&	4.49	&	600	&	1315	&	682	&	6	&	983.9	\\
6346.6	&	24.4	&	2.9	&	6.8	&	3.9	&	600	&	753	&	441	&	6	&	983.9	\\
6356	&	15	&	2.9	&	6.8	&	3.9	&	600	&	753	&	441	&	3.3	&	983.3	\\
6356	&	15	&	2.6	&	5.8	&	3.2	&	600	&	520	&	266	&	3.3	&	983.3	\\
6368	&	3	&	2.6	&	5.8	&	3.2	&	600	&	520	&	266	&	0.3	&	982.2	\\
6368	&	3	&	1.02	&	1.45	&	0	&	0	&	21	&	0	&	0.2	&	982.2	\\
6371	&	0	&	1.02	&	1.45	&	0	&	0	&	21	&	0	&	0	&	981.5	\\
 \hline 
\end{tabular}
\end{center}
\caption{Par\'ametros del modelo PREM (cont.)}
\label{tabla_PREM_cont2}
\end{table}

\chapter{Integrales I}
\label{apendice_2}

Tenemos que resolver esta ecuaci\'on diferencial:

\beq
d \phi = -\ { M\ \OmegaL \over H_0^2\ \OmegaM}\ {\left( \tau - {H_0\ A
    \over M} \right)\over \sinh^2\left[ {3\over 2}\sqrt{\OmegaL} \tau
    + C' \right]}\ d\tau
\eeq
es decir, algo de la forma:
\beq
\int {a + x\over \sinh^2 \left( c + bx\right)} dx = \underbrace{ a \int{dx\over
  \sinh^2\left( c +bx\right)}}_{I}\ + \ \underbrace{\int {x\ dx\over \sinh^2\left( c +bx\right)}}_{II}
\eeq

Sea 
\bdm
y=c+bx \quad \Rightarrow dy=b\ dx \quad \Rightarrow dx = {1\over
  b}\ dy
\edm

\beq
I = a \int {dx\over \sinh^2\left(c+b x \right)} = {a\over b} \int
{dy\over \sinh^2 y} = - {a\over b} \coth y= - {a\over b} \coth \left(c+bx \right)
\eeq
ya que 
\beq
{d \left( - \coth y \right)\over dy} = {d \over dy} \left(- {\cosh
  y\over \sinh y}\right) = - {\sinh^2 y - \cosh^2 y \over \sinh^2 y} =
{1\over \sinh^2 y}
\eeq

Por otra parte,

\bea
II = \int {x\ dx\over \sinh^2\left(c+b x \right)}  &=& {1\over b^2} \int
{(y-c) dy\over \sinh^2 y} = {1\over b^2} \left[ \int{y\ dy\over
    \sinh^2 y} -c \underbrace{\int{dy\over \sinh^2 y}}_{-\coth
    y}\right]\nonumber\\
&=& {1\over b^2} \int{y\ dy\over \sinh^2 y} + {c \over b^2} \coth y
\eea

Calculamos ahora
\beq
\int {y\ dy\over \sinh^2 y} = - y\ \coth y + \int\coth y dy
\eeq

donde hemos integrado por partes. Adem\'as:

\beq
\int \coth y\ dy = \int {\cosh y\ dy \over \sinh y} = \ln\left( \sinh y\right) 
\eeq

Entonces,

\beq
\int {y\ dy \over \sinh^2 y} = - y\ \coth y + \ln\left( \sinh y\right) 
\eeq

Por lo tanto:

\bea
II &=& {1\over b^2} \left( - y\ \coth y + \ln\left( \sinh y\right)
\right) + {c \over b^2} \coth y \nonumber\\
&=& - {\left(c + bx\right)\over b^2} \coth \left(c + bx\right)
+{1\over b^2} \ln\left( \sinh \left(c + bx\right) \right)+ {c \over
  b^2} \coth \left(c + bx\right) \nonumber\\
{   }
\eea

En definitiva:
\beq
\int {a + x\over \sinh^2 \left( c + bx\right)} dx = - {\left( x+a
  \right)\over b} \coth \left(c + bx\right) + {1 \over b^2}  \ln\left( \sinh \left(c + bx\right) \right)
\eeq

Reemplazando las cantidades originales, llegamos a esta expresi\'on:

\bea
&& \phi ^{\Lambda}- \phi_0 = { M\ \OmegaL \over H_0^2\ \OmegaM} \times \nonumber\\
&&\times
 \left[ {2 \left(
    \tau - {H_0\ A\over M} \right) \over 3 \sqrt{\OmegaL} } \coth
  \left( C' + {3\over 2} \sqrt{\OmegaL}\tau \right) - {4 \over
  9 \OmegaL} \ln \left( \sinh \left(C' + {3\over 2} \sqrt{\OmegaL}
  \tau \right) \right) - \right. \nonumber\\
&&\left. - {2 \left( \tau_0 - {H_0\ A\over M} \right) \over 3 \sqrt{\OmegaL} } \coth
  \left( C' + {3\over 2} \sqrt{\OmegaL} \tau_0 \right) + {4 \over
  9 \OmegaL} \ln \left( \sinh \left(C' + {3\over 2} \sqrt{\OmegaL}
  \tau_0 \right) \right) 
\right]\nonumber\\
&& {}\label{solucion_phi_MCC}
\eea

Reemplazando el valor de $\tau_0$, reordenando t\'erminos y llamando
$\gamma= \sqrt{\OmegaL/\OmegaM}$, la soluci\'on para la \'epoca de
dominio de materia m\'as constante cosmol\'ogica es:

\begin{multline}
 \phi ^{\Lambda} (\tau) = \phi_0 + \frac{M}{H_0^2}\, \frac{2}{3\OmegaM}\left[ \sqrt{\OmegaL} \tau \coth \left( C' +  \frac{3}{2}\sqrt{\OmegaL}\tau  \right) \right.  \nonumber\\
- \frac{2}{3} \ln \left(\sinh \left( C' + \frac{3}{2}\sqrt{\OmegaL}\tau \right) \right)  
 \left. +  \frac{2}{3}  \left( \ln\gamma - \frac{\sqrt{1 + \gamma^2}}{\gamma} \left[ C' + \ln \left(\gamma + \sqrt{1+\gamma^2}\right) \right]\right)   \right] + \nonumber\\
 + \frac{A}{H_0}\,\frac{2\sqrt{\OmegaL}}{3\OmegaM}\left[-\coth\left(C' +   \frac{3}{2}\sqrt{\OmegaL}\tau \right) +   \frac{\sqrt{1 + \gamma^2}}{\gamma} \right] \nonumber\\
 {}
\end{multline}

Esta soluci\'on vale desde el momento de ajuste, $\tau_A$ (donde los
dos reg{\'\i}menes se superponen), hasta hoy,
$\tau_0$. Evaluamos la Ec.~(\ref{solucion_phi_MCC}) en $\tau_A$:

\bea 
\phi_A &=& \phi_0 + { M\ \OmegaL \over H_0^2\ \OmegaM} \left[ {2
\over 3\sqrt{\OmegaL} }\left( \tau_A - {H_0\ A\over M}
  \right)\sqrt{\xi +1} - {4 \over 9 \OmegaL} \ln{(\xi^{-1/2})} - \right. \nonumber\\
&&\left. -  {2 \over 3\sqrt{\OmegaL} }\left( \tau_0 - {H_0\ A\over M} \right)
	{\sqrt{1+\gamma^2}\over \gamma} + {4 \over 9 \OmegaL}
\ln \gamma \right]
\label{phi_A}
\eea

\chapter{Integrales II}
\label{apendice_1}

La ecuaci\'on a integrar es:
\beq
\int_{\phi_A}^{\phi} d\phi'  = \int_{\eta_A}^{\eta} {-{ M \over 2
      H_0} \eta'^2 \left({\OmegaM \over 6} \eta' + \sqrt{\OmegaR}
    \right) + A \over H_0 \eta'^2 \left( {\OmegaM \over 4} \eta' +
    \sqrt{\OmegaR}\right)^2} d\eta' 
\eeq
Por lo tanto, integramos la siguiente integral indefinida:
\bea
I &=& \int{{-{M\over 2\ H_0}\ \eta^2 \left( {\OmegaM \over 6} \eta +
    \sqrt{\OmegaR} \right) +  A \over H_0\ \eta^2 \left( {\OmegaM \over 4} \eta +
    \sqrt{\OmegaR} \right)^2} }\ d\eta 
\nonumber\\
&=&{-{M \over 2\ H_0^2 \ \sqrt{\OmegaR} }\int {\left( {\OmegaM \over
      6\ \sqrt{\OmegaR}} \eta + 1 \right) \over \left( {\OmegaM \over 4\
      \sqrt{\OmegaR} } \eta + 1 \right)^2}} \ d\eta \nonumber\\
&+& {A \over H_0
  \OmegaR} \int{{1 \over \eta^2 \left( {\OmegaM \over 4\
      \sqrt{\OmegaR}} \eta + 1\right)^2}}\ d\eta
\eea

Sea $\beta = \OmegaM / \sqrt{\OmegaR}$, $a=\beta/6$ y $b=\beta/4$
($a=2\ b/3$),

\beq
I =  -{M \over 2\ H_0^2 \ \sqrt{\OmegaR} } \int {a \eta + 1\over
  \left( b \eta +1 \right)^2} d\eta + {A \over H_0 \OmegaR} \int
{d\eta \over \eta^2 \left( b \eta + 1\right)^2}
\eeq

Primero resolvemos:

\bea
\int {a \eta + 1\over \left(b \eta + 1 \right)^2}\ d\eta &=& {a \over b}
\int {b\eta + b/a + 1 - 1 \over \left(b \eta + 1 \right)^2}\ d\eta \nonumber\\
&=& {a \over b} \left[ \int {d\eta \over \left( b \eta + 1\right)} +
       \left({b\over a} - 1\right)\int{d\eta \over \left(b \eta + 1
	 \right)^2} \right] \nonumber\\
&=& {a \over b^2} \ln \left( b\eta + 1\right) + {a - b \over b^2}\ {1
  \over \left( b\eta + 1\right)}\nonumber\\
&=& {2\over 3b} \ln \left(b \eta +1 \right) -{1\over 3 b} {1\over
  \left( b\eta + 1\right)} \nonumber\\
&=& {4\over 3 \beta} \left[ 2 \ln \left( {1\over 4} \beta \eta +
  1\right) - {1\over \left( {1\over 4} \beta \eta + 1\right)} \right]
\eea

La otra integral que se debe resolver es:

\bea
\int {d \eta \over \eta^2 \left( b \eta + 1\right)^2} &=& -2 b
\int{d\eta \over \eta} + \int{d\eta \over \eta^2} + 2 b^2 \int {d \eta
  \over \left( b \eta + 1\right)} + b^2 \int{d\eta \over\left( b \eta
  + 1\right)^2 }\nonumber\\
&=& -2b \ln \eta -{1\over \eta} + 2 b \ln \left( b \eta + 1 \right) -
b {1\over \left(b \eta + 1 \right)}
\eea

Por lo tanto,

\bea
I &=& - {M \over 6\ b H_0^2 \sqrt{\OmegaR}} \left[  2 \ln \left( b
  \eta + 1 \right) - {1\over \left(b \eta + 1 \right)}\right]
  +\nonumber\\
&+& {A\ b\over H_0 \OmegaR} \left[-2 \ln{\eta} - {1\over b \eta} + 2
    \ln{\left(b \eta + 1\right)} - {1 \over \left( b \eta + 1\right)}  \right]
\eea

Reemplanzando las cantidades originales, la soluci\'on es:
\bea
\phi^{R} - \phi_A &=& - {2\over 3} {M\over H_0^2 \OmegaM}\left( 2
\ln(b\eta +1) -{1\over (b\eta +1) }\right) + \nonumber\\ 
&+&  { A\ \OmegaM \over 4 H_0 \OmegaR^{3/2}} \left[ -2 \ln \eta - {1\over
  b\eta} + 2 \ln(b\eta +1) - {1\over (b\eta +1)}\right] + \nonumber\\ 
&+&   {2\over 3} {M\over H_0^2 \OmegaM} \left( 2
\ln(b\eta_A +1) - {1\over (b\eta_A +1) }\right) - \nonumber\\
&-&  { A\ \OmegaM \over 4 H_0 \OmegaR^{3/2}} \left[ -2 \ln \eta_A - {1\over
  b\eta_A} + 2 \ln(b\eta_A +1) - {1\over (b\eta_A +1)}\right]
  \nonumber\\
{  }
\label{phirad_int}
\eea
con
\bdm
b\eta_A = {1\over 2} \left( \sqrt{1 + \xi} - 1\right), \quad
b={1\over 4} {\OmegaM \over \sqrt{\OmegaR}}\Rightarrow \quad
\eta_A = 2 {\sqrt{\OmegaR}\over \OmegaM} \left[ \sqrt{1 + \xi} - 1
  \right]
\edm

Considerando que $\phi$ es continuo, y reemplazando $\phi_A$ (de la
Ec.~(\ref{phi_A})) en la ec.~(\ref{phirad_int}), el campo $\phi$ para
la \'epoca dominada por materia y radiaci\'on es:
\bea
&&\phi^R = - {2\over 3} {M\over H_0^2}{1\over \OmegaM}\left( 2
\ln(b\eta +1) -{1\over (b\eta +1) }\right) + \nonumber\\ 
&+& {1 \over 4} { A \over H_0} {\OmegaM \over  \OmegaR^{3/2}} \left[ -2 \ln \eta - {1\over
  b\eta} + 2 \ln(b\eta +1) - {1\over (b\eta +1)}\right] + \nonumber\\ 
&+&   {2\over 3}{M\over H_0^2}{1\over \OmegaM}  \left[ 2
\ln(b\eta_A +1) - {1\over (b\eta_A +1) } + {2\over 3} f(\xi) \sqrt{\xi
+ 1} - {2\over 3} \ln(\xi^{-1/2}) - \right. \nonumber \\
&&\left. - {2\over 3} \left( \sinh^{-1}(\gamma) -
\sinh^{-1}(\xi^{-1/2}) + f(\xi) \right) {\sqrt{1 + \gamma^2}\over
  \gamma} + {2\over 3}\ln(\gamma)  \right] - \nonumber\\
&-&  { A \over H_0}{\OmegaM \over \OmegaR^{3/2}} \left[ -{1 \over 2} \ln \eta_A - {1\over
 4 b\eta_A} + {1\over 2} \ln(b\eta_A +1) - {1\over 4(b\eta_A +1)} +
  {2\over 3} { \sqrt{\xi +1} - {\sqrt{\gamma^2 +1}\over \gamma}\over
    \xi^2}\right] \nonumber\\
&& {  }\nonumber\\
&& {  }
\label{phirad_final}
\eea
donde $\gamma$ se defini\'o en el Ap\'endice \ref{apendice_2} y 
\beq
\xi = \OmegaM \OmegaR^{-3/4} \OmegaL^{-1/4} \qquad f\left( \xi \right) = {2 + (\xi - 2) \sqrt{1+ \xi} \over \xi^2}
\eeq

Adem\'as:

\beq
\sinh^{-1} x = \ln \left( x + \sqrt{x^2 +1}\right)
\eeq

$A$ es la constante de integraci\'on, y  
\beq
M= \frac{\rho_{e0}}{\omega} c^4
\eeq

\bibliographystyle{aa}

\bibliography{bibliografia_tesis}

\begin{thebibliography}{139}
\expandafter\ifx\csname natexlab\endcsname\relax\def\natexlab#1{#1}\fi

\bibitem[{{Albrecht} \& {Magueijo}(1999)}]{AlbMag99}
{Albrecht}, A. \& {Magueijo}, J.~. 1999, \prd, 59, 043516

\bibitem[{{Avelino} {et al.}(2001){Avelino}, {Esposito}, {Mangano}, {Martins},
  {Melchiorri}, {Miele}, {Pisanti}, {Rocha}, \& {Viana}}]{av01}
{Avelino}, P.~P., {Esposito}, S., {Mangano}, G., {et al.} 2001, \prd, 64,
  103505

\bibitem[{{Avelino} {et al.}(2000){Avelino}, {Martins}, {Rocha}, \&
  {Viana}}]{AV00}
{Avelino}, P.~P., {Martins}, C.~J.~A.~P., {Rocha}, G., \& {Viana}, P. 2000,
  \prd, 62, 123508

\bibitem[{{Bahcall} {et al.}(2004){Bahcall}, {Steinhardt}, \&
  {Schlegel}}]{Bahcall04}
{Bahcall}, J., {Steinhardt}, C.~L., \& {Schlegel}, D. 2004, Astrophys.J., 600,
  520

\bibitem[{{Barr} \& {Mohapatra}(1988)}]{Barr88}
{Barr}, S.~M. \& {Mohapatra}, P.~K. 1988, \prd, 38, 3011

\bibitem[{{Barrow} \& {Magueijo}(2005)}]{BM05}
{Barrow}, J.~D. \& {Magueijo}, J. 2005, \prd, 72, 043521

\bibitem[{{Battye} {et al.}(2001){Battye}, {Crittenden}, \& {Weller}}]{BCW01}
{Battye}, R.~A., {Crittenden}, R., \& {Weller}, J. 2001, \prd, 63, 043505

\bibitem[{Bekenstein(1982)}]{Bekenstein82}
Bekenstein, J.~D. 1982, \prd, 25, 1527

\bibitem[{{Beno{\^i}t}(2004)}]{ARCHEOPS}
{Beno{\^i}t}, A. 2004, Advances in Space Research, 34, 479

\bibitem[{{Bize et al.}(2003)}]{Bize03}
{Bize et al.} 2003, Physical Review Letters, 90, 150802

\bibitem[{Braginskii \& Panov(1972)}]{Braginski72}
Braginskii, V.~B. \& Panov, V.~I. 1972, Sov. Phys. JETP, 34, 463

\bibitem[{{Brax} \& {Davis}(2001)}]{BraxDavis01}
{Brax}, P. \& {Davis}, A.~C. 2001, Journal of High Energy Physics, 5, 7

\bibitem[{{Brax} {et al.}(2003){Brax}, {van de Bruck}, {Davis}, \&
  {Rhodes}}]{branes03b}
{Brax}, P., {van de Bruck}, C., {Davis}, A.-C., \& {Rhodes}, C.~S. 2003,
  Astrophysics and Space Science, 283, 627

\bibitem[{{Breit} \& {Teller}(1940)}]{BreitTeller40}
{Breit}, G. \& {Teller}, E. 1940, \apj, 91, 215

\bibitem[{{Burigana} {et al.}(2004){Burigana}, {Finelli}, {Salvaterra}, {Popa},
  {Mandolesi}, {De Zotti}, {Butler}, {Cuttaia}, {Franceschi}, {Gruppuso},
  {Malaspina}, {Morgante}, {Morigi}, {Sandri}, {Terenzi}, {Valenziano}, \&
  {Villa}}]{Planck_mission}
{Burigana}, C., {Finelli}, F., {Salvaterra}, R., {et al.} 2004, Memorie della
  Societa Astronomica Italiana Supplement, 5, 415

\bibitem[{{Caldwell}(2002)}]{caldwell02}
{Caldwell}, R.~R. 2002, Physics Letters B, 545, 23

\bibitem[{{Clayton} \& {Moffat}(1999)}]{ClaytonMoffat99}
{Clayton}, M.~A. \& {Moffat}, J.~W. 1999, Physics Letters B, 460, 263

\bibitem[{{Clayton} \& {Moffat}(2000)}]{ClaytonMoffat00}
{Clayton}, M.~A. \& {Moffat}, J.~W. 2000, Physics Letters B, 477, 269

\bibitem[{{Clayton} \& {Moffat}(2001)}]{ClaytonMoffat01}
{Clayton}, M.~A. \& {Moffat}, J.~W. 2001, Physics Letters B, 506, 177

\bibitem[{Cole {et al.}(2005)}]{2dF05}
Cole, S. {et al.} 2005, Mon. Not. Roy. Astron. Soc., 362, 505

\bibitem[{Cram\'er(1972)}]{cramer}
Cram\'er, H. 1972, Teor{\'\i}a de Probabilidades (Aguilar ediciones)

\bibitem[{{Damour} \& {Dyson}(1996)}]{DD96}
{Damour}, T. \& {Dyson}, F. 1996, Nuclear Physics B, 480, 37

\bibitem[{{Damour} {et al.}(2002{\natexlab{a}}){Damour}, {Piazza}, \&
  {Veneziano}}]{DPV2002a}
{Damour}, T., {Piazza}, F., \& {Veneziano}, G. 2002{\natexlab{a}}, \prl, 89,
  081601

\bibitem[{{Damour} {et al.}(2002{\natexlab{b}}){Damour}, {Piazza}, \&
  {Veneziano}}]{DPV2002b}
{Damour}, T., {Piazza}, F., \& {Veneziano}, G. 2002{\natexlab{b}}, \prd, 66,
  046007

\bibitem[{{Damour} \& {Polyakov}(1994)}]{DP94}
{Damour}, T. \& {Polyakov}, A.~M. 1994, Nuclear Physics B, 95, 10347

\bibitem[{{Dixit} \& {Sher}(1988)}]{dixit88}
{Dixit}, V.~V. \& {Sher}, M. 1988, \prd, 37, 1097

\bibitem[{{Drake} \& {Morton}(2007)}]{DM07}
{Drake}, G.~W.~F. \& {Morton}, D.~C. 2007, Astrophys.J.Suppl.Ser., 170, 251

\bibitem[{{Drake} {et al.}(1969){Drake}, {Victor}, \& {Dalgarno}}]{Drake69}
{Drake}, G.~W.~F., {Victor}, G.~A., \& {Dalgarno}, A. 1969, Physical Review,
  180, 25

\bibitem[{{Dunkley} {et al.}(2009){Dunkley}, {Komatsu}, {Nolta}, {Spergel},
  {Larson}, {Hinshaw}, {Page}, {Bennett}, {Gold}, {Jarosik}, {Weiland},
  {Halpern}, {Hill}, {Kogut}, {Limon}, {Meyer}, {Tucker}, {Wollack}, \&
  {Wright}}]{wmap5}
{Dunkley}, J., {Komatsu}, E., {Nolta}, M.~R., {et al.} 2009, \apjs, 180, 306

\bibitem[{{Dupac}(2007)}]{planck}
{Dupac}, X. 2007, ArXiv Astrophysics e-prints

\bibitem[{{Fischer et al.}(2004)}]{Fischer04}
{Fischer et al.} 2004, Physical Review Letters, 92, 230802

\bibitem[{{Fixsen} {et al.}(1996){Fixsen}, {Cheng}, {Gales}, {Mather},
  {Shafer}, \& {Wright}}]{Fixsen96}
{Fixsen}, D.~J., {Cheng}, E.~S., {Gales}, J.~M., {et al.} 1996, \apj, 473, 576

\bibitem[{{Flambaum} \& {Kozlov}(2007)}]{FK07}
{Flambaum}, V.~V. \& {Kozlov}, M.~G. 2007, \prd, 98, 240801

\bibitem[{{Fraisse} {et al.}(2008){Fraisse}, {Brown}, {Dobler}, {Dotson},
  {Draine}, {Frisch}, {Haverkorn}, {Hirata}, {Jansson}, {Lazarian},
  {Magalh{\~a}es}, {Waelkens}, \& {Wolleben}}]{cmbpol}
{Fraisse}, A.~A., {Brown}, J.~.~C., {Dobler}, G., {et al.} 2008, ArXiv e-prints

\bibitem[{{Freedman} {et al.}(2001){Freedman}, {Madore}, {Gibson}, {Ferrarese},
  {Kelson}, {Sakai}, {Mould}, {Kennicutt}, {Ford}, {Graham}, {Huchra},
  {Hughes}, {Illingworth}, {Macri}, \& {Stetson}}]{hst01}
{Freedman}, W.~L., {Madore}, B.~F., {Gibson}, B.~K., {et al.} 2001,
  Astrophys.J., 553, 47

\bibitem[{{Fujii} {et al.}(2000){Fujii}, {Iwamoto}, {Fukahori}, {Ohnuki},
  {Nakagawa}, {Hidaka}, {Oura}, \& {M{\" o}ller}}]{Fujii00}
{Fujii}, Y., {Iwamoto}, A., {Fukahori}, T., {et al.} 2000, Nuclear Physics B,
  573, 377

\bibitem[{Gelman(1996)}]{Gelman96}
Gelman, A. 1996, in Markov Chain Monte Carlo in Practice, ed. W.~Gilks,
  S.~Richardson, \& D.~J. Spiegelhalter (Chapman and Hall), 131

\bibitem[{Gilks {et al.}(1996)Gilks, Richardson, \& Spiegelhalter}]{GRS96}
Gilks, W., Richardson, S., \& Spiegelhalter, D.~J. 1996, in Markov Chain Monte
  Carlo in Practice, ed. W.~Gilks, S.~Richardson, \& D.~J. Spiegelhalter
  (Chapman and Hall), 1

\bibitem[{{Gleiser} \& {Taylor}(1985)}]{GT85}
{Gleiser}, M. \& {Taylor}, J.~G. 1985, \prd, 31, 1904

\bibitem[{{Goldman}(1989)}]{Goldman89}
{Goldman}, S.~P. 1989, \pra, 40, 1185

\bibitem[{{Grainge} {et al.}(2003){Grainge}, {Carreira}, {Cleary}, {Davies},
  {Davis}, {Dickinson}, {Genova-Santos}, {Guti{\'e}rrez}, {Hafez}, {Hobson},
  {Jones}, {Kneissl}, {Lancaster}, {Lasenby}, {Leahy}, {Maisinger}, {Pooley},
  {Rebolo}, {Rubi{\~n}o-Martin}, {Sosa Molina}, {{\"O}dman}, {Rusholme},
  {Saunders}, {Savage}, {Scott}, {Slosar}, {Taylor}, {Titterington}, {Waldram},
  {Watson}, \& {Wilkinson}}]{VSA}
{Grainge}, K., {Carreira}, P., {Cleary}, K., {et al.} 2003, \mnras, 341, L23

\bibitem[{{Halverson} {et al.}(2002){Halverson}, {Leitch}, {Pryke}, {Kovac},
  {Carlstrom}, {Holzapfel}, {Dragovan}, {Cartwright}, {Mason}, {Padin},
  {Pearson}, {Readhead}, \& {Shepherd}}]{DASI}
{Halverson}, N.~W., {Leitch}, E.~M., {Pryke}, C., {et al.} 2002, \apj, 568, 38

\bibitem[{{Hannestad}(1999)}]{Hannestad99}
{Hannestad}, S. 1999, \prd, 60, 023515

\bibitem[{{Harder} \& {Schubert}(2001)}]{harder_schubert}
{Harder}, H. \& {Schubert}, G. 2001, Icarus, 151, 118

\bibitem[{{Hinshaw} {et al.}(2003){Hinshaw}, {Spergel}, {Verde}, {Hill},
  {Meyer}, {Barnes}, {Bennett}, {Halpern}, {Jarosik}, {Kogut}, {Komatsu},
  {Limon}, {Page}, {Tucker}, {Weiland}, {Wollack}, \& {Wright}}]{wmap1}
{Hinshaw}, G., {Spergel}, D., {Verde}, L., {et al.} 2003, \apjs, 148, 135

\bibitem[{{Hirata} \& {Switzer}(2008)}]{SH08b}
{Hirata}, C.~M. \& {Switzer}, E.~R. 2008, \prd, 77, 083007

\bibitem[{{Hummer}(1994)}]{Hummer94}
{Hummer}, D.~G. 1994, \mnras, 268, 109

\bibitem[{{Hummer} \& {Storey}(1998)}]{HummerStorey98}
{Hummer}, D.~G. \& {Storey}, P.~J. 1998, \mnras, 297, 1073

\bibitem[{{Ichikawa} {et al.}(2006){Ichikawa}, {Kanzaki}, \&
  {Kawasaki}}]{Ichi06}
{Ichikawa}, K., {Kanzaki}, T., \& {Kawasaki}, M. 2006, \prd, 74, 023515

\bibitem[{{Ivanchik et al.}(2003)}]{Ivanchik03}
{Ivanchik et al.} 2003, Astrophysics and Space Science, 283, 583

\bibitem[{{Ivanchik et al.}(2005)}]{Ivanchik05}
{Ivanchik et al.} 2005, \aap, 440, 45

\bibitem[{Jones {et al.}(2006)}]{BOOM05_temp}
Jones, W.~C. {et al.} 2006, \apj, 647, 823

\bibitem[{Kaluza(1921)}]{Kaluza}
Kaluza, T. 1921, Sitzungber. Preuss. Akad. Wiss.K, 1, 966

\bibitem[{{Kaplinghat} {et al.}(1999){Kaplinghat}, {Scherrer}, \&
  {Turner}}]{Turner}
{Kaplinghat}, M., {Scherrer}, R., \& {Turner}, M. 1999, \prd, 60, 023516

\bibitem[{{Kehagias} \& {Kiritsis}(1999)}]{KehagiasKiritsis99}
{Kehagias}, A.~A. \& {Kiritsis}, E. 1999, Journal of High Energy Physics, 11,
  22

\bibitem[{{Keiser} \& {Faller}(1982)}]{KeiFall82}
{Keiser}, G.~M. \& {Faller}, J.~E. 1982, in Proceedings of the Second Marcel
  Grossman Meeting on General Relativity, ed. R.~{Ruffin}, "969--976"

\bibitem[{{Kholupenko} {et al.}(2007){Kholupenko}, {Ivanchik}, \&
  {Varshalovich}}]{KIV07}
{Kholupenko}, E.~E., {Ivanchik}, A.~V., \& {Varshalovich}, D.~A. 2007, \mnras,
  378, L39

\bibitem[{{King} {et al.}(2008){King}, {Webb}, {Murphy}, \&
  {Carswell}}]{King08}
{King}, J.~A., {Webb}, J.~K., {Murphy}, M.~T., \& {Carswell}, R.~F. 2008,
  Physical Review Letters, 101, 251304

\bibitem[{Klein(1926)}]{Klein}
Klein, O. 1926, Z. Phys., 37, 895

\bibitem[{{Kujat} \& {Scherrer}(2000)}]{KS00}
{Kujat}, J. \& {Scherrer}, R.~J. 2000, \prd, 62, 023510

\bibitem[{{Kuo} {et al.}(2002){Kuo}, {Ade}, {Bock}, {Daub}, {Goldstein},
  {Holzapfel}, {Lange}, {Newcomb}, {Peterson}, {Ruhl}, {Runyan}, \&
  {Torbet}}]{ACBAR}
{Kuo}, C.-L., {Ade}, P., {Bock}, J.~J., {et al.} 2002, in Bulletin of the
  American Astronomical Society, Vol.~34, Bulletin of the American Astronomical
  Society, 1324--+

\bibitem[{Kuo {et al.}(2004)}]{ACBAR02}
Kuo, C.-l. {et al.} 2004, Astrophys. J., 600, 32

\bibitem[{{Kuskov} \& {Kronrod}(1998)}]{Kuskov_kronrod}
{Kuskov}, O.~L. \& {Kronrod}, V.~A. 1998, Physics of the Earth and Planetary
  Interiors, 107, 285

\bibitem[{{Landau} {et al.}(2008){Landau}, {Mosquera}, {Sc{\'o}ccola}, \&
  {Vucetich}}]{landau08}
{Landau}, S.~J., {Mosquera}, M.~E., {Sc{\'o}ccola}, C.~G., \& {Vucetich}, H.
  2008, \prd, 78, 083527

\bibitem[{{{Lay}, T. and {Wallace}, T.}(1995)}]{lay_wallace}
{{Lay}, T. and {Wallace}, T.} 1995, Modern Global Seismology., Vol.~58
  (International Geophysics series. Academic Press)

\bibitem[{{Lee} {et al.}(1999){Lee}, {Ade}, {Balbi}, {Bock}, {Borrill},
  {Boscaleri}, {Crill}, {de Bernardis}, {del Castillo}, {Ferreira}, {Ganga},
  {Hanany}, {Hristov}, {Jaffe}, {Lange}, {Mauskopf}, {Netterfield}, {Oh},
  {Pascale}, {Rabii}, {Richards}, {Ruhl}, {Smoot}, \& {Winant}}]{MAXIMA}
{Lee}, A.~T., {Ade}, P., {Balbi}, A., {et al.} 1999, in American Institute of
  Physics Conference Series, Vol. 476, 3K cosmology, ed. L.~{Maiani},
  F.~{Melchiorri}, \& N.~{Vittorio}, 224--+

\bibitem[{{Levshakov et al.}(2002)}]{LE02}
{Levshakov et al.} 2002, Mon.Not.R.Astron.Soc., 333, 373

\bibitem[{{Lewis} \& {Bridle}(2002)}]{LB02}
{Lewis}, A. \& {Bridle}, S. 2002, \prd, 66, 103511

\bibitem[{{Lewis} {et al.}(2000){Lewis}, {Challinor}, \& {Lasenby}}]{LCL00}
{Lewis}, A., {Challinor}, A., \& {Lasenby}, A. 2000, \apj, 538, 473

\bibitem[{{Maeda}(1988)}]{Maeda88}
{Maeda}, K. 1988, Modern Physics. Letters A, 31, 243

\bibitem[{{Marion} {et al.}(2003){Marion}, {Pereira Dos Santos}, {Abgrall},
  {Zhang}, {Sortais}, {Bize}, {Maksimovic}, {Calonico}, {Gr{\" u}nert},
  {Mandache}, {Lemonde}, {Santarelli}, {Laurent}, {Clairon}, \&
  {Salomon}}]{Marion03}
{Marion}, H., {Pereira Dos Santos}, F., {Abgrall}, M., {et al.} 2003, \prl, 90,
  150801

\bibitem[{{Mart{\'{\i}}nez Fiorenzano} {et al.}(2003){Mart{\'{\i}}nez
  Fiorenzano}, {Vladilo}, \& {Bonifacio}}]{MVB04}
{Mart{\'{\i}}nez Fiorenzano}, A.~F., {Vladilo}, G., \& {Bonifacio}, P. 2003,
  Societa Astronomica Italiana Memorie Supplement, 3, 252

\bibitem[{{Martins} {et al.}(2002){Martins}, {Melchiorri}, {Trotta}, {Bean},
  {Rocha}, {Avelino}, \& {Viana}}]{Martins02}
{Martins}, C.~J.~A.~P., {Melchiorri}, A., {Trotta}, R., {et al.} 2002,
  Phys.Rev.D, 66, 023505

\bibitem[{{Mason} {et al.}(2001){Mason}, {Pearson}, {Readhead}, {Shepherd},
  {Sievers}, {Udomprasert}, {Cartwright}, \& {Padin}}]{CBI}
{Mason}, B.~S., {Pearson}, T.~J., {Readhead}, A.~C.~S., {et al.} 2001, in
  American Institute of Physics Conference Series, Vol. 586, 20th Texas
  Symposium on relativistic astrophysics, ed. J.~C. {Wheeler} \& H.~{Martel},
  178--+

\bibitem[{{Mather} {et al.}(1994){Mather}, {Cheng}, {Cottingham}, {Eplee},
  {Fixsen}, {Hewagama}, {Isaacman}, {Jensen}, {Meyer}, {Noerdlinger}, {Read},
  {Rosen}, {Shafer}, {Wright}, {Bennett}, {Boggess}, {Hauser}, {Kelsall},
  {Moseley}, {Silverberg}, {Smoot}, {Weiss}, \& {Wilkinson}}]{FIRAS}
{Mather}, J.~C., {Cheng}, E.~S., {Cottingham}, D.~A., {et al.} 1994, \apj, 420,
  439

\bibitem[{{McElhinny} {et al.}(1978){McElhinny}, {Taylor}, \&
  {Stevenson}}]{McElhinny}
{McElhinny}, M.~W., {Taylor}, S.~R., \& {Stevenson}, D.~J. 1978, Nature, 271,
  316

\bibitem[{{Mel{\' e}ndez} \& {Ram{\'{\i}}rez}(2004)}]{MR04}
{Mel{\' e}ndez}, J. \& {Ram{\'{\i}}rez}, I. 2004, Astrophys.J.Lett., 615, L33

\bibitem[{{Mosquera} {et al.}(2008){Mosquera}, {Sc{\'o}ccola}, {Landau}, \&
  {Vucetich}}]{Mosquera07}
{Mosquera}, M.~E., {Sc{\'o}ccola}, C.~G., {Landau}, S.~J., \& {Vucetich}, H.
  2008, Astronomy and Astrophysics, 478, 675

\bibitem[{{Murphy} {et al.}(2008){Murphy}, {Flambaum}, {Muller}, \&
  {Henkel}}]{Murphy08}
{Murphy}, M.~T., {Flambaum}, V.~V., {Muller}, S., \& {Henkel}, C. 2008,
  Science, 320, 1611

\bibitem[{{Murphy} {et al.}(2003){Murphy}, {Webb}, \& {Flambaum}}]{Murphy03b}
{Murphy}, M.~T., {Webb}, J.~K., \& {Flambaum}, V.~V. 2003,
  Mon.Not.R.Astron.Soc., 345, 609

\bibitem[{{Murphy} {et al.}(2001{\natexlab{a}}){Murphy}, {Webb}, {Flambaum},
  {Dzuba}, {Churchill}, {Prochaska}, {Barrow}, \& {Wolfe}}]{Murphy01a}
{Murphy}, M.~T., {Webb}, J.~K., {Flambaum}, V.~V., {et al.} 2001{\natexlab{a}},
  Mon.Not.R.Astron.Soc., 327, 1208

\bibitem[{{Murphy} {et al.}(2001{\natexlab{b}}){Murphy}, {Webb}, {Flambaum},
  {Prochaska}, \& {Wolfe}}]{Murphy01b}
{Murphy}, M.~T., {Webb}, J.~K., {Flambaum}, V.~V., {Prochaska}, J.~X., \&
  {Wolfe}, A.~M. 2001{\natexlab{b}}, Mon.Not.R.Astron.Soc., 327, 1237

\bibitem[{{Nakashima} {et al.}(2008){Nakashima}, {Nagata}, \&
  {Yokoyama}}]{nakashima08}
{Nakashima}, M., {Nagata}, R., \& {Yokoyama}, J. 2008, Progress of Theoretical
  Physics, 120, 1207

\bibitem[{{Netterfield} {et al.}(2002){Netterfield}, {Ade}, {Bock}, {Bond},
  {Borrill}, {Boscaleri}, {Coble}, {Contaldi}, {Crill}, {de Bernardis},
  {Farese}, {Ganga}, {Giacometti}, {Hivon}, {Hristov}, {Iacoangeli}, {Jaffe},
  {Jones}, {Lange}, {Martinis}, {Masi}, {Mason}, {Mauskopf}, {Melchiorri},
  {Montroy}, {Pascale}, {Piacentini}, {Pogosyan}, {Pongetti}, {Prunet},
  {Romeo}, {Ruhl}, \& {Scaramuzzi}}]{BOOMERANG}
{Netterfield}, C.~B., {Ade}, P.~A.~R., {Bock}, J.~J., {et al.} 2002, \apj, 571,
  604

\bibitem[{{North} {et al.}(2008){North}, {Johnson}, {Ade}, {Audley}, {Baines},
  {Battye}, {Brown}, {Cabella}, {Calisse}, {Challinor}, {Duncan}, {Ferreira},
  {Gear}, {Glowacka}, {Goldie}, {Grimes}, {Halpern}, {Haynes}, {Hilton},
  {Irwin}, {Jones}, {Lasenby}, {Leahy}, {Leech}, {Maffei}, {Mauskopf},
  {Melhuish}, {O'Dea}, {Parsley}, {Piccirillo}, {Pisano}, {Reintsema},
  {Savini}, {Sudiwala}, {Sutton}, {Taylor}, {Teleberg}, {Titterington},
  {Tsaneva}, {Tucker}, {Watson}, {Withington}, {Yassin}, \& {Zhang}}]{clover}
{North}, C.~E., {Johnson}, B.~R., {Ade}, P.~A.~R., {et al.} 2008, ArXiv
  e-prints

\bibitem[{{Olive et al.}(2004)}]{Olive04b}
{Olive et al.} 2004, \prd, 69, 027701

\bibitem[{{Overduin} \& {Wesson}(1997)}]{OW97}
{Overduin}, J.~M. \& {Wesson}, P.~S. 1997, Phys.Rep., 283, 303

\bibitem[{{Palma} {et al.}(2003){Palma}, {Brax}, {Davis}, \& {van de
  Bruck}}]{branes03a}
{Palma}, G.~A., {Brax}, P., {Davis}, A.~C., \& {van de Bruck}, C. 2003, \prd,
  68, 123519

\bibitem[{{Peebles}(1968)}]{Peebles68}
{Peebles}, P.~J.~E. 1968, \apj, 153, 1

\bibitem[{{Peik et al.}(2004)}]{Peik04}
{Peik et al.} 2004, Physical Review Letters, 93, 170801

\bibitem[{{Pequignot} {et al.}(1991){Pequignot}, {Petitjean}, \&
  {Boisson}}]{Pequignot91}
{Pequignot}, D., {Petitjean}, P., \& {Boisson}, C. 1991, \aap, 251, 680

\bibitem[{Piacentini {et al.}(2006)}]{BOOM05_polar}
Piacentini, F. {et al.} 2006, \apj, 647, 833

\bibitem[{{Potekhin et al.}(1998)}]{P98}
{Potekhin et al.} 1998, Astrophys.J., 505, 523

\bibitem[{{Prestage} {et al.}(1995){Prestage}, {Tjoelker}, \& {Maleki}}]{PTM95}
{Prestage}, J.~D., {Tjoelker}, R.~L., \& {Maleki}, L. 1995, \prl, 74, 3511

\bibitem[{{Prodanovi{\'c}} \& {Fields}(2007)}]{PF07}
{Prodanovi{\'c}}, T. \& {Fields}, B.~D. 2007, \prd, 76, 083003

\bibitem[{{Quast} {et al.}(2004){Quast}, {Reimers}, \& {Levshakov}}]{QRL04}
{Quast}, R., {Reimers}, D., \& {Levshakov}, S.~A. 2004, \aap, 415, L7

\bibitem[{Raftery \& Lewis(1992)}]{Raftery&Lewis}
Raftery, A.~E. \& Lewis, S.~M. 1992, in Bayesian Statistics, ed. J.~M. Bernado
  (OUP), 765

\bibitem[{Readhead {et al.}(2004)}]{CBI04}
Readhead, A. C.~S. {et al.} 2004, Astrophys. J., 609, 498

\bibitem[{{Richard} {et al.}(2005){Richard}, {Michaud}, \&
  {Richer}}]{richard05}
{Richard}, O., {Michaud}, G., \& {Richer}, J. 2005, Astrophys.J., 619, 538

\bibitem[{{Rocha} {et al.}(2003){Rocha}, {Trotta}, {Martins}, {Melchiorri},
  {Avelino}, \& {Viana}}]{Rocha03}
{Rocha}, G., {Trotta}, R., {Martins}, C.~J.~A.~P., {et al.} 2003, New Astronomy
  Review, 47, 863

\bibitem[{Roll {et al.}(1964)Roll, Krotkov, \& Dicke}]{RKD64}
Roll, P.~G., Krotkov, R., \& Dicke, R.~H. 1964, Ann. Phys., 26, 442

\bibitem[{Sakurai(1967)}]{Sakurai}
Sakurai, J.~J. 1967, Advanced Quantum Mechanics (Addison-Wesley Publishing
  Company)

\bibitem[{{Samtleben}(2008)}]{quiet}
{Samtleben}, D. e.~a. 2008, ArXiv e-prints

\bibitem[{{Schramm} \& {Turner}(1998)}]{SchrammTurner98}
{Schramm}, D. N. \& {Turner}, M.~S. 1998, Rev. Mod. Phys., 70, 303

\bibitem[{{Sc{\'o}ccola} {et al.}(2008{\natexlab{a}}){Sc{\'o}ccola}, {Landau},
  \& {Vucetich}}]{Scoccola08b}
{Sc{\'o}ccola}, C.~G., {Landau}, S.~J., \& {Vucetich}, H. 2008{\natexlab{a}},
  Physics Letters B, 669, 212

\bibitem[{{Sc{\'o}ccola} {et al.}(2008{\natexlab{b}}){Sc{\'o}ccola},
  {Mosquera}, {Landau}, \& {Vucetich}}]{Scoccola08}
{Sc{\'o}ccola}, C.~G., {Mosquera}, M.~E., {Landau}, S.~J., \& {Vucetich}, H.
  2008{\natexlab{b}}, Astrophysical Journal, 681, 737

\bibitem[{{Seager} {et al.}(1999{\natexlab{a}}){Seager}, {Sasselov}, \&
  {Scott}}]{seager99}
{Seager}, S., {Sasselov}, D.~D., \& {Scott}, D. 1999{\natexlab{a}}, \apjl, 523,
  L1

\bibitem[{{Seager} {et al.}(1999{\natexlab{b}}){Seager}, {Sasselov}, \&
  {Scott}}]{recfast}
{Seager}, S., {Sasselov}, D.~D., \& {Scott}, D. 1999{\natexlab{b}},
  Astrophys.J.Lett., 523, L1

\bibitem[{{Seager} {et al.}(2000){Seager}, {Sasselov}, \& {Scott}}]{seager00}
{Seager}, S., {Sasselov}, D.~D., \& {Scott}, D. 2000, \apjs, 128, 407

\bibitem[{{Seaton}(1959)}]{seaton59}
{Seaton}, M.~J. 1959, Mon. Not. Roy. Astron. Soc., 119, 81

\bibitem[{{Sisterna} \& {Vucetich}(1990)}]{SV90}
{Sisterna}, P. \& {Vucetich}, H. 1990, Phys.Rev.D, 41, 1034

\bibitem[{{Smoot}(1993)}]{DMR}
{Smoot}, G.~F. 1993, in Astronomical Society of the Pacific Conference Series,
  Vol.~51, Observational Cosmology, ed. G.~L. {Chincarini}, A.~{Iovino},
  T.~{Maccacaro}, \& D.~{Maccagni}, 477--+

\bibitem[{{Sortais} {et al.}(2000){Sortais}, {Bize}, {Abgrall}, {Zhang},
  {Nicolas}, {Mandache}, P, {Laurent}, {Santarelli}, {Dimarcq}, {Petit},
  {Clairon}, {Mann}, {Luiten}, {Chang}, \& {Salomon}}]{Sortais00}
{Sortais}, Y., {Bize}, S., {Abgrall}, M., {et al.} 2000, Physica Scripta, T95,
  50

\bibitem[{{Spergel} {et al.}(2007){Spergel}, {Bean}, {Dor{\'e}}, {Nolta},
  {Bennett}, {Dunkley}, {Hinshaw}, {Jarosik}, {Komatsu}, {Page}, {Peiris},
  {Verde}, {Halpern}, {Hill}, {Kogut}, {Limon}, {Meyer}, {Odegard}, {Tucker},
  {Weiland}, {Wollack}, \& {Wright}}]{wmap3a}
{Spergel}, D., {Bean}, R., {Dor{\'e}}, O., {et al.} 2007,
  Astrophys.J.Suppl.Ser., 170, 377

\bibitem[{{Spitzer} \& {Greenstein}(1951)}]{SpitzerGreenstein51}
{Spitzer}, L.~J. \& {Greenstein}, J.~L. 1951, \apj, 114, 407

\bibitem[{{Spohn} {et al.}(2001){Spohn}, {Sohl}, {Wieczerkowski}, \&
  {Conzelmann}}]{spohn}
{Spohn}, T., {Sohl}, F., {Wieczerkowski}, K., \& {Conzelmann}, V. 2001,
  Planetary and Space Science, 49, 1561

\bibitem[{{Srianand} {et al.}(2004){Srianand}, {Chand}, {Petitjean}, \&
  {Aracil}}]{Srianand04}
{Srianand}, R., {Chand}, H., {Petitjean}, P., \& {Aracil}, B. 2004, \prl, 92,
  121302

\bibitem[{{Stefanescu}(2007)}]{stefanescu07}
{Stefanescu}, P. 2007, New Astronomy, 12, 635

\bibitem[{Su {et al.}(1994)}]{Su94}
Su, Y. {et al.} 1994, Phys. Rev., D50, 3614

\bibitem[{{Switzer} \& {Hirata}(2008{\natexlab{a}})}]{SH08a}
{Switzer}, E.~R. \& {Hirata}, C.~M. 2008{\natexlab{a}}, \prd, 77, 083006

\bibitem[{{Switzer} \& {Hirata}(2008{\natexlab{b}})}]{SH08c}
{Switzer}, E.~R. \& {Hirata}, C.~M. 2008{\natexlab{b}}, \prd, 77, 083008

\bibitem[{{Takahashi} {et al.}(2008){Takahashi}, {Barkats}, {Battle},
  {Bierman}, {Bock}, {Chiang}, {Dowell}, {Hivon}, {Holzapfel}, {Hristov},
  {Jones}, {Kaufman}, {Keating}, {Kovac}, {Kuo}, {Lange}, {Leitch}, {Mason},
  {Matsumura}, {Nguyen}, {Ponthieu}, {Rocha}, {Yoon}, {Ade}, \&
  {Duband}}]{bicep}
{Takahashi}, Y.~D., {Barkats}, D., {Battle}, J.~O., {et al.} 2008, in Society
  of Photo-Optical Instrumentation Engineers (SPIE) Conference Series, Vol.
  7020, Society of Photo-Optical Instrumentation Engineers (SPIE) Conference
  Series

\bibitem[{{Tzanavaris et al.}(2007)}]{Tzana07}
{Tzanavaris et al.} 2007, \mnras, 374, 634

\bibitem[{{Varshalovich} \& {Levshakov}(1993)}]{VL93}
{Varshalovich}, D.~A. \& {Levshakov}, S.~A. 1993, Soviet Journal of
  Experimental and Theoretical Physics Letters, 58, 237

\bibitem[{{Verner} \& {Ferland}(1996)}]{VernerFerland96}
{Verner}, D.~A. \& {Ferland}, G.~J. 1996, \apjs, 103, 467

\bibitem[{{Villa} {et al.}(2003){Villa}, {Mandolesi}, \& {Butler}}]{Villa03}
{Villa}, F., {Mandolesi}, N., \& {Butler}, R.~C. 2003, Memorie della Societa
  Astronomica Italiana, 74, 223

\bibitem[{{Webb} {et al.}(1999){Webb}, {Flambaum}, {Churchill}, {Drinkwater},
  \& {Barrow}}]{Webb99}
{Webb}, J.~K., {Flambaum}, V.~V., {Churchill}, C.~W., {Drinkwater}, M.~J., \&
  {Barrow}, J.~D. 1999, \prl, 82, 884

\bibitem[{{Webb} {et al.}(2001){Webb}, {Murphy}, {Flambaum}, {Dzuba}, {Barrow},
  {Churchill}, {Prochaska}, \& {Wolfe}}]{Webb01}
{Webb}, J.~K., {Murphy}, M.~T., {Flambaum}, V.~V., {et al.} 2001, \prl, 87,
  091301

\bibitem[{{Weinberg}(1983)}]{Weinberg83}
{Weinberg}, S. 1983, Physics Letters B, 125, 265

\bibitem[{{White}(2006)}]{White06}
{White}, M. 2006, New Astronomy Review, 50, 938

\bibitem[{{Wieczorek} {et al.}(2006){Wieczorek}, L., {Khan}, {Pritchard},
  {Weiss}, {Williams}, {Righter}, {Neal}, {Shearer}, {McCallum}, {Tompkins},
  {Hawke}, {Peterson}, {Gillis}, \& {Bussey}}]{Wieczorek}
{Wieczorek}, M.~A., L., J.~B., {Khan}, A., {et al.} 2006, Reviews in Mineralogy
  and Geochemistry, 60, 221

\bibitem[{{Will}(1981)}]{will}
{Will}, C.~M. 1981, Theory and Experiment in Gravitational Physics. (Cambridge
  U. Press)

\bibitem[{{Wilson} \& {Penzias}(1967)}]{PW68}
{Wilson}, R.~W. \& {Penzias}, A.~A. 1967, Science, 156, 1100

\bibitem[{{Wong} {et al.}(2008){Wong}, {Moss}, \& {Scott}}]{wong08}
{Wong}, W.~Y., {Moss}, A., \& {Scott}, D. 2008, \mnras, 386, 1023

\bibitem[{{Wu} \& {Wang}(1986)}]{Wu86}
{Wu}, Y. \& {Wang}, Z. 1986, \prl, 57, 1978

\bibitem[{{Yoo} \& {Scherrer}(2003)}]{YS03}
{Yoo}, J.~J. \& {Scherrer}, R.~J. 2003, Phys.Rev.D, 67, 043517

\bibitem[{{Youm}(2001{\natexlab{a}})}]{Youm2001a}
{Youm}, D. 2001{\natexlab{a}}, \prd, 63, 125011

\bibitem[{{Youm}(2001{\natexlab{b}})}]{Youm2001b}
{Youm}, D. 2001{\natexlab{b}}, \prd, 64, 085011

\bibitem[{{Zeldovich} {et al.}(1968){Zeldovich}, {Kurt}, \&
  {Syunyaev}}]{zeldovich68}
{Zeldovich}, Y.~B., {Kurt}, V.~G., \& {Syunyaev}, R.~A. 1968, Zhurnal
  Eksperimental noi i Teoreticheskoi Fiziki, 55, 278

\end{thebibliography}


\end{document}